\documentclass[11pt]{article} 
\usepackage[margin=1in]{geometry} 
\pdfoutput=1 

\newif\ifanon
\anonfalse  
\def\papertitle{The Metacomplexity Structure of Search Problems} 
\def\authornames{Shalev Ben{-}David and M. H. Ebtehaj} 

\usepackage{silence} 

\usepackage[dvipsnames,svgnames,table]{xcolor} 
\usepackage[obeyFinal]{todonotes}

\usepackage{amsmath,amsfonts,amssymb,amsthm}
\usepackage{mathtools,thmtools}

\usepackage{graphicx}
\usepackage{tikz,tikz-qtree} 
\usetikzlibrary{cd}

\usepackage{mmap}
\usepackage[T1]{fontenc}

\usepackage[utf8]{inputenc}
\usepackage{csquotes}
\usepackage[english]{babel}

\usepackage[
    activate={true,nocompatibility},
    final, 
    tracking=true,
    kerning=true,
    spacing=true,
    factor=1100,
    stretch=12,
    shrink=10
    ]{microtype}
\microtypecontext{spacing=nonfrench}

\usepackage[backend=biber,
              style=alphabetic,
        maxbibnames=99,
      minalphanames=3,
      maxalphanames=4,
            backref=true]{biblatex}

\ifanon
\usepackage{hyperref}
\hypersetup{
    pdftitle={\papertitle}, 
    pdfauthor={Anonymous submission}, 
    colorlinks=true, 
    linkcolor=NavyBlue, 
    citecolor=Mahogany, 
    urlcolor=ForestGreen 
}
\else
\usepackage{hyperref}
\hypersetup{
    pdftitle={\papertitle}, 
    pdfauthor={\authornames}, 
    colorlinks=true, 
    linkcolor=NavyBlue, 
    citecolor=Mahogany, 
    urlcolor=ForestGreen 
}
\fi

\usepackage[capitalise,nameinlink]{cleveref}

\AtEveryBibitem{
 \clearlist{address}
 \clearfield{date}
 \clearlist{location}
 \clearfield{month}
 \clearfield{day}
 \clearfield{series}
 \clearfield{pages}
 \clearlist{organization}
 \clearfield{number}
 \clearlist{language}

 \ifentrytype{book}{}{ 
  \clearlist{publisher}
  \clearname{editor}
  \clearfield{issn}
  \clearfield{isbn}
  \clearfield{volume}
 }
}

\DeclareFieldFormat{eprint:eccc}{
\mkbibacro{ECCC}\addcolon\space\ifhyperref
    {\href{http://eccc.hpi-web.de/report/#1/}{\nolinkurl{#1}}}
    {\nolinkurl{#1}}}
\DeclareFieldAlias{eprint:ECCC}{eprint:eccc}
\DeclareFieldFormat{eprint:iacr}{Cryptology ePrint\addcolon\space\ifhyperref
    {\href{https://eprint.iacr.org/#1}{\nolinkurl{#1}}}
    {\nolinkurl{#1}}}
\DeclareFieldAlias{eprint:IACR}{eprint:iacr}
\DeclareFieldFormat{eprint:arxiv}{arXiv\addcolon\space\ifhyperref
    {\href{https://arxiv.org/abs/#1}{\nolinkurl{#1}}}
    {\nolinkurl{#1}}}

\DeclareFieldFormat[article]{title}{#1}
\DeclareFieldFormat[inproceedings]{title}{#1}
\DeclareFieldFormat[incollection]{title}{#1}
\DeclareFieldFormat[online]{title}{#1}

\renewbibmacro{in:}{}

\newcommand{\lName}{1}

\newcommand{\donothing}[1]{#1}

\newcommand{\JACM}{\if\lName1\donothing{Journal of the {ACM}}\else{JACM}\fi}
\newcommand{\SICOMP}{\if\lName1\donothing{{SIAM} Journal on Computing}\else{SICOMP}\fi}
\newcommand{\ToC}{\if\lName1\donothing{Theory of Computing}\else{ToC}\fi}
\newcommand{\ToCGS}{\if\lName1\donothing{Theory of Computing Graduate Surveys}\else{ToC}\fi}
\newcommand{\TOCT}{\if\lName1\donothing{{ACM} Transactions on Computation Theory}\else{TOCT}\fi}
\newcommand{\ToIT}{\if\lName1\donothing{{IEEE} Transactions on Information Theory}\else{TOCT}\fi}
\newcommand{\CCjournal}{\if\lName1\donothing{Computational Complexity}\else{CC}\fi}
\newcommand{\CJTCS}{\if\lName1\donothing{Chicago Journal of Theoretical Computer Science}\else{CJTCS}\fi}

\newcommand{\TCS}{\if\lName1\donothing{Theoretical Computer Science}\else{TCS}\fi}
\newcommand{\IPL}{\if\lName1\donothing{Information Processing Letters}\else{IPL}\fi}
\newcommand{\JCSS}{\if\lName1\donothing{Journal of Computer and System Sciences}\else{JCSS}\fi}

\newcommand{\RSA}{\if\lName1\donothing{Random Structures and Algorithms}\else{RSA}\fi}
\newcommand{\JCTA}{\if\lName1\donothing{Journal of Combinatorial Theory, Series A}\else{JCTA}\fi}
\newcommand{\JCTB}{\if\lName1\donothing{Journal of Combinatorial Theory, Series B}\else{JCTB}\fi}
\newcommand{\PJM}{\if\lName1\donothing{Pacific Journal of Mathematics}\else{PJM}\fi}
\newcommand{\QICjournal}{\if\lName1\donothing{Quantum Information and Computation}\else{QIC}\fi}
\newcommand{\IJQI}{\if\lName1\donothing{International Journal of Quantum Information}\else{IJQI}\fi}
\newcommand{\PRA}{\if\lName1\donothing{Physical Review A}\else{PRA}\fi}
\newcommand{\PRL}{\if\lName1\donothing{Physical Review Letters}\else{PRL}\fi}
\newcommand{\VLDB}{\if\lName1\donothing{International Journal on Very Large Data Bases}\else{VLDB}\fi}
\newcommand{\SIDMA}{\if\lName1\donothing{{SIAM} Journal on Discrete Mathematics}\else{SIDMA}\fi}

\DefineBibliographyStrings{english}{%
  backrefpage = {p{.}},
  backrefpages = {pp{.}},
}

\newif\ifbibnonempty
\bibnonemptyfalse
\AtEveryCitekey{\global\bibnonemptytrue}

\declaretheorem[numberwithin=section]{theorem}

\declaretheorem[sibling=theorem]{corollary}
\declaretheorem[sibling=theorem]{lemma}

\declaretheorem[sibling=theorem,style=definition]{definition}

\declaretheorem[name=Question,sibling=theorem,style=definition]{question}

\DeclareMathAlphabet{\mathbbold}{U}{bbold}{m}{n}

\DeclareMathOperator{\bN}{\mathbb{N}}

\DeclareMathOperator{\downcl}{\downarrow}

\newcommand{\printArxivBibliography}{%
  \renewcommand{\UrlFont}{\ttfamily\small}
  \let\oldpath\path
  \renewcommand{\path}[1]{\small\oldpath{##1}}
  \emergencystretch=2em 

  \ifbibnonempty{ 
  \phantomsection\addcontentsline{toc}{section}{References} 
  \printbibliography 
  }\else{
  \nocite{AA05} 
  }
  \fi
}

\begin{document}

\title{\papertitle}

\ifanon
\author{Anonymous Submission}
\else
\author{
    Shalev Ben{-}David\\
    \small Institute for Quantum Computing\\
    \small University of Waterloo\\
    \small \texttt{shalev.b@uwaterloo.ca}
    \and
    M. H. Ebtehaj\\
    \small Institute for Quantum Computing\\
    \small University of Waterloo\\
    \small \texttt{mhebtehaj@uwaterloo.ca}
}
\fi

\date{}
\maketitle

\begin{abstract}
We study the metacomplexity tasks of refuting an algorithm
for a given search problem or refuting a mapping reduction
between two search problems. We show that these tasks
are deeply related to the structure of the lattice of search
problems under mapping reductions: the ``refutation of a reduction''
task amounts to a certain Heyting-type arrow in this lattice.
We then give several structural complexity results about search
problems and show they have implications for metacomplexity.

First, by relating metacomplexity to a Heyting algebra structure,
we are able to translate results from intuitionistic logic
into complexity theoretic consequences. For example,
the propositional logic identity $x\wedge \neg x=\bot$ translates to
the following complexity-theoretic fact: if a search problem
$R$ reduces to both $S$ and to the refutation task of $S$,
then $R$ is in $\mathsf{FP}$. All such logical identities
that hold in intuitionistic logic translate to complexity in a
black box way.

We find prime ideals of the reducibility lattice of search problems;
as a consequence, we show that for any verifiable search problem
$S$, either $S$ is in $(\mathsf{FP/poly})_{io}$ or else the refutation
task of $S$ is in $(\mathsf{FP/poly})_{io}$. Similarly,
either $S$ has infinitely-often heuristic algorithms or else
the refutation task of $S$ has infinitely often heuristic algorithms.

Finally, we give a structural result that may be of independent
interest: any ``tamely semantic'' class of search problems
(one which can be enumerated using a Halting oracle) is characterized
by its decidably verifiable hard problems.
\end{abstract}

\clearpage
{\small\tableofcontents}
\clearpage

\section{Introduction}

Our work connects structural complexity and metacomplexity:
we show that metacomplexity refutation tasks naturally arise
in the study of structural complexity of search problems.
We start with some background on structural complexity;
in \cref{sec:StructuralSearch}, we make the simple but perhaps
surprising observation that structural complexity is much
nicer for search problems than it is for decision problems.
The main thrust of this work is the observation in
\cref{sec:IntroMetacomplexity}, where we show that a natural
structural complexity definition has a characterization in terms
of refutation tasks from metacomplexity.

\subsection{Background: structural complexity of languages}

Consider mapping reductions as a binary relation on languages.
This relation is reflexive and transitive, making it a 
preorder.\footnote{Preorders are the same as partial orders,
except without the antisymmetry axiom: in a preorder, we may have
$x\le y$ and $y\le x$ but $x\ne y$. Every partial order
is a preorder, and every preorder can be converted to a partial
order by moving to the set of equivalence classes (under
the equivalence relation of ``$x\le y$ and $y\le x$'');
the two notions are therefore essentially the same. We stick to
the term preorder for mapping reductions
since it is more precise.}

When we have a preorder or partial order, a natural question is
whether it satisfies other properties, such as having
meets and joins.\footnote{A meet of two elements $x,y$ in a preorder
is a maximum lower bound for them, that is, an element $z$
which lower bounds both $x$ and $y$ and which is satisfies $w\le z$
for all other lower bounds $w$ on $x$ and $y$. A join
is a minimum upper bound: an element $z$ which upper bounds both
$x$ and $y$ and which satisfies $z\le w$ for all other upper bounds
$w$ on $x$ and $y$. In a partial order, meets and joins
are unique if they exist,
and are denoted $x\wedge y$ and $x\vee y$ respectively.}
It turns out the languages under mapping reductions have joins,
but they are not believed to have meets in general. Explicitly,
a join of languages $L_0$ and $L_1$ is the disjoint union:
the set of strings $\langle x,b\rangle$ where $b\in\{0,1\}$
and $x\in L_b$. If we exclude the empty language
and the set of all strings,
the preorder has bottom elements, which are the languages
in $\mathsf{P}$.
Note that the existence of binary joins and bottom
implies the existence of arbitrary finite
joins, since a bottom element is a join of the empty set.

In any preorder $(L,\le)$ that has all finite joins,
we can define an ideal $I$ to be a subset of $L$ which is nonempty,
downwards closed (meaning $x\in I$ and $y\le x$ implies $y\in I$),
and closed under finite joins
(so $x,y\in I$ implies $(x\vee y)\in I$).
The set of ideals $\overline{L}$ is then partially
ordered under $\subseteq$. This partial order turns out to have
\emph{arbitrary} meets and joins, giving what's called a complete
lattice. Indeed, if $A\subseteq\overline{L}$ is a set of ideals,
then $\bigcap A$ is an ideal which is the meet of the ideals in $A$
(the maximum ideal which is a subset of each element of $A$).
For the join, start with $\bigcup A$, take the closure
under finite joins, and then take the downwards closure; the resulting
ideal $\bigvee A$ is the join of the ideals in $A$.
Note that a complete lattice is necessarily bounded (having
a top element and a bottom element): the top is the join of all
elements, which in this case is the ideal of all elements $L$;
the bottom is the meet of all elements, which is
the set of bottom elements of $L$.

The pair $(\overline{L},\subseteq)$ is called the ideal completion
of $(L,\le)$. It upgrades $L$ from a preorder with finite joins
into a partial order with arbitrary joins and arbitrary meets.

For the preorder of languages under reductions, ideals correspond
to complexity classes. Indeed, complexity classes are nonempty
and downwards closed; moreover, virtually all complexity classes
of interest are closed under joins (for example,
if $L_0$ and $L_1$ are in $\mathsf{NP}$, then
a nondeterministic machine for $L_0\vee L_1$ can first check
that the input is formatted like $\langle b,x\rangle$, then fork
on the bit $b$: if $b=0$, run the $\mathsf{NP}$ for $L_0$ on $x$,
and if $b=1$, run the $\mathsf{NP}$ for $L_1$ on $x$).

An ideal is called \emph{principal} if it is the downwards closure
of a single element of $L$; for languages, this means an ideal
is principal if and only if it has a complete problem
as a complexity class.

Structural complexity studies the preorder of search problems
under mapping reductions and the resulting partial order of
ideals (complexity classes) under subset. Famous results include
Ladner's theorem \cite{Lad75} (which gives intermediate elements between
$x$ and $y$ when $x<y$ under some conditions)
and the exact pair phenomenon of Ambos-Spies
\cite{Amb84}
and Merkle \cite{Mer97} (which says certain syntactic ideals
are an intersection of two
principal ideals; i.e. such classes may not have complete problems,
but they have the next best thing, a pair of hard problems characterizing the class).

\subsection{Structural complexity of search problems}
\label{sec:StructuralSearch}

Instead of languages, consider search problems.
Formally, a search problem $S$ is a set of pairs of strings,
with each pair in the set corresponding to a valid input/output
pair. We generally restrict to verifiable search problems,
which are those that have a polynomial-time verifier for checking
pairs of strings (that is, $S$ is verifiable if it is in $\mathsf{P}$
when considered as a language).
A mapping reduction from search problem $R$ to $S$ is
a pair of polynomial-time computable functions $F$ and $G$
such that for all strings $x,y$, we have
$(F(x),y)\in S\Rightarrow (x,G(x,y))\in R$. Here $F$ converts
an $R$-input into an $S$-input, and $G$ converts back from an
output to $S$ to an output to $R$ (and $G$ is allowed
access to the original $R$-input $x$).
Mapping reducibility is reflexive and transitive,
so search problems form a preorder under reducibility.

\paragraph{First observation:
search problems have both meets and joins.}
Unlike languages, the preorder of search
problems has meets, not just joins.
For search problems $S_0$ and $S_1$, the join will be
``solve either $S_0$ or $S_1$, as the adversary specifies,''
just as for languages. The meet will be
``solve either $S_0$ or $S_1$,
as the algorithm chooses''.
That is, the join will be the set of pairs
\[S_0\vee S_1\coloneqq \{(\langle x,b\rangle,y) :
    b\in\{0,1\}, (x,y)\in S_b\},\]
and the meet will be the set of pairs
\[S_0\wedge S_1\coloneqq
    \{(\langle x_0,x_1\rangle,\langle y,b\rangle) :
        b\in\{0,1\},(x_b,y)\in S_b\}.\]
(For technical reasons, we will also add to these sets
all the pairs of the form
$(w,z)$ whenever $w$ is incorrectly formatted as an input;
this makes poorly-formatted inputs trivial rather than impossible.)
It is not hard to check that the join and meet properties
are respectively satisfied by these constructions. Moreover,
if $S_0$ and $S_1$ are verifiable, so are their meet and join.
Further, one can check that the meet and join operations
satisfy the distributivity property:
$R\wedge (S\vee T)$ is equivalent to $(R\wedge S)\vee (R\wedge T)$
(with reductions in both directions), and similarly,
$R\vee (S\wedge T)$ is equivalent to $(R\vee S)\wedge (R\vee T)$.

\paragraph{Ideal completion.}
As with languages, ideals in the reducibility
preorder of search problems correspond to complexity classes of
search problems. However, this time, the initial preorder has both
meets and joins and satisfies the distributivity property.
For such a preorder $(L,\le)$, the ideal completion
$(\overline{L},\subseteq)$ is even more structured than before:
instead of just being a complete lattice, it is now a complete
Heyting algebra.

A complete Heyting algebra is an interesting
mathematical object that shows up in several different fields.
Most notably, it shows up in mathematical logic, where Heyting
algebras are used to model intuitionistic/constructivist logic.
It also shows up in pointless topology, where a complete Heyting
algebra is known as a \emph{frame} or \emph{locale};\footnote{Frames
and locales mean the same thing, though a morphism of frames
is different from a morphism of locales.} it is the main
object of study in pointless topology, acting as a generalization
of topological spaces.

\paragraph{Second observation:
universe of complexity classes is a complete Heyting algebra.}
As mentioned, the ideals in the reducibility preorder of
search problems are the complexity classes of search problems.
Since search problems have distributive meets and joins,
the collection of complexity classes of search problems
forms a complete Heyting algebra under the $\subseteq$ partial order.
This allows us to employ intuition and tools from fields such as
intuitionistic logic and pointless topology.

\paragraph{Logic interpretation.}
A complete Heyting algebra is a partial order with arbitrary meets,
arbitrary joins, and the following special property:
for any elements $x,y$, the set $\{z:x\wedge z\le y\}$ has a
maximum element, denoted $x\to y$. One can interpret
the elements of a complete Heyting algebra as truth values,
with the bottom element $\bot$ representing absolute falsehood,
the top element $\top$ representing absolute truth, and intermediate
elements representing intermediate truth values (partially ordered).
The meet $x\wedge y$ is then interpreted as a logical AND,
while the join $x\vee y$ is a logical OR. The Heyting arrow
$x\to y$ acts as a logical implication operator on these intermediate
truth values; $(x\to y)$ itself is an element, and hence is itself
interpreted as a truth value, which becomes more true as $y$
gets more true and less true as $x$ gets more true.
Negations are also defined: we use $\neg x$ to mean $x\to \bot$.
Heyting algebras then follow the rules of intuitionistic logic:
for example, $x\le\neg\neg x$ (``$p$ implies $\neg\neg p$''),
but equality need not hold. On the other hand,
$\neg x=\neg\neg\neg x$
(``$\neg p$ is equivalent to $\neg\neg\neg p$'').

In the context of complexity, the elements of our Heyting algebra
will be complexity classes of verifiable search problems. Elements that are ``more true''
will be complexity classes that are
more intractable; the bottom element will
be the class $\mathsf{FP}$ of trivial search
problems, and the top element will be the class
of all search problems.
We are then guaranteed to get operations $\neg\mathsf{C}$
and $\mathsf{A}\to\mathsf{B}$ on complexity classes
which satisfy all the rules of intuitionistic logic.
As we will see in the next section, these operations have
a natural interpretation in terms of \emph{refutation problems},
connecting structural complexity to metacomplexity.

\paragraph{Topology interpretation.}
In pointless topology, we try to do topology by keeping the
notion of open sets but discarding the points inside them.
The elements of a complete Heyting algebra then thought of as
the collection of open sets of a topological space, but for
which we've forgotten what the points were
(and instead, only remember
which open sets are subsets of which other ones and how to take
unions and finite intersections, which are now the joins and meets).

In the context of complexity, the complexity classes become
the ``open sets'' in a locale. Assuming the axiom of choice,
the resulting locale is \emph{spatial},
meaning it comes from an actual topological space.
The points of this topological space are not the search
problems, however; instead, they correspond to \emph{prime ideals}
in the preorder lattice of search problems. That is, points
correspond to complexity classes $\mathsf{C}$ with the property
that if $R\wedge S\in \mathsf{C}$ then either $R\in\mathsf{C}$
or $S\in\mathsf{C}$.
Localic motivation leads us to examine the prime ideals
and dense sublocales, which lead to complexity theoretic consequences
for refutation problems.

\subsection{The metacomplexity angle}
\label{sec:IntroMetacomplexity}

Our original motivation for this project actually came from
metacomplexity and the study of refutation problems,
with the structural connection arising later.

In metacomplexity, an important recent notion is that of
constructing explicit refutations. For example, given
a polynomial time algorithm that claims to solve SAT,
can one construct an explicit input on which it fails? 
We call this the refutation ask,
and denote it by $\neg T$ for a task $T$.
Refutation problems have been studied in several previous works \cite{Kab01,GST07,Ats06,DFG13,CJSW24,FGJLRShY26}, 
and in relation with derandomization \cite{CTW23,CRT25,MM25} 
and in other settings such as bounded arithmetic and proof complexity \cite{LLR24,GC25}.

The refutation problem for a given task is always a search
problem (e.g. given an algorithm for SAT, find an input on which it
fails). This motivates the study of refutation problems for
search problems themselves: this way, we start
and end in the same space of search problems.

\paragraph{The simulation issue.}
Attempting to formalize the refutation task immediately leads
one to encounter an annoying technical problem:
it's not possible to take a polynomial time machine as input
and simulate it in polynomial time. This may sound contradictory
at first glance, because a universal Turing machine $U$ can
simulate a given input machine $\langle M\rangle$ with minimal
overhead cost. The problem is the order of quantifiers.
Let us suppose that the machine $M$ comes attached with its runtime
bound, represented by an integer $k$ such that the number
of steps of $M$ is at most $p_k(n)=n^k+k$ on all inputs of
length $n$. On input $\langle M,k,x\rangle$,
the universal machine can simulate $M(x)$ in $\tilde{O}(n^k)$ time.
The problem is that this is exponential in $k$,
which was part of the input! $k$ is constant from the perspective
of $M$ but not from the perspective of $U$.

A few natural attempts to mitigate this all fail. First,
we can try to represent $k$ in unary, like $\langle M,0^k,x\rangle$
or even $\langle M,0^{2^k},x\rangle$. This doesn't help:
the runtime of $U$ is $\tilde{O}(n^k)$, so in terms of its input size
$N$ it is $N^N$ or $N^{\log N}$ depending
on the encoding of $k$. No matter how we represent $k$,
the $k$ in the exponent will be some slow-growing function of
the input size $N$, so not polynomial time.

Another attempt: if $M$ is polynomial time and $U$ runs in
time $N^{O(\log N)}$, we could simply allow quasipolynomial
time algorithms. This works if $M$ is polynomial time and
$U$ is quasipolynomial time, but if we
want them to be of the same type (i.e. to feed $U$-type algorithms
to $U$ as input), this will still fail;
if $M$ is quasipolynomial time,
$U$ will need to be quasiquasipolynomial time.
More generally, one can show that for any subexponential family 
$\{\alpha_k\}_k$ of increasing growth rates that we could
consider ``efficient'', and any encoding $\langle k\rangle$ of $k$,
the runtime of $U(\langle M,\langle k\rangle,x\rangle)=M(x)$
(where $M(x)$ takes at most $\alpha_k(|x|)$ steps)
effectively diagonalizes across the
entire set $\{\alpha_k\}_k$ and is guaranteed not to lie below any
of the $\alpha_k$ growth rates!\footnote{The impossibility
result requires the assumption that the family $\{\alpha_k\}_k$
is invariant to constant factors, that is, for each $k$ there exists
$\ell$ such that $2\alpha_k$ lies below $\alpha_{\ell}$. We do not prove this
impossibility result, but it is not hard.}

\paragraph{The workaround.}
We address the simulation issue by defining a hierarchy of
refutation tasks for each search problem $S$: instead of a single
task $\neg S$, we define a refutation task $\neg_k S$ for each $k$.

As we will see shortly, collecting $\{\neg_k S:k\in\mathbb{N}\}$
into a class will be a natural approach, and the $\neg$ operation
will behave well when acting on \emph{classes}
of search problems rather than on individual search problems;
indeed, it will be exactly the Heyting negation from the previous
section.
We start with the definition of $\neg_k S$ for a single search
problem $S$.

\begin{definition}[Graded refutation tasks]
For any search problem $S$ and any $k\in\mathbb{N}$,
the search problem $\neg_k S$ is defined as follows:
given as input
an algorithm $\langle A\rangle$, find a string $x$ such that
either $A(x)$ runs for more than $O(|x|^k)$ steps,\footnote{Formally,
the bound on the number of steps will actually be
$p_k(|x|+|\langle A\rangle|)$, where recall $p_k(n)=n^k+k$.} or else
$A(x)$ outputs a string $y$ which is not a valid solution for $x$
(with respect to the original search problem $S$).
\end{definition}

A nice property of these ``graded'' refutation tasks $\neg_k S$
is that if $S$ is a verifiable search problem, the tasks
$\neg_k S$ are all also verifiable search problems. Indeed, to verify
that $x$ is a solution to input $\langle A\rangle$ for task
$\neg_k S$, we just run $A(x)$ for $O(|x|^k)$ steps,
check if it halts, and if it outputs a string $y$ we run the
verifier of $S$ on the pair $(x,y)$.

Let $\downcl S$ denote the complexity class of verifiable
search problems for which $S$ is complete: that is,
$\downcl S\coloneqq \{R:R\le S\}$, where $\le$ denotes mapping
reducibility.
We can then define the negation (refutation) operation on
the complexity class $\downcl S$, and subsequently on more
general classes $\mathsf{C}$, as follows:
\[\neg\downcl S\coloneqq \bigcup_{k\in\mathbb{N}}\downcl(\neg_k S),
\qquad
\neg\mathsf{C}\coloneqq \bigcap_{S\in \mathsf{C}}\neg \downcl S.\]
In other words, the negation of a class $\mathsf{C}$
is the class of tasks which are reducible to the refutation
of anything in $\mathsf{C}$.

\paragraph{Central insight:
refutation classes are the Heyting negations.}
Surprisingly, this definition of $\neg\mathsf{C}$
as class-level refutation
turns out to be exactly the same as the negation operation
in the Heyting algebra of complexity classes,
defined in the previous section. Concretely, this means
that $\neg\mathsf{C}$ is the largest ideal whose intersection
with $\mathsf{C}$ is the bottom ideal $\mathsf{FP}$.
In other words, the structural analysis of the reducibility
partial order for search problems naturally
gave rise to the definition of the refutation tasks in metacomplexity.

In particular, this means that refutations obey the rules of
intuitionistic logic! For example, for any complexity class
$\mathsf{C}$, we must have $\mathsf{C}\subseteq \neg\neg\mathsf{C}$
and $\neg\neg\neg\mathsf{C}=\neg\mathsf{C}$. If $S$ is complete for
$\mathsf{C}$, we get the conclusion that for all $k\in\mathbb{N}$
there exists $\ell\in\mathbb{N}$ such that $S$ reduces to
$\neg_\ell\neg_k S$ (and a similar statement for
$\neg_k S$ and $\neg_\ell\neg_m\neg_n S$, in both directions).
We give additional examples in \cref{sec:IntroLogicResults}.

\paragraph{Interpreting the arrows.}
For the Heyting arrows,
we need the following generalization of the graded refutation tasks.
Instead of refuting an \emph{algorithm} for a search problem $S$,
a more general notion is that of refuting a \emph{reduction}
between search problems $R$ and $S$.
We denote the reduction refutation
problem from $R$ to $S$ (time-bounded to $O(n^k)$) by $R\to_k S$.

To formally define the reduction-refutation task, we recall
how mapping reductions work for search problems.
If $R$ and $S$ are search problems, a mapping reduction
is a pair of efficiently computable functions $(F,G)$
such that $(F(x),y)\in S\Rightarrow (x,G(x,y))\in R$
for all strings $x,y$ (i.e. if $y$ is a valid output for $F(x)$,
then $G$ converts it back to a valid output $G(x,y)$ for $x$).

\begin{definition}[Graded arrows]\label{def:IntroGradedArrow}
For any search problems $R$ and $S$ and any $k\in\mathbb{N}$,
the search problem $R\to_k S$ is defined as follows:
given as input a pair $\langle f,g\rangle$ of algorithms
representing a claimed reduction from $R$ to $S$,
find a pair $\langle x,y\rangle$ of strings such that
either $f(x)$ fails to run in $O(|x|^k)$,
or $g(x,y)$ fails to run in $O((|x|+|y|)^k)$,\footnote{Formally,
the runtime bound for $f$ will be $p_k(|x|+|\langle f\rangle|)$
and the runtime bound for $g$ will be
$p_k(|x|+|y|+|\langle g\rangle|)$.}
or else the reduction property fails,
meaning $(f(x),y)\in S$ but $(x,g(x,y))\notin R$.
\end{definition}

It is not hard to see that if $R$ and $S$ are verifiable,
then $R\to_k S$ is also verifiable for any fixed $k$. We then define
\[\downcl R\to\downcl S\coloneqq
    \bigcup_{k\in\mathbb{N}} \downcl (R\to_k S),
\qquad
\mathsf{A}\to\mathsf{B}\coloneqq
    \bigcap_{R\in \mathsf{A}}\bigcup_{S\in\mathsf{B}}
    (\downcl R\to\downcl S).
\]
In other words, the class $\mathsf{A}\to\mathsf{B}$ is the class
of search problems which reduce to the task of refuting
a reduction from anything in $\mathsf{A}$ to something
in $\mathsf{B}$.

This definition of $\mathsf{A}\to\mathsf{B}$
turns out to exactly correspond to the Heyting arrows
in the ideal completion of the reducibility partial order,
as defined in the previous section. Concretely,
this means that $\mathsf{A}\to\mathsf{B}$ is the largest
ideal whose intersection with $\mathsf{A}$
is contained in $\mathsf{B}$. Once again, this implies
the rules of intuitionistic logic hold, giving conclusions
such as
$((\mathsf{A}\cap\mathsf{B})\to\mathsf{C})
=(\mathsf{A}\to(\mathsf{B}\to\mathsf{C}))$.

\subsection{Our results}

\subsubsection{Basic structural results}

We summarize basic structural results we discussed 
in the previous sections. Recall a search problem is a set
of pairs of strings (representing valid input/output pairs),
and such a set is called verifiable if it is in $\mathsf{P}$
as a language. Let $\mathsf{Verifiable}$ be the class
of all verifiable search problems, and recall that this
includes impossible search problems. $\mathsf{Verifiable}$
is a preorder under the reducibility relation $\le$.

\begin{theorem}[Likely folklore]
The preorder $\mathsf{Verifiable}$ has joins, meets, bottom elements,
top elements,\footnote{The top elements are the impossible search
problems.} and satisfies distributivity; this makes it
a bounded distributive preorder lattice.
\end{theorem}

The ideal completion of a distributive preorder lattice is always a
complete Heyting algebra; our central insight in this work
is that the resulting Heyting algebra has a natural interpretation
in terms of refutation problems. We summarize this in the following
theorem.

\begin{theorem}\label{thm:main}
The ideal completion of $\mathsf{Verifiable}$
is the partial order of complexity classes of search problems
under $\subseteq$. This partial order is a complete Heyting algebra.
Its properties are as follows:
\begin{enumerate}
\item The bottom element is the class $\mathsf{FP}$
\item The top element is the class $\mathsf{Verifiable}$
\item The meet of two classes $\mathsf{A}$ and $\mathsf{B}$ is
$\mathsf{A}\wedge\mathsf{B}=\mathsf{A}\cap\mathsf{B}$;
this generalizes to infinite meets
\item The join of two classes $\mathsf{A}$ and $\mathsf{B}$,
denoted $\mathsf{A}\vee\mathsf{B}$,
is formed by taking $\mathsf{A}\cup\mathsf{B}$, taking the closure
under finite joins of search problems,
and then taking the downwards closure under
reductions; this generalizes to infinite joins of classes
\item The Heyting negation $\neg \mathsf{A}$ of a class $\mathsf{A}$
consists of all search problems $R$ such that for each
$S\in\mathsf{A}$, there exists $k\in\mathbb{N}$ such that
$R$ reduces to $\neg_k S$ (the task of refuting
an $O(n^k)$-time algorithm claiming to solve $S$)
\item The Heyting arrow $\mathsf{A}\to\mathsf{B}$ between
classes $\mathsf{A}$ and $\mathsf{B}$ consists of all search problems
$T$ such that for each $R\in\mathsf{A}$ there exists $S\in\mathsf{B}$
and $k\in\mathbb{N}$ such that $T$ reduces to $R\to_k S$
(the task of refuting an $O(n^k)$-time reduction claiming to
reduce $R$ to $S$)
\item In the special case where the classes have complete problems
$R$ and $S$, we have the equations
\[(\downcl R)\wedge (\downcl S)=\downcl(R\wedge S),
\qquad (\downcl R)\vee(\downcl S)=\downcl(R\vee S),\]
\[\neg \downcl S=\downcl\{\neg_k S\}_k,
\qquad (\downcl R)\to(\downcl S)=\downcl\{R\to_k S\}_k,\]
where we use $\downcl$ to denote the downward closure under
mapping reductions.
\end{enumerate}
\end{theorem}

\subsubsection{Consequences of intuitionistic logic}
\label{sec:IntroLogicResults}

\begin{corollary}\label{cor:intuitionistic}
Complexity classes of search problems obey the laws of intuitionistic
logic. That is, if we take any valid deduction in intuitionistic
logic, replace the falsehood/contradiction symbol $\bot$
by $\mathsf{FP}$, the
absolute truth symbol $\top$ by $\mathsf{Verifiable}$,
propositional variables by complexity classes,
the logical AND by the intersection,
the logical OR by the join,
the logical negation by the refutation operator $\neg$ on classes,
the logical implication by the reduction-refutation operator
$\to$ on classes, and the ``deduces'' relation $\vdash$ by
$\subseteq$, we get a valid universal statement for complexity
classes.
\end{corollary}

\paragraph{Easy examples.} In intuitionistic logic,
$\neg \bot=\top$ and
$\neg \top=\bot$. This translates to
$\neg \mathsf{FP}=\mathsf{Verifiable}$
and $\neg\mathsf{Verifiable}=\mathsf{FP}$.
Since $\neg$ means refutation,
$\mathsf{FP}$ has the trivial problem as a complete problem,
and $\mathsf{Verifiable}$ has the impossible problem as a
complete problem, the statements become
``refuting an algorithm for the trivial search problem
is impossible''
and ``refuting an algorithm for the impossible search problem
is trivial''.

Another easy example is $\neg (p\vee q)=\neg p\wedge \neg q$,
which holds in intuitionistic logic. This gives
$\neg(\mathsf{A}\vee\mathsf{B})=\neg\mathsf{A}\cap\neg\mathsf{B}$
for all classes $\mathsf{A}$ and $\mathsf{B}$. If these classes
have complete problems $R$ and $S$ respectively, this can simplify
to
\[\downcl\{\neg_k(R\vee S)\}_k
=\downcl\{\neg_\ell R\wedge\neg_\ell S\}_\ell;\]
in other words, the claim is that each $\neg_k(R\vee S)$
reduces to some $\neg_\ell R\wedge\neg_\ell S$ (for sufficiently
large $\ell$ dependent on $k$) and conversely,
each $\neg_\ell R\wedge\neg_\ell S$ reduces to some
$\neg_k(R\vee S)$ (for sufficiently large $k$ dependent on $\ell$).
A direct argument for this is as follows: $\neg_k(R\vee S)$
asks to refute an input algorithm $\langle A\rangle$ which
claims to be able to solve both $R$ and $S$. On the other hand,
$\neg_\ell R \wedge \neg_\ell S$ asks to take in a pair of
algorithms $\langle A_R,A_S\rangle$, one of which claims to solve
$R$ and the other of which claims to solve $S$, and refute at least
one of them. It's easy to convert an algorithm
that claims to solve both $R$ and $S$ to a pair of algorithms,
one for $R$ and one for $S$ (just use the same algorithm twice);
similarly, it's easy to convert back from such a pair of algorithms
to a single algorithm which solves both $R$ and $S$
(the single algorithm just calls either $A_R$ or $A_S$ as subroutines,
as appropriate). This completes the argument.

A perhaps more surprising consequence is the following
non-intermediacy result.

\begin{corollary}[Non-intermediacy result]\label{cor:nonitermediacy}
For all complexity classes $\mathsf{C}$ of verifiable
search problems (i.e. ideals of the reducibility partial order),
$\mathsf{C}\cap\neg\mathsf{C}=\mathsf{FP}$.

In particular, suppose $\mathsf{C}\ne\mathsf{FP}$
has a complete problem $S$;
concretely, perhaps $\mathsf{C}=\mathsf{PPAD}$ and its complete
problem $S$ is end-of-line, and assume $\mathsf{PPAD}\ne\mathsf{FP}$.
Then for each $k\in\mathbb{N}$, we get a non-intermediacy result:
the $k$-th refutation task $\neg_k S$ is either in $\mathsf{FP}$
or outside of $\mathsf{PPAD}$, but it is not
$\mathsf{PPAD}$-intermediate (nor $\mathsf{PPAD}$-complete).
\end{corollary}

\begin{proof}
Given \cref{cor:intuitionistic},
this follows from $p\wedge\neg p=\bot$,
which holds in intuitionistic logic.
\end{proof}

Somewhat magically, intuitionistic logic gave a proof of a
non-intermediacy result for refutation tasks! One could also
give a direct proof, similar to the types of arguments used above;
we find working with intuitionistic logic to be simpler, in that
it can easily predict which results we should expect and which
results we should not expect.

\paragraph{Further examples.}
\begin{enumerate}
\item $p\to\bot=\neg p$ gives the important property
$(\mathsf{C}\to\mathsf{FP})=\neg\mathsf{C}$.
\item $p\vdash\neg\neg p$ gives
$\mathsf{C}\subseteq\neg\neg\mathsf{C}$. The reverse direction
need not hold in general.
\item $\neg\neg\neg p=\neg p$ gives
$\neg\neg\neg\mathsf{C}=\neg\mathsf{C}$. This is quite
cumbersome to prove directly, yet we think it is an
intuitively interesting result: while double refutation
(as a class) may be harder than the original task,
triple refutation simplifies to single refutation.
\item $p\vee \neg p$ may not equal $\top$, and indeed
$\mathsf{C}\vee\neg\mathsf{C}$ need not contain all problems
(in particular, $S\vee \neg_k S$ may be non-impossible for all $k$).
\item In intuitionistic logic, if $\phi\vdash \psi$
then $\neg\psi\vdash\neg\phi$. This gives:
if $\mathsf{A}\subseteq\mathsf{B}$, then
$\neg\mathsf{B}\subseteq\neg\mathsf{A}$. For search problems,
we get the consequence that if $R\le S$, then for all $k$
there exists $\ell$ such that $\neg_k S\le \neg_\ell R$.
Note the reduction flipped directions for the refutation tasks.
\item Modus ponens, $p\wedge (p\to q)\vdash q$, gives
$\mathsf{A}\cap (\mathsf{A}\to\mathsf{B})\subseteq\mathsf{B}$.
This generalizes \cref{cor:nonitermediacy} when considering
$\mathsf{B}=\mathsf{FP}$, since
$(\mathsf{A}\to\mathsf{FP})=\neg\mathsf{A}$, and gives
the following non-intermediacy result: if $R\le S$, then
for each $k$, the reduction refutation task
$S\to_k R$ either reduces to $R$ or does not reduce to $S$,
but is not intermediate between $R$ and $S$.
\end{enumerate}

\subsubsection{Tractability of refutations}

A central question in the metacomplexity study of refutations is:
when an algorithm doesn't exist, can we explicitly find refutations?
For search problems, the question becomes: if $S$ is intractable,
is $\neg_k S$ tractable for each $k$?

One elegant way to rephrase the question in the language of order
theory is to ask whether $\mathsf{FP}$ is a prime ideal in the
preorder lattice of verifiable search problems. A \emph{prime ideal}
is a proper ideal $I$ with the property that if $(x\wedge y)\in I$,
then either $x\in I$ or $y\in I$. Note that if $\mathsf{FP}$
was a prime ideal, then since $S\wedge \neg_k S\in\mathsf{FP}$
for all search problems $S$ and all $k$
(see \cref{cor:nonitermediacy}), we'd get that
for each search problem $S$, either $S\in\mathsf{FP}$
or else for all $k$, $\neg_k S\in\mathsf{FP}$.

However, $\mathsf{FP}$ being a prime ideal is too much to hope for,
because it is possible to construct a search problem $R$
that has superpolynomial output size on some input sizes
but is easy on others, and a search
problem $S$ which has superpolynomial output size
on a disjoint set of input sizes,
such that their meet $R\wedge S$ becomes unconditionally easy.
Perhaps a better way to ask the question is as follows.

\begin{question}
Does $\mathsf{FP_{io}}$ contain a prime ideal?
\end{question}

Here $\mathsf{FP_{io}}$ is the set of all verifiable search problems
which are solvable in polynomial time on infinitely many
input sizes.

We view this as an interesting question which we are unable
to settle. However, we are able to find two other prime
ideals which are relatively ``small'' (they represent tractable
search problems). Our prime ideals require the notion of
\emph{nonprincipal ultrafilters}, which we introduce in
\cref{sec:nonprincipal}, and are a little hard to define. The first
leads to the following theorem.

\begin{theorem}\label{thm:FPpolyPrime}
There exists a prime ideal of $\mathsf{Verifiable}$
which is a subset of the set
$\mathsf{(FP/poly)_{io}}$ of verifiable search problems
which can be solved by polynomial circuits on infinitely many
input sizes.
\end{theorem}

\begin{corollary}\label{cor:FPio}
For each verifiable search problem $S$,
either $S\in\mathsf{(FP/poly)_{io}}$ or else for all
$k\in\mathbb{N}$ we have $\neg_k S\in \mathsf{(FP/poly)_{io}}$.
\end{corollary}

\begin{proof}
Let $I$ be the prime ideal from \cref{thm:FPpolyPrime}.
Since $\mathsf{FP}$ is the bottom of the ideal completion,
$\mathsf{FP}\subseteq I$. For any search problem $S$
and any $k\in\mathbb{N}$, we have $S\wedge \neg_k S\in\mathsf{FP}$,
so $S\wedge \neg_k S\in I$. Since $I$ is prime,
either $S\in I$ or $\neg_k S\in I$, and since 
$I\subseteq\mathsf{(FP/poly)_{io}}$ the desired result follows.
\end{proof}

Another prime ideal comes from heuristic search problems.
Let $\mathsf{HeurBPP_{io}}$ denote the set of verifiable
search problems such that for any samplable distribution $D$
(meaning $D(0^n)$ samples a string for each $n\in\mathbb{N}$
in polynomial time)
and any $k\in\mathbb{N}$, there exists a polynomial time
randomized algorithm
$M$ and infinitely many sizes $n$ such that $M$ solves
inputs from $D(0^n)$ with probability at least $1-1/p_k(n)$
(over both the distribution $D(0^n)$
and the internal randomness of $M$).

\begin{theorem}\label{thm:HeurBPPprime}
There exists a prime ideal of $\mathsf{Verifiable}$
which is a subset of $\mathsf{HeurBPP_{io}}$.
\end{theorem}

\begin{corollary}
For each verifiable search problem $S$, either
$S\in\mathsf{HeurBPP_{io}}$ or else for each $k\in\mathbb{N}$
we have $\neg_k S\in\mathsf{HeurBPP_{io}}$.
\end{corollary}

\subsection{Closure properties of regular elements}

In a Heyting algebra, elements satisfying $\neg\neg x=x$ are called
regular. Since $\neg\neg\neg x=\neg x$,
an element is regular if and only if it is a negation
of some element: the regular elements are the negated ones.

In pointless topology, the regular elements in a locale
form a special sublocale, known as the minimal dense
sublocale.\footnote{A
dense sublocale is analogous to a dense subset in a topological
space, meaning a subset that intersects every open set; however,
unlike for topological spaces, locales have the property
that the intersection of dense sublocales is dense, and the
intersection of all dense sublocale forms the minimal dense sublocale.
For each open set $A$ in a subspace $Y$ of a topological space $X$,
consider the largest possible open set $B_A\supseteq A$ such that
$B_A\cap Y=A$; doing this to each ``open set'' in the minimal
dense sublocale gives a collection of ``open sets'' in the original
locale, which turn out to exactly be the regular elements of the
locale.} The following theorem follows from a general property
of all locales.

\begin{theorem}\label{thm:prenucleus}
Let $\nu$ be an operator on the ideals of $\mathsf{Verifiable}$
(meaning $\nu$ sends complexity classes to complexity classes).
Suppose $\nu$ satisfies the following for all
complexity classes $\mathsf{A}$ and $\mathsf{B}$:
\begin{enumerate}
\item Inflationary: $\mathsf{A}\subseteq\nu(\mathsf{A})$
\item Monotonicity: if $\mathsf{A}\subseteq\mathsf{B}$ then
$\nu(\mathsf{A})\subseteq\nu(\mathsf{B})$
\item Intersection property: 
$\nu(\mathsf{A})\cap\mathsf{B}\subseteq\nu(\mathsf{A}\cap\mathsf{B})$
\item Density: $\nu(\mathsf{FP})=\mathsf{FP}$.
\end{enumerate}
Then $\nu$ fixes every regular class. That is,
$\nu(\neg\mathsf{C})=\neg\mathsf{C}$ for all complexity
classes $\mathsf{C}$ of search problems.
\end{theorem}

This theorem from pointless topology allows us to prove a variety
of closure properties for the refutation classes $\neg\mathsf{C}$.

\begin{corollary}
For any class $\mathsf{C}$ of search problems,
$\neg\mathsf{C}$ is closed under Turing reductions
(not merely mapping reductions).
\end{corollary}

There are a couple of small subtleties in the definition
of a Turing reduction for search problems, which we defer to
\cref{sec:TuringReductionClosure}.

\begin{proof}[Proof sketch.]
Let $\nu(\mathsf{A})=\mathsf{FP}^{\mathsf{A}}$ for any complexity
class $\mathsf{A}$. It is not hard to show that 
$\mathsf{FP}^{\mathsf{A}}$ is nonempty, downwards closed under
mapping reductions, and closed under joins, so that $\nu$
is a well-defined operator on ideals of the reducibility lattice.
Moreover, the inflationary, monotonicity, and density
properties are easy to see.
It remains to show the intersection property.

If $S\in \mathsf{FP}^\mathsf{A}\cap\mathsf{B}$,
then $S$ is in $\mathsf{B}$ and Turing reduces to problems
in $\mathsf{A}$; when this Turing reduction makes an oracle
call to a problem $R\in\mathsf{A}$, we can convert this
to an oracle call to $R\wedge S$ by asking the oracle
``either solve the desired problem $R$, or solve the original
task $S$''. Since $R\wedge S\in \mathsf{A}\cap\mathsf{B}$
for all $R\in\mathsf{A}$, this gives an
$\mathsf{FP}^{\mathsf{A}\cap\mathsf{B}}$ algorithm for $S$,
as desired.
\end{proof}

We give the following further closure property, which we discuss
in more detail in the subsequent section.

\begin{theorem}\label{thm:RegularHardnessClosure}
For every complexity class $\mathsf{C}\subseteq\mathsf{Verifiable}$
of polynomially verifiable search problems,
$\neg\mathsf{C}$ is characterized
by its decidably verifiable hard problems.
\end{theorem}

\subsubsection{Characterization of classes by their
hard problems}

Along the way to \cref{thm:RegularHardnessClosure},
we establish some structural results that we believe
are of independent interest. To describe them,
we first require the notion
of \emph{decidably} verifiable search problems, rather than
just polynomial-time verifiable search problems.
For a class $\mathsf{C}\subseteq\mathsf{Verifiable}$ of
polynomially-verifiable search problems, we define
$\mathrm{Hard}(\mathsf{C})$ to be the set of all decidably-verifiable
search problems $T$ such that $S\le T$ for all $S\in\mathsf{C}$.
(We always require polynomial verifiability for elements
in a complexity class and decidable verifiability for upper bounds
on a complexity class.)

We say a class $\mathsf{C}$ of polynomially-verifiable
search problems is \emph{characterized by its decidably verifiable
hard problems} if every polynomially-verifiable $S$
which reduces to all search problems in $\mathrm{Hard}(\mathsf{C})$
is in $\mathsf{C}$. This means that $\mathsf{C}$ can be defined
by its set of decidably-verifiable hard problems.

Note that since the collection of decidably-verifiable problems
is countable, we may enumerate them, writing
$\mathrm{Hard}(\mathsf{C})=\{T_1,T_2,\dots\}$.
Moreover, since decidably verifiable problems have finite meets,
we can take $R_1=T_1$, $R_2=T_1\wedge T_2$,
$R_3=T_1\wedge T_2\wedge T_3$, and so on, and by the meet property
these will all be in $\mathrm{Hard}(\mathsf{C})$. In other words,
there is necessarily a decreasing sequence $R_1\ge R_2\ge\dots$
of hard problems for $\mathsf{C}$ such that every hard problem
$T$ for $\mathsf{C}$ is above some sufficiently large $R_n$.
One can similarly use joins to form an increasing sequence
of problems in $\mathsf{C}$ such that any problem $S$ in
$\mathsf{C}$ is eventually below some sequence element.

A class $\mathsf{C}$ which is characterized by its
decidably verifiable
hard problems is then simultaneously characterized by both
an increasing sequences of (polynomially verifiable)
problems in $\mathsf{C}$
and a decreasing sequence of (decidably verifiable)
hard problems for $\mathsf{C}$; the mental picture should be
one of a ``missing'' complete problem that ought to be
sandwiched between these sequences.
\cref{thm:RegularHardnessClosure} asserts that
$\neg\mathsf{C}$ is characterized in this way for
every ideal $\mathsf{C}\subseteq\mathsf{Verifiable}$.

In fact, we will establish a general structural result.
Call a subclass of $\mathsf{Verifiable}$ \emph{syntactic}
if it is possible to enumerate clocked verifiers
for its languages. Non-syntactic classes are often called
semantic, but some semantic classes are simpler than others;
call a class \emph{tamely semantic} if it is possible
to enumerate clocked verifiers for its languages using
an enumerator with access to an oracle for the Halting problem.
Several traditional semantic classes
such as $\mathsf{FBPP}\cap\mathsf{Verifiable}$
are actually tamely semantic. Then the following holds.

\begin{theorem}\label{thm:TamelySemantic}
Every tamely semantic ideal $\mathsf{C}$ of $\mathsf{Verifiable}$
is characterized by its decidably-verifiable hard problems.
\end{theorem}

Note that we are able to establish \cref{thm:RegularHardnessClosure}
using \cref{thm:TamelySemantic} even though the regular
classes $\neg\mathsf{C}$ are not generally tamely semantic
as far as we can tell.

\subsection{Our techniques}

Our main result, \cref{thm:main}, follows from just a few
simple claims:
\begin{enumerate}
\item $S\to_k \mathtt{triv}$ is equivalent to $\neg_k S$
up to a change in $k$,
\item $R\wedge (R\to_k S)\le S$ for all $k$, and
\item $(R\wedge S)\le T$ implies the existence of $k$
such that $R\le (S\to_k T)$.
\end{enumerate}
These are relatively simple to prove.
We consider our biggest contribution to be the definitions
and setup involved: the idea to consider reduction-refutation problems in the first place; working with search problems instead of decision problems, and Turing machines instead of circuits;
requiring verifiability, but not totality, and handling non-total problems by allowing impossible problems instead of via promises;
working with a hierarchy of refutation problems (one for each $k$)
instead of a single one;
and using the ideal completion instead of the Dedekind-MacNeille completion.\footnote{Indeed, it took us years of thinking
about refutation tasks before we came up with the clean formulation
of \cref{thm:main}, even after we've had the intuition that
something like it should hold.}

\paragraph{Prime ideals.} Recall that a proper ideal $\mathsf{C}$
is prime if $R\wedge S\in \mathsf{C}$ implies $R\in\mathsf{C}$
or $S\in\mathsf{C}$. Recall also that $R\wedge S$ is the task
``given $\langle x_0,x_1\rangle$, solve $x_0$ for $R$ or
$x_1$ for $S$''.

The underlying intuition behind
the $\mathsf{HeurBPP}$ type of prime ideal is that if we have
hard distributions for $R$ and for $S$ against which heuristic
algorithms fail, the product distribution over inputs pairs
should be a hard distribution for $R\wedge S$.

The underlying intuition behind the $\mathsf{FP/poly}$ type
of prime ideal is that if we have a circuit for
$R\wedge S$ which solves either $R$ or $S$ on each input pair,
we can consider a $2$-player game in which one player
chooses an $R$ input and the other chooses an $S$ input,
and both try to get the circuit to solve the other's input.
The minimax theorem says that one of the players has a mixed strategy
that succeeds with probability at least $1/2$, no matter
the other player's choice of input. This mixed strategy
turns into a distribution over circuits whose mixture
solves one of the tasks with probability at least $1/2$
on worst-case inputs. The tasks are verifiable, so we can
amplifying to success probability $>1-1/2^n$ and then derandomize
by fixing a single circuit in the support of the distribution
that succeeds for all inputs.

Both of these approaches require some work to turn them into
actual prime ideals. The definition of each of these
prime ideals requires the use of a nonprincipal ultrafilter
$\mathcal{U}$ on $\mathbb{N}$. This is a collection of subsets
of $\mathbb{N}$ we consider to be ``large'', which satisfies
the properties (a) finite sets are not large, (b) for each subset
of $\mathbb{N}$, either it is large or its complement is large,
(c) the intersection of large sets is large.
Proving the existence of a nonprincipal ultrafilter requires
the axiom of choice.

Once we have $\mathcal{U}$, we define the prime ideal
$\mathsf{(FP/poly)}_{\mathcal{U}}$ to be the set of
search problems $S\in\mathsf{Verifiable}$ such that
for each $k\in\mathbb{N}$ there exists $U\in\mathcal{U}$
and $\ell\in\mathbb{N}$ such that for each $n\in U$ and
$m\le p_k(n)$, there is a circuit $C_{n,m}$ taking
$m$ input bits and having size at most $p_\ell(n)$
such that $C_{n,m}$ solves $S$ on all inputs of size $m$.
This definition ensures downwards closure under reductions
and closure under joins
(due to the flexibility of $k$ and $\ell$)
and meet primality (due to the properties of the ultrafilter).
The definition of the $\mathsf{HeurBPP}$ type prime ideal
is similar, but with additional quantifiers to handle
the samplable distribution and error level.

\paragraph{Characterization by hard problems.}
\cref{thm:TamelySemantic} uses a somewhat intricate slow 
diagonalization argument. Roughly speaking,
we enumerate the search problems $S_1,S_2,\dots$
in our tamely semantic class $\mathsf{C}$, then consider
for slow-growing $\alpha$ a
search problem $\mathsf{C}_\alpha$ which, on input size $n$,
behaves like the join of the first $\alpha(n)$ search
problems $S_1\vee S_2\vee\dots\vee S_{\alpha(n)}$.
If $\alpha(n)\to\infty$, this will upper bound every search
problem in $\mathsf{C}$. We'll pick $\alpha$ so that
it tends to infinity extremely slowly, too slowly
for any polynomial time algorithm to notice it is not constant.
Then if $S$ is any search problem that reduces to all upper
bounds on $\mathsf{C}$, it in particular reduces to
$\mathsf{C}_\alpha$, but since this reduction is polynomial
time, it fails to notice that $\alpha$ grows beyond some
constant $k$; hence $S$ reduces to
$S_1\vee\dots\vee S_k\in\mathsf{C}$, so $S\in\mathsf{C}$.
To implement this strategy for tamely semantic ideals
(for which we cannot enumerate the elements of $\mathsf{C}$
without a Halting oracle), we will need to employ
Shoenfield's limit lemma and a finite injury argument.

To get \cref{thm:RegularHardnessClosure}, we will define
$\nu(\mathsf{C})$ to be the set of verifiable search problems
reducing to all problems in $\mathrm{Hard}(\mathsf{C})$,
and show that $\nu$ satisfies the properties of
\cref{thm:prenucleus}. The only difficulty is showing
the intersection property
$\nu(\mathsf{A})\cap \mathsf{B}
\subseteq \nu(\mathsf{A}\cap\mathsf{B})$.
To see this, let $S\in \nu(\mathsf{A})\cap \mathsf{B}$ and
let $T\in\mathrm{Hard}(\mathsf{A}\cap\mathsf{B})$;
we wish to show $S\le T$.

Since $S\in\mathsf{B}$, for any $R\in\mathsf{A}$ we have
$R\wedge S\in\mathsf{A}\cap\mathsf{B}$ and hence
$R\wedge S\le T$. A Heyting property gives
$R\le (S\to_k T)$, which we show holds even though $T$
is not polynomial time verifiable; since this holds for all $R$,
we get
$\mathsf{A}\subseteq\downcl\{S\to_k T\}_k\cap\mathsf{Verifiable}$.
But $\downcl\{S\to_k T\}_k\cap\mathsf{Verifiable}$ is tamely
semantic, so $\nu$ fixes it; since $\nu$ is monotone,
$\nu(\mathsf{A})\subseteq
\downcl\{S\to_k T\}_k\cap\mathsf{Verifiable}$,
from which we get $S\le (S\to_k T)$ for some $k$.
Another Heyting property (which again must be shown
when $T$ is not verifiable) then gives $S\le T$, as desired.

\subsection{Extensions and generalizations}

Our setting involves polynomial-time verifiable search problems
under polynomial-time mapping reductions. There are several
modifications one might like to make to the setting.

First, our definition of polynomial-time verifiability allows
the verifier to depend polynomially on both the input $x$
and the output $y$; we do not have a guarantee that the output
size $|y|$ is polynomially related to the input size $|x|$,
so we technically allow search problems for which valid solutions
are exponentially long in the input size. It is natural
to want to restrict to search problems for which a
polynomially-sized valid output string exists for each input;
this is the class $\mathsf{TFNP}$, which is an ideal of
$\mathsf{Verifiable}$.

If $R,S\in\mathsf{TFNP}$ are search problems, the reduction
refutation task $R\to_k S$ need not be in $\mathsf{TFNP}$
in general, since it may be impossible to solve, or may only
have solutions that are superpolynomially long in the input size
(though note $R\to_k S$ is still verifiable, so still in
$\mathsf{Verifiable}$). This means we cannot define $\to_k$
as a binary operation on $\mathsf{TFNP}$. However, we can still
define $\to$ as a binary operation on the subclasses of 
$\mathsf{TFNP}$
(that is, on the ideals contained in $\mathsf{TFNP}$) in a way
that gives a complete Heyting algebra. A simple way to do this
is for any two ideals $\mathsf{A},\mathsf{B}\subseteq\mathsf{TFNP}$,
define the new arrow to be
$(\mathsf{A}\to\mathsf{B})\cap\mathsf{TFNP}$
(the set of total search problems which reduce to a problem
in $\mathsf{A}\to\mathsf{B}$). This amounts to restricting
to the \emph{open sublocale} of $\mathsf{TFNP}$ within the
original locale. This sublocale inherits most of the results
above. In other words, we can restrict our story
only to total search problems if need be, and everything still
works out (however, it is easier to work with non-total verifiable
problems, which are more general).

Generalizing in a different direction, one may wish to remove
the verifiability condition from the search problems we consider.
This can be done if one switches to ``list computations,''
a setting in which to solve an input $x$ to a search problem $S$
we may provide a list of possible solutions $y_1,y_2,\dots,y_k$,
at least one of which is a valid output for $x$. If we modify
mapping reductions to this setting, most of our results go through
without the verifiability assumption (in particular, the ideals
of this preorder lattice of non-verifiable search problems
still forms a complete Heyting algebra).

Similarly, it is also possible to change the notion of
efficiency (e.g.
allow randomized or quantum reductions) and the main results
still go through. Our results in this work also relativize:
one could equip all reductions with an oracle $\mathcal{O}$
and all the results still work relative to $\mathcal{O}$.

With sufficient care, we believe one could get the Heyting
algebra story to also work with non-uniform algorithms (circuit
classes) and in concrete models of computation such as query
or communication complexity; in other words, we believe
the basic story of ``metacomplexity refutation tasks
are Heyting negations in a complete Heyting algebra, and obey
the laws of intuitionistic logic'' is a fairly robust story
that is true in a variety of models of computation if set up
correctly. However, we have not done this setup in these other models.

\subsection{Related work}

A literature search revealed three previous
instances in which something like search problems gave
rise to a Heyting algebra or similar structure. These are
the Medvedev lattice, and Muchnik lattice, and Weihrauch
lattice. We refer to \cite{Pau17,HP13,BG20}. These are
all lattices in computability theory; they involve multi-output
problems (essentially search problems)
and Turing reductions between them.
The difference between the two lattices comes from how Turing
reductions are defined for search problems; the Weihrauch
lattice involves real-number computations.

In all of these cases, a Heyting structure is only
present when there is no resource bound
(so no need for graded negations or graded arrows like
$R\to_k S$). There is also generally
no verifiability, and the Heyting structure
applies to the reducibility order itself, not its ideal
completion. Our work bears some similarity to these lattices,
but we use a resource-bounded setting, which requires
moving from search problems to classes of search problems
(the ideal completion) to get a Heyting algebra
and works better in the presence of verifiers.

\subsection{AI statement}

Most of this work was done without assistance from AI
(except for literature search and for studying
related mathematics, as well as proof reading).
This is because this work has been
in progress for at least 3 years. In particular,
the main result (\cref{thm:main}) relating structural
complexity and metacomplexity was not AI assisted
(we attempted to use older AI tools such as ChatGPT o3 mini
and Gemini 2.5 pro, and later Gemini 3.0 pro, but this did
not lead to significant progress).

The main exception is the prime ideal in
$\mathsf{(FP/poly)_{io}}$ (\cref{thm:FPpolyPrime}),
which was due to more recent discussions
with ChatGPT 5.4; before these discussions, we only had
the prime ideal from \cref{thm:HeurBPPprime}
and some related results. All other results in this work
are due to human insight,
with at most minor contributions from older AI models
(Gemini 3.0 pro or earlier).

\section{The lattice of search problems}

\subsection{Preliminaries: search problems and reductions}\label{sec:FNPprelim}

\paragraph{Notation for polynomials.}
We will use the notation $p_k(n)=n^k+k$ to indicate
the $k$-th polynomial growth rate; note that for every polynomial
growth rate, there exists some $k$ such that $p_k$ dominates
that growth rate on all natural numbers $n$. Hence we may restrict
polynomials to be $p_k$ growth rates without loss of generality.

\paragraph{Power set notation.} We will use $\mathcal{P}(X)$
to denote the power set of $X$.

\paragraph{Search problems.}
A search problem is a relation from strings to strings,
$S\subseteq\{0,1\}^*\times\{0,1\}^*$.
We interpret the task associated
with the relation $S$ to be the problem of taking an input
$x\in\{0,1\}^*$ and producing an output $y\in \{0,1\}^*$
such that $(x,y)\in S$.
We denote the set of search problems by
$\mathsf{Search}=\mathcal{P}(\{0,1\}^*\times\{0,1\}^*)$.
Note if there exists $x$ such that
$(x,y)\notin S$ for all $y$, then the search problem is impossible
to solve; we allow such impossible problems.

\paragraph{Verifiability.} A search problem $S$ is
\emph{polytime verifiable} if $S\in\mathsf{P}$ as a language.
In other words, $S$ is verifiable in polynomial time if there
exists a polynomial time Turing machine $V$ which accepts
exactly the pairs $(x,y)\in S$ and rejects everything else.
We will sometimes also need the notion of decidably verifiable
search problems; we say $S$ is \emph{decidably verifiable} if
$S\subseteq\mathsf{Decidable}$ as a language.
We denote the class of polytime verifiable search problems by
$\mathsf{Verifiable}=\mathsf{Search}\cap\mathsf{P}$.
Note that impossible problems may still be in
$\mathsf{Verifiable}$.

\paragraph{Totality.} A search problem $S$ is called \emph{total}
if there exists a polynomial bound $p_k$ such that
for all $x\in\{0,1\}^*$ there exists $y\in\{0,1\}^*$
such that $|y|\le p_k(|x|)$ and $(x,y)\in S$.
In other words, a search problem is total if every input
has at least one polynomially-sized valid output.
Impossible problems cannot be total; however,
note that a non-total problem is not necessarily impossible,
since we define a problem to be impossible
if for some input it has no valid outputs of any size
(instead of restricting to polynomial size).

The class of search problems which are both total and
polytime verifiable is denoted $\mathsf{TFNP}$.

\paragraph{Tractability.} The class $\mathsf{FP}$ consists of all
polynomial time verifiable search problems $S$ for
which there exists a polynomial time Turing
machine mapping each $x$ to some string $y$ such that $(x,y)\in S$.
$\mathsf{FP}$ problems are necessarily total. Observe that
\[\mathsf{FP}\subseteq\mathsf{TFNP}\subseteq\mathsf{Verifiable}
\subseteq\mathsf{Search}.\]

\paragraph{Reductions.} For two search problems $R$ and $S$,
we define a reduction from $R$ to $S$ as follows: it is a pair
$(F,G)$ of functions such that
\begin{enumerate}
\item $F$ maps strings to strings
\item $G$ maps pairs of strings to strings
\item $F$ and $G$ are computable in polynomial time
\item for all $x,y\in\{0,1\}^*$, we have
$(F(x),y)\in S\Rightarrow (x,G(x,y))\in R$.
\end{enumerate}

We denote the property that $R$ reduces to $S$ via the reduction
$(F,G)$ using the notation $R\le_{F,G} S$, or simply
$R\le S$ if we wish to omit the name of the reduction.

\paragraph{Important technicality: reductions do not preserve
verifiability.} We warn that $\mathsf{Verifiable}$
is not a lower set (i.e. not a downwards closed
set) in the reducibility preorder $\le$ on $\mathsf{Search}$.
In other words, reductions do not preserve verifiability: it is
possible for a problem to be easy to solve but hard to verify,
since verification requires verifying \emph{all} input-output
pairs while solving only requires producing \emph{some} valid output
for each input. Reductions preserve ``easy to solve'' in an existential
sense: there is at least one solution that is easy to find for each
input. Reductions do not preserve ``easy to verify'',
which has a universal quantifier over the output strings.

\paragraph{Identity reduction.} We define a special reduction
$I=(F_I,G_I)$ called the identity reduction, as follows:
$F_I(x)=x$ and $G_I(x,y)=y$. Note that $F_I$ and $G_I$ have
the correct signatures and are computable in polynomial time.
The condition that the identity reduction reduces $R$ to $S$
then becomes the condition $(x,y)\in S\Rightarrow (x,y)\in R$,
so $R\le_I S\Leftrightarrow R\supseteq S$. In other words,
the partial order $\le_I$ is precisely the partial order $\supseteq$,
and the preorder $\le$ is an extension that relates more
pairs of search problems.

\begin{lemma}[Reducibility is reflexive]
For any $S\in\mathsf{Search}$, we have $S\le S$.
\end{lemma}

\begin{proof}
The reduction we need is just the identity reduction;
it reduces $S$ to $S$ because $S\subseteq S$.
\end{proof}

\begin{lemma}[Reducibility is transitive]\label{lem:transitivity}
For any search problems $R,S,T\in\mathsf{Search}$,
if $R\le S$ and $S\le T$ then $R\le T$.
\end{lemma}

\begin{proof}
Let $(F_1,G_1)$ reduce $R\le S$ and let $(F_2,G_2)$ reduce $S\le T$.
Define the composition of these two reductions to be
$(F,G)$ with $F(x)=F_2(F_1(x))$ and $G(x,y)=G_1(x,G_2(F_1(x),y))$.
Note that $F$ and $G$ have the correct signatures, and since
$F_1,G_1,F_2,G_2$ all run in polynomial time, so do $F$ and $G$.
We show that $(F,G)$ reduces $R$ to $T$.

Suppose $(F(x),y)\in T$. Then $(F_2(F_1(x)),y)\in T$,
so by definition of the $(F_2,G_2)$ reduction from $S$ to $T$,
we have $(F_1(x),G_2(F_1(x),y))\in S$. Then by definition of the
$(F_1,G_1)$ reduction from $R$ to $S$, we have
$(x,G_1(x,G_2(F_1(x),y)))\in R$, or $(x,G(x,y))\in R$, as desired.
\end{proof}

This formally shows that $\le$ is a preorder on $\mathsf{Search}$,
and hence also on $\mathsf{Verifiable}$.

\paragraph{$\mathtt{triv}$ and $\mathtt{imp}$.}
Let $\mathtt{triv}=\{0,1\}^*\times\{0,1\}^*$ be the trivial
search problem in which all outputs are accepted for every input,
and let $\mathtt{imp}=\emptyset$ be the fully impossible search problem
in which no outputs are accepted for any input. Then
it is clear that $\mathtt{triv}$ is the bottom of the $\supseteq$
partial order, and $\mathtt{imp}$ is the top. In particular,
since $\mathtt{triv}\le_I S\le_I \mathtt{imp}$ for all
$S\in\mathsf{Search}$, we also have
$\mathtt{triv}\le S\le \mathtt{imp}$ for all $S\in\mathsf{Search}$
(though since $\le$ does not satisfy antisymmetry,
these bottom and top elements are not unique).

\begin{lemma}
The bottom elements of $\mathsf{Search}$ are those that have
polynomial time algorithms; in particular, the bottom
elements of $\mathsf{Verifiable}$ form the set $\mathsf{FP}$.
\end{lemma}

\begin{proof}
Note that since $\mathtt{triv}$ is a bottom element,
the bottom elements are precisely those search problems $S$
that reduce to $\mathtt{triv}$. This means there are polynomial
time computable functions $F$ and $G$ with 
$(F(x),y)\in\mathtt{triv}\Rightarrow (x,G(x,y))\in S$.
Since all pairs of strings are in $\mathtt{triv}$, this condition
is equivalent to $(x,G(x,y))\in S$ for all $x,y\in\{0,1\}^*$.
Note that if we had a polynomial time algorithm $A$ for $S$,
we could set $G(x,y)=A(x)$ to get a reduction from $S$ to
$\mathtt{triv}$; conversely, if we have a reduction from $S$ to
$\mathtt{triv}$, setting $A(x)=G(x,\varepsilon)$ (where $\varepsilon$
is the empty string) gives a polynomial time algorithm
for $S$. Hence the bottom elements $S$ are exactly those with
polynomial time algorithms.

Recall that a search problem with a polynomial time algorithm
is not necessarily verifiable (since there may exist some valid outputs
that are hard to verify, different from the outputs produced
by the algorithm). However, we defined $\mathsf{FP}$ to consist
of only polytime verifiable polytime solvable search problems.
It follows that $\mathsf{FP}$ is the set of bottoms of the preorder
$(\mathsf{Verifiable},\le)$.
\end{proof}

\begin{lemma}
The top elements of $\mathsf{Search}$ are the impossible search
problems.
\end{lemma}

\begin{proof}
Since $\mathtt{imp}$ is a top element, top elements are precisely
those search problems $S$ with $\mathtt{imp}\le S$.
This means there are polynomial time computable functions $F$ and $G$
with $(F(x),y)\in S\Rightarrow (x,G(x,y))\in\mathtt{imp}$.
Since $\mathtt{imp}$ contains no pairs of strings,
this is equivalent to the condition $(F(x),y)\notin S$ for all
$x,y\in\{0,1\}^*$. If $S$ is impossible, it means there is some
fixed string $z\in\{0,1\}^*$ such that $(z,y)\notin S$ for all $y$;
in that case we can set $F(x)=z$ for all $x$ to get a reduction
from $\mathtt{imp}$ to $S$. Conversely, if we have a reduction
from $\mathtt{imp}$ to $S$, it must hold for the empty string 
$\varepsilon$ that $(F(\varepsilon),y)\notin S$ for all $y$,
so $F(\varepsilon)$ has no valid output for $S$, and $S$ is impossible.
\end{proof}

So far, we've seen that reducibility gives a bounded preorder
on search problems, and we've characterized the top and bottom
elements. We conclude by noting that $\mathsf{TFNP}$
is a lower set in the preorder $(\mathsf{Verifiable},\le)$.

\begin{lemma}
$\mathsf{TFNP}$ is downwards closed under reductions within
$\mathsf{Verifiable}$: that is, if $R\le S$ with $S\in\mathsf{TFNP}$
and $R\in\mathsf{Verifiable}$, then $R\in\mathsf{TFNP}$.
\end{lemma}

\begin{proof}
Recall that $\mathsf{TFNP}$ is not merely the set of non-impossible
problems; it consists of search problems $S$ such that there exists
a polynomial growth rate $p_k$ so that for each $x\in\{0,1\}^*$
there exists $y\in\{0,1\}^*$ with $|y|\le p_k(|x|)$ such that
$(x,y)\in S$. Let $S$ and $p_k$ satisfy this condition,
and let $R\le_{F,G} S$, with $R,S\in\mathsf{Verifiable}$.

By the reduction property, $(F(x),y)\in S\Rightarrow (x,G(x,y))\in R$.
Since $F$ runs in polynomial time, the maximum possible size of
$F(x)$ is polynomial in the size of $x$, say at most $p_i(|x|)$
for some $i$; then there exists $y_x\in\{0,1\}^*$ of size at most
$p_k(p_i(|x|))$ such that $(F(x),y_x)\in S$. Next, since $G$
runs in polynomial time, the maximum possible size of $G(x,y_x)$
is polynomial in $|x|$ and $|y_x|$, so there is some $j$
for which $|G(x,y_x)|$ is bounded by $p_j(p_k(p_i(|x|)))$.
Since this is polynomial in $|x|$, there is some $\ell\in\mathbb{N}$
such that $|G(x,y_x)|\le p_\ell(|x|)$ for all $x\in\{0,1\}^*$.
Since $(x,G(x,y_x))\in R$ for all $x$, we conclude that $R$
has valid outputs that grow in size according to $p_\ell$
of the input, and hence $R$ is in $\mathsf{TFNP}$.
\end{proof}

\subsection{Join and Meet}\label{sec:FNPjoinmeet}

We now characterize the join of the partial order.

\begin{definition}
For search problems $S_0,S_1$, define $S_0\vee S_1$
to be the search problem with inputs $\langle x,b\rangle$
with $x\in\{0,1\}^*$
and $b\in\{0,1\}$, and with valid outputs being strings $y$
satisfying $(x,y)\in S_b$.
Inputs not of the form $\langle x,b\rangle$ are considered
poorly formatted, and all strings $y$ are valid outputs for
poorly formatted strings. In other words, $S_0\vee S_1$
consists of all pairs $(\langle x,b\rangle,y)$ with
$(x,y)\in S_b$ as well as all
pairs $(z,y)$ where $z$ is not formatted like $\langle x,b\rangle$
and $y$ is arbitrary.
\end{definition}

\begin{lemma}\label{lem:join1}
If $S_0,S_1$ are verifiable then
so is $S_0\vee S_1$. In other words, $\mathsf{Verifiable}$ is closed under $\vee$.
\end{lemma}

\begin{proof}
Let $V_0$ be a verifier for $S_0$ and let $V_1$ be a verifier
for $S_1$. Define $V(z,y)$ to accept if $z$ is not formatted
like $\langle x,b\rangle$ for some $b\in\{0,1\}$, and otherwise
$V(\langle x,b\rangle,y)=V_b(x,y)$. It is clear that $V$ is
as efficient as $V_0$ and $V_1$
and accepts exactly the pairs in $S_0\vee S_1$.
\end{proof}

\begin{lemma}\label{lem:join2}
For all search problems $S_0,S_1$, we have
$S_0\le S_0\vee S_1$ and $S_1\le S_0\vee S_1$.
\end{lemma}

\begin{proof}
Let $G(x,y)=y$ and let $F_b(x)=\langle x,b\rangle$ for $b\in\{0,1\}$.
We show $(F_b,G)$ is a reduction from $S_b$ to $S_0\vee S_1$.
Efficiency is clear, so suppose $(F_b(x),y)\in S_0\vee S_1$.
Then $(\langle x,b\rangle,y)\in S_0\vee S_1$, so by definition,
$(x,y)\in S_b$. Hence $(x,G(x,y))\in S_b$, as desired.
\end{proof}

\begin{lemma}\label{lem:join3}
For all search problems $S_0,S_1,T$, if $S_0\le T$
and $S_1\le T$ then $S_0\vee S_1\le T$. In other words,
$S_0\vee S_1$ is a join of $S_0$ and $S_1$ in the preorder
$\le$ on $\mathsf{Search}$.
\end{lemma}

\begin{proof}
Let $(F_b,G_b)$ reduce $S_b$ to $T$ for $b=0,1$.
Define $(F,G)$ by $F(\langle x,b\rangle)=F_b(x)$
and $G(\langle x,b\rangle,y)=G_b(x,y)$, with $F(z)$
and $G(z,y)$ acting arbitrarily if $z$ is not formatted like
$\langle x,b\rangle$.
We show $(F,G)$ reduces $S_0\vee S_1$ to $T$.

Suppose $(F(z),y)\in T$. If $z$ is poorly formatted,
then by definition $(z,w)\in S_0\vee S_1$ for all $w$,
so in particular $(z,G(z,y))\in S_0\vee S_1$ and we are done.
Otherwise, if $z=\langle x,b\rangle$, our assumption gives
$(F(\langle x,b\rangle),y)\in T$,
which then gives $(F_b(x),y)\in T$. By definition of the reduction
$(F_b,G_b)$ from $S_b$ to $T$, we get $(x,G_b(x,y))\in S_b$.
By definition of $S_0\vee S_1$, we get
$(\langle x,b\rangle,G_b(x,y))\in S_0\vee S_1$. Finally,
by definition of $G$, we get
$(\langle x,b\rangle,G(\langle x,b\rangle,y))\in S_0\vee S_1$,
so $(z,G(z,y))\in S_0\vee S_1$, as desired.
\end{proof}

We now characterize the meet of the preorder and conclude
it is a preorder lattice.

\begin{definition}
For search problems $S_0,S_1$, define $S_0\wedge S_1$
to be the search problem with inputs $\langle x_0,x_1\rangle$ with
$x_0,x_1\in\{0,1\}^*$, and with valid outputs being pairs
$\langle y,b\rangle$ with $y\in\{0,1\}^*$ and $b\in\{0,1\}$ such that
$(x_b,y)\in S_b$. Inputs not of the form $\langle x_0,x_1\rangle$
for strings $x_0,x_1$ are considered poorly formatted,
and all output strings are valid for poorly formatted
input strings. Outputs not of the form $\langle y,b\rangle$
for a string $y$ and $b\in\{0,1\}$ are considered poorly formatted,
and poorly formatted outputs are never valid except when the
input is also poorly formatted. In other words, $S_0\wedge S_1$
consists of all pairs of strings $(z,w)$ where either
$z=\langle x_0,x_1\rangle$
and $w=\langle y,b\rangle$ with $(x_b,y)\in S_b$,
or else $z$ is not
of the form $\langle x_0,x_1\rangle$ for any pair of strings
(and $w$ may then be arbitrary).
\end{definition}

\begin{lemma}
If $S_0,S_1$ are verifiable then
so is $S_0\wedge S_1$. In other words
$\mathsf{Verifiable}$ is closed under $\wedge$.
\end{lemma}

\begin{proof}
Let $V_0$ and $V_1$ be verifiers for $S_0$ and $S_1$ respectively.
Define $V(z,w)$ to accept if $z$ is poorly formatted, and otherwise,
reject is $w$ is poorly formatted. If both are correctly
formatted, set
$V(\langle x_0,x_1\rangle,\langle y,b\rangle)=V_b(x_b,y)$.
It is clear that
$V$ as efficient as $V_0,V_1$ and accepts $(z,w)$ if and only if
$(z,w)\in S_0\wedge S_1$.
\end{proof}

\begin{lemma}
For all $S_0,S_1$, we have $S_0\wedge S_1\le S_0$
and $S_0\wedge S_1\le S_1$.
\end{lemma}

\begin{proof}
For $b\in\{0,1\}$ let $G_b(z,y)=\langle y,b\rangle$.
Let $F_b(z)$ output arbitrarily if $z$
is not of the form $\langle x_0,x_1\rangle$,
and otherwise $F_b(\langle x_0,x_1\rangle)=x_b$.
We show $(F_b,G_b)$ reduces $S_0\wedge S_1$ to $S_b$.
Efficiency is clear, so suppose $(F_b(z),y)\in S_b$. If $z$
is poorly formatted, then $(z,w)\in S_0\wedge S_1$ for all $w$,
and in particular $(z,G(z,y))\in S_0\wedge S_1$, so we are done.
Otherwise, if $z=\langle x_0,x_1\rangle$, then $(F_b(z),y)\in S_b$
means that $(x_b,y)\in S_b$. By definition of $S_0\wedge S_1$,
this means
$(\langle x_0,x_1\rangle,\langle y,b\rangle)\in S_0\wedge S_1$,
so $(z,G(z,y))\in S_0\wedge S_1$, as desired.
\end{proof}

\begin{lemma}
For all $S_0,S_1,R$, if $R\le S_0$ and $R\le S_1$,
then $R\le S_0\wedge S_1$. In other words, $S_0\wedge S_1$
is a meet of $S_0$ and $S_1$ in the preorder $\le$ on $\mathsf{Search}$.
\end{lemma}

\begin{proof}
Let $(F_b,G_b)$ reduce $R$ to $S_b$ for $b=0,1$.
Define $(F,G)$ as follows. Set $F(x)=\langle F_0(x),F_1(x)\rangle$
and $G(x,\langle y,b\rangle)=G_b(x,y)$, with $G(x,w)$
acting arbitrarily if $w$ is not of the form $\langle y,b\rangle$.
We show $(F,G)$ reduces $R$ to $S_0\wedge S_1$.

Suppose $(F(x),w)\in S_0\wedge S_1$. Since $F(x)$ always returns
a pair of strings, it is well-formatted for $S_0\wedge S_1$;
since $(F(x),w)\in S_0\wedge S_1$, it must be the case that $w$
is well-formatted as an output, so $w=\langle y,b\rangle$
for $b\in\{0,1\}$.
Now $(F(x),\langle y,b\rangle)\in S_0\wedge S_1$ means that
$(\langle F_0(x),F_1(x)\rangle,\langle y,b\rangle)\in S_0\wedge S_1$ 
(by definition of $F$),
and hence $(F_b(x),y)\in S_b$ (by definition of $S_0\wedge S_1$).
By the reduction property of $(F_b,G_b)$ from $R$ to $S_b$,
we conclude $(x,G_b(x,y))\in R$. By definition of $G$, we get
$(x,G(x,\langle y,b\rangle))\in R$, so $(x,G(x,w))\in R$, as desired.
\end{proof}

\begin{lemma}[Distributivity]\label{lem:distributive}
Let $R,S,T$ be search problems. Then
$R\wedge(S\vee T)\le (R\wedge S)\vee (R\wedge T)$.
In particular, $\mathsf{Verifiable}$
and $\mathsf{Search}$ are distributive bounded
preorder lattices under $\le$.
\end{lemma}

\begin{proof}
We define a reduction $(F,G)$ as follows. Let
$F(\langle u,\langle x,b\rangle\rangle)=\langle\langle u,x\rangle,b\rangle$,
with $F(z)$ acting arbitrarily
if $z$ is not formatted like $\langle u,\langle x,b\rangle\rangle$
for $u,x\in\{0,1\}^*$ and $b\in\{0,1\}$.
Let
$G(\langle u,\langle x,b\rangle\rangle,\langle y,b'\rangle)
    =\langle y,b'\rangle$,
with $G(z,w)$ returning $\langle s,1\rangle$ for an arbitrary string $s$
if $z$ is not formatted like
$\langle u,\langle x,b\rangle\rangle$ or $w$ is not formatted
like $\langle y,b'\rangle$. It is clear that these are efficient.

Suppose $(F(z),w)\in (R\wedge S)\vee (R\wedge T)$.
If $z$ is not formatted like a pair of strings, then
$(z,G(z,w))\in R\wedge(S\vee T)$ always holds (regardless of the
string $G(z,w)$). If $z$ is formatted like a pair of strings,
say $\langle u,v\rangle$, but $v$ is not formatted like
$\langle x,b\rangle$,
then $G(z,w)$ returns a pair $\langle s,1\rangle$,
so the statement $(z,G(z,w))\in R\wedge(S\vee T)$
is equivalent to $(v,s)\in S\vee T$, which holds due to
the poor formatting of $v$. Suppose then that
$z=\langle u,\langle x,b\rangle\rangle$. In this case, the assumption
$(F(z),w)\in (R\wedge S)\vee (R\wedge T)$ gives us
$(\langle\langle u,x\rangle,b\rangle,w)\in (R\wedge S)\vee (R\wedge T)$.
If $b=0$, this gives $(\langle u,x\rangle,w)\in R\wedge S$,
and if $b=1$, this gives $(\langle u,x\rangle,w)\in R\wedge T$.
Since $\langle u,x\rangle$ is correctly formatted for both of these,
$w$ must also be correctly formatted, so $w=\langle y,b'\rangle$ for some
string $y$ and bit $b'$. If $b'=0$, then $(u,y)\in R$
(regardless of the value of $b$). If $b'=1$, then
either $(x,y)\in S$ (if $b=0$) or $(x,y)\in T$
(if $b=1$), which means $(\langle x,b\rangle,y)\in S\vee T$.
The two cases for $b'$
together give
$(\langle u,\langle x,b\rangle\rangle,\langle y,b'\rangle)
\in R\wedge (S\vee T)$,
so $(z,G(z,w))\in R\wedge (S\vee T)$, as desired.
\end{proof}

\subsection{Ideal completion}

We've seen that $\mathsf{Verifiable}$ is a bounded
distributive preorder lattice.
We now discuss the ideal completion of such lattices.
Most of this section states standard results from
order theory (with proofs included for completeness).

\begin{definition}[Ideal]
Consider any preorder $(L,\le)$. An \emph{ideal}
of $L$ is a subset $I\subseteq L$ which satisfies
three properties:
\begin{enumerate}
\item nonempty: $I\ne\emptyset$,
\item lower set: if $x\le y$ and $y\in I$ then $x\in I$,
\item upwards directed: if $x,y\in I$
then there exists $z\in I$ with $x\le z$ and $y\le z$.
\end{enumerate}
\end{definition}
Note that if $L$ has bottom elements, the nonempty
and lower set properties imply that every ideal $I$ of $L$
contains all bottom elements of $L$.
If $L$ has joins, the directedness property implies
that every ideal $I$ is closed under joins
(if $x,y\in I$ then $I$ contains all joins of $x$ and $y$).

Let $\overline{L}$ denote the set of all ideals of $(L,\le)$,
and partially order this set by $\subseteq$. This is a partial
order.

\begin{lemma}
If the preorder $(L,\le)$ has bottom elements and finite
joins, then $(\overline{L},\subseteq)$ is a complete lattice:
a partial order with infinite meets, infinite joins,
and unique top and bottom elements.
\end{lemma}

\begin{proof}
If $L$ has bottoms and joins
it is not hard to see that an arbitrary intersection of
ideals of $L$ is an ideal, and intersection gives the meet
for the $\subseteq$ partial order. To construct
the join in $\overline{L}$ of a set
of ideals, we can take their union, add all finite joins of
elements to that union, and then take the downwards closure
under $\le$ to get the smallest ideal containing all of them.
The set of bottom
elements of $L$ is an ideal and forms the
unique bottom element of $\overline{L}$; the set $L$
itself is an ideal which is the top element of $\overline{L}$.
\end{proof}

\begin{lemma}
Let the preorder $(L,\le)$ have bottom elements and finite
joins. Then this preorder embeds into 
$(\overline{L},\subseteq)$ by mapping each $x\in L$
to the ideal $\downcl x\coloneqq\{z\in L:z\le x\}$ generated
by it. This mapping preserves order:
$x\le y$ if and only if $\downcl x\subseteq\downcl y$.
Moreover, $\downcl x=\downcl y$ if and only if $x\le y$
and $y\le x$.
\end{lemma}

\begin{proof}
This is straightforward.
\end{proof}

The partial order $(\overline{L},\subseteq)$ is therefore
an extension of (the equivalence classes of) $L$, and is
called the ideal completion of $L$. An ideal of $L$ is called
\emph{principal} if it is of the form $\downcl x$ for some
$x\in L$.

In general, a complete lattice like $\overline{L}$ need
not be distributive. However, if the original preorder $L$
 was distributive,
then $\overline{L}$ becomes distributive, and even more,
it satisfies an infinite distributivity property.

\begin{lemma}
Let $(L,\le)$ be a preorder which has bottom elements,
finite joins, finite meets, and the distributivity property.
Then its ideal completion $(\overline{L},\subseteq)$
is a complete lattice with the infinite distributivity
property finite meets over infinite joins:
\[I\wedge \bigvee_{a\in A} J_a=\bigvee_{a\in A} I\wedge J_a.\]
\end{lemma}

\begin{proof}
First note that the right hand side
is always less than or equal to the left: each $J_b$
is at most the join $\bigvee_a J_a$, so each meet
$I\wedge J_b$ is at most $I\wedge \bigvee_a J_a$,
and since $I\wedge \bigvee_a J_a$ upper bounds all
$I\wedge J_b$, it upper bounds their join.

For the reverse direction,
recall our partial order is $\subseteq$,
so it suffices to take $x\in I\wedge \bigvee_{a\in A} J_a$
and show that $x$ is in $\bigvee_{a\in A} I\wedge J_a$.
Since the meet is the intersection, we have
$x\in I$ and $x\in \bigvee_{a\in A} J_a$. Since the join
is the downwards closure of the set of finite joins of
elements in the union, we get
$x\le y_1\vee y_2\vee\dots\vee y_k$ with each $y_i\in J_{a_i}$
for $i=1,2,\dots,k$. Since both $x$ and
$y_1\vee y_2\vee\dots\vee y_k$ upper bound $x$,
so does their meet, so
$x\le x\wedge (y_1\vee y_2\vee\dots\vee y_k)$.
Finite distributivity in $L$ then gives
$x\le (x\wedge y_1)\vee \dots \vee (x\wedge y_k)$.
Since $x\wedge y_i$ is below $x$, it must be in $I$
since $x\in I$. Since $x\wedge y_i$ is below $y_i$,
it must be in $J_{a_i}$ since $y_i\in J_{a_i}$.
Hence $x\wedge y_i\in I\cap J_{a_i}$.
This means that $(x\wedge y_1)\vee \dots \vee (x\wedge y_k)$
is a finite join of elements which are each in some
$I\cap J_{a_i}=I\wedge J_{a_i}$, so it must be in
the join $\bigvee_{a\in A} I\wedge J_a$. Since
$x$ is below $(x\wedge y_1)\vee \dots \vee (x\wedge y_k)$,
we get $x\in \bigvee_{a\in A} I\wedge J_a$, as desired.
\end{proof}

\begin{lemma}
A complete lattice has the infinite distributivity
property (of finite meets over infinite joins)
if and only if it is a Heyting algebra.
\end{lemma}

\begin{proof}
Recall that a Heyting algebra is a bounded distributive
lattice such that for any pair of elements $a,b$
the set $\{c:a\wedge c\le b\}$ has a maximum element
(denoted $a\to b$).

First suppose we have infinite distributivity, and fix $a,b$.
A complete lattice has arbitrary joins, so let
$z=\bigvee\{c:a\wedge c\le b\}$. We need to show $z$
is in the set, that is, $a\wedge z\le b$.
But by infinite distributivity,
\[a\wedge z=a\wedge \bigvee\{c:a\wedge c\le b\}
=\bigvee \{a\wedge c:a\wedge c\le b\}.\]
All elements in the latter join are at most $b$,
so the join must be at most $b$ as well;
hence $a\wedge z\le b$ as desired.

In the other direction, suppose we have arrows $a\to b$.
Recall that we always have
$\bigvee_{b\in B}(a\wedge b)\le a\wedge\bigvee_{b\in B} b$,
so we must show
$a\wedge \bigvee_{b\in B} b\le \bigvee_{b\in B}a\wedge b$.
Note that for any $\hat{b}\in B$, we have
$a\wedge \hat{b}\le \bigvee_{b\in B} a\wedge b$.
Thus $\hat{b}\le a\to\left(\bigvee_{b\in B} a\wedge b\right)$
for all $\hat{b}\in B$. Since
$a\to\left(\bigvee_{b\in B} a\wedge b\right)$ upper bounds
all of $B$, it upper bounds the join $\bigvee B$ as well,
so $\bigvee_{b\in B} b\le a\to\left(\bigvee_{b\in B} a\wedge b\right)$.
By definition of the arrow, this means
$a\wedge \bigvee_{b\in B} b\le \bigvee_{b\in B} a\wedge b$,
as desired.
\end{proof}

\begin{corollary}
The ideal completion of $\mathsf{Verifiable}$
is a complete Heyting algebra (also known as a frame or
locale). Its elements (the ideals of $\mathsf{Verifiable}$)
correspond to complexity classes of verifiable search
problems: each of these is a nonempty set of polytime verifiable search problems which is downwards closed under
mapping reductions and closed under finite joins
of search problems.

The bottom element of this completion is the ideal
$\mathsf{FP}$ of polynomial-time solvable search problems.
\end{corollary}

We note that the requirement that ideals be closed under
finite joins is satisfied by most complexity class
one might consider: solving the binary join just requires
solving the underlying two search problems and the ability
to fork on an input bit in order
to choose which of the two to solve. Ideals are therefore
a good representation of complexity classes.

\section{Metacomplexity refutation and Heyting arrows}

\subsection{Graded Heyting algebra}\label{sec:GradedHeyting}

A graded Heyting algebra is not a standard object, but a new
definition we introduce in this work. We define it as a tuple
\[(L,\le,\wedge,\vee,\bot,\top,\{\to_k\}_{k\in\mathbb{N}})\]
where $L$ is a set of elements, $\le$ is a preorder on $L$,
$\wedge$ and $\vee$ are binary operations satisfying
the meet and join properties for $\le$ respectively,
and $\bot$ and $\top$ are a bottom and a top element for $\le$
respectively (note that in a preorder, meets, joins, bottom, and top
are not unique; the definition asks us to pick specific choices
for these). Moreover, for each $k\in\mathbb{N}$, we have a binary
operation $\to_k$ on $L$ satisfying the following conditions:
\begin{enumerate}
\item Monotonicity: if $x,y\in L$ and $k,\ell\in\mathbb{N}$ with
$k\le \ell$, then $(x\to_k y)\le (x\to_\ell y)$.
\item Modus ponens: $x\wedge (x\to_k y)\le y$ for all $x,y\in L$
and $k\in\mathbb{N}$.
\item Deduction: For all $x,y,z\in L$, if $x\wedge z\le y$,
then there exists $k\in \mathbb{N}$ such that $z\le (x\to_k y)$.
\end{enumerate}

A graded Heyting algebra is not quite a Heyting algebra, for two
reasons. First, we did not require antisymmetry: the preorder
is not a partial order, as two elements may satisfy $x\le y$ and
$y\le x$ without being equal. Second, instead of having
a single operation $\to$, we have an increasing sequence of them,
$\to_k$. To address the first issue, one may be tempted
to move to the equivalence classes of the partial order;
the problem is that $\to_k$ does not descend to a well-defined
binary operation on the equivalence classes.

However, many properties of Heyting algebras are still satisfied
by Heyting algebras with the addition of quantifiers over $k$.
The structure is therefore still useful. For example, every graded
Heyting algebra satisfies distributivity.

\begin{lemma}\label{lem:GradedHeytingDistributive}
If $L$ is a graded Heyting algebra, then
$x\wedge (y\vee z)\le (x\wedge y)\vee (x\wedge z)$,
so the preorder lattice $(L,\le)$ is distributive.
\end{lemma}

\begin{proof}
Let $w=(x\wedge y)\vee (x\wedge z)$.
We know $x\wedge y\le w$ and $x\wedge z\le w$ by the join
property of $\vee$. Deduction gives $z\le (x\to_{k_1} w)$
and $y\le (x\to_{k_2} w)$ for some $k_1,k_2\in\mathbb{N}$.
Letting $k=\max\{k_1,k_2\}$ and using monotonicity
(and transitivity),
we get $z\le (x\to_k w)$ and $y\le (x\to_k w)$.
By the join property, $(y\vee z)\le (x\to_k w)$.
Applying meet with $x$ to both sides (meet is monotone),
we get $x\wedge (y\vee z)\le x\wedge (x\to_k w)$.
Using modus ponens, we get $x\wedge (x\to_k w)\le w$,
so by transitivity $x\wedge (y\vee z)\le w$,
as desired.
\end{proof}

Our main result in this section is that the set
of polytime verifiable search problems,
ordered by mapping reducibility $\le$,
is a graded Heyting algebra. Note that together with
\cref{lem:GradedHeytingDistributive}, this gives
a reproof of \cref{lem:distributive} for the fact that
$(\mathsf{Verifiable},\le)$ is distributive.
The graded arrows $\to_k$ will then allow us to interpret
the ideal completion of $\mathsf{Verifiable}$
(which is fully a Heyting algebra rather than merely
a graded one).

\begin{definition}\label{def:tok}
Let $R,S\in\mathsf{Search}$ and let $k\in\mathbb{N}$.
Define the search problem $(R\to_k S)\subseteq\{0,1\}^*\times\{0,1\}^*$
as follows. The inputs are pairs $\langle f,g\rangle$
of Turing machines (with $f$ intended to compute a function
from strings to strings and $g$ intended to compute a function
from pairs of strings to strings);
inputs $z$ not of this $\langle f,g\rangle$ form are considered
poorly-formatted and match to all possible output strings
(meaning $(z,w)\in (R\to_k S)$ for all strings $w$ whenever $z$
is poorly-formatted).
For a well-formatted input $\langle f,g\rangle$ to $R\to_k S$
valid outputs are defined to be pairs
$\langle x,y\rangle$ of strings with any of the following properties:
\begin{enumerate}
\item $f(x)$ takes more than $p_k(|x|+|\langle f\rangle|)$ steps, or
\item $g(x,y)$ takes more than $p_k(|x|+|y|+|\langle g\rangle|)$
steps, or
\item the output strings $f(x)$ and $g(x,y)$ satisfy
$(f(x),y)\in S$ and $(x,g(x,y))\notin R$.
\end{enumerate}
\end{definition}

(Recall that $p_k$ is the polynomial $p_k(n)=n^k+k$.)

\begin{lemma}\label{lem:toVerifiable}
If $R,S$ are verifiable then
so is $(R\to_k S)$ for any $k\in\mathbb{N}$.
In other words, $\mathsf{Verifiable}$ is closed under $\to_k$.
\end{lemma}

\begin{proof}
We construct a verifier $V$ for $R\to_k S$. Let $V_R$ and $V_S$
be verifiers for $R$ and $S$ respectively.
$V(z,w)$ first checks if $z$ is formatted like $\langle f,g\rangle$
with $f$ and $g$ Turing machines of the correct form; if not,
it accepts. Next, $V(z,w)$ checks if $w$ is formatted like
$\langle x,y\rangle$ for strings $x$ and $y$; if not, it rejects.
Next, $V(\langle f,g\rangle,\langle x,y\rangle)$ computes
the bound $p_k(|x|+|\langle f\rangle|)$ and simulates $f(x)$
for that many steps; if it fails to halt, $V$ accepts. Similarly,
it computes the bound $p_k(|x|+|y|+|\langle g\rangle|)$
and simulates $g(x,y)$ for that many steps; if it fails to halt,
$V$ accepts. Finally, if both runs halted, $V$ is in possession
of the output strings $f(x)$ and $g(x,y)$, and it runs $V_S(f(x),y)$
and $V_R(x,g(x,y))$, accepting if both
(a) the former accepts and (b) the
latter rejects; otherwise, $V$ rejects.

The correctness of $V$ should be clear: it follows precisely
the definition of $R\to_k S$ in deciding to accept or reject
a pair $(z,w)$. The efficiency of $V$ follows from the fact
that it only needs to simulate the input machines for $p_k$
steps, and it does so on strings that were part of the input
to $V$. Additionally, $V$ simulates $V_R$ and $V_S$,
and performs formatting checks.
\end{proof}

\begin{lemma}\label{lem:monotonicity}
If $R,S$ are search problems and $k,\ell\in \mathbb{N}$
with $k\le \ell$, then $(R\to_k S)\le (R\to_{\ell} S)$.
\end{lemma}

\begin{proof}
We use the identity reduction. We need to show
$(R\to_k S)\le_I (R\to_\ell S)$, which is equivalent to showing
$(R\to_k S)\supseteq (R\to_\ell S)$. Suppose $(z,w)\in (R\to_\ell S)$.
If $z$ is poorly formatted, then $(z,w)\in (R\to_k S)$ and we are
done, so suppose
$z=\langle f,g\rangle$; since $(z,w)\in(R\to_\ell S)$
and $z$ is well-formatted, we must have $w=\langle x,y\rangle$.
If $f$ or $g$ take too long with respect to $k$,
then $(\langle f,g\rangle,\langle x,y\rangle)\in (R\to_k S)$
and we are done. So suppose both $f$ and $g$ don't take too
long with respect to $k$. Since $\ell\ge k$ and
$p_\ell(n)\ge p_k(n)$ for all $n\in\mathbb{N}$, it must
be the case that they also don't take too long with respect to $\ell$.
Since $(\langle f,g\rangle,\langle x,y\rangle)\in (R\to_\ell S)$,
the only remaining possibility is that $(f(x),y)\in S$
and $(x,g(x,y))\notin R$, in which case we once again
get $(\langle f,g\rangle,\langle x,y\rangle)\in (R\to_k S)$,
as desired.
\end{proof}

\begin{lemma}[Graded modus ponens]\label{lem:GradedModusPonens}
Let $R,S$ be search problems and let $k\in\mathbb{N}$.
Assume that $R\in\mathsf{Verifiable}$.
Then $R\wedge (R\to_k S)\le S$.
\end{lemma}

\begin{proof}
The idea is simple: the input to $R\wedge (R\to_k S)$
gives us an input $x$ to $R$ and a claimed reduction
$(f,g)$ from $R$ to $S$, so we can simply run $f(x)$
to get an $S$ input and then use $g$ to convert a solution
to that $S$-input back to an output for $R$;
this either solves the $R$-input $x$, or else it leads
to a refutation of the reduction $(f,g)$, solving
the $(R\to_k S)$ input. We now go through the details.

We define a reduction $(F,G)$ from $R\wedge (R\to_k S)$ to $S$
as follows.
On input $z=\langle x,\langle f,g\rangle\rangle$,
the machine $F(z)$ simulates $f(x)$ for
$p_k(|x|+|\langle f\rangle|)$ steps, and if it halts,
$F(z)$ outputs $f(x)$. If $f(x)$ doesn't halt in that time,
or if the input $z$ is not formatted this way, then
$F(z)$ outputs arbitrarily.
$G(z,y)$ outputs $\langle y',1\rangle$ for an arbitrary $y'$
if $z$ is not formatted like $\langle x,\langle f,g\rangle\rangle$.
Otherwise, it simulates $g(x,y)$ for
$p_k(|x|+|y|+|\langle g\rangle|)$ steps, and if $g$ fails
to halt $G(z,y)$ outputs
$\langle\langle x,y\rangle,1\rangle$.
If $g(x,y)$ halts in that time, $G$ simulates
$V_R(x,g(x,y))$, where $V_R$ is the verifier for $R$.
If $V_R(x,g(x,y))$ accepts, $G(z,y)$ returns
$\langle g(x,y),0\rangle$.
If it rejects, $G(z,y)$ returns
$\langle \langle x,y\rangle,1\rangle$.

$F$ and $G$ simulate the input programs $f$ and $g$,
but since this simulation is bounded by $p_k(\cdot)$ and
$k$ is constant, this simulation is efficient. $G$ also
runs the verifier $V_R$ for $R$, which is efficient by
our assumption. Other than this, they only do basic formatting
checks and some string manipulation, so $F$ and $G$ run in
polynomial time.

Suppose $(F(z),y)\in S$. If $z$ is poorly formatted,
then $G(z,y)=\langle y',1\rangle$. In this case, consider the case
in which $z$ is not
formatted as a pair of strings, so it is poorly formatted
even for $\wedge$ of two search problems. In that case,
$(z,G(z,w))\in R\wedge (R\to_k S)$ regardless of the output
of $G(z,w)$. Suppose then that $z=\langle x,w\rangle$ for
some pair of strings $x$ and $w$, with $w$ poorly formatted
for $R\to_k S$. Then $(w,y')\in (R\to_k S)$ for any $y'$.
By definition of $\wedge$, we then get
$(\langle x,w\rangle, \langle y',1\rangle)\in R\wedge (R\to_k S)$,
which gives $(z,G(z,y))\in R\wedge (R\to_k S)$ as desired.

Suppose then that $z=\langle x,\langle f,g\rangle\rangle$.
Note that $G(z,y)$ can only return the tuple
$\langle g(x,y),0\rangle$
if $V_R(x,g(x,y))$ accepted. Therefore, if $G(z,y)$ returns
$\langle g(x,y),0\rangle$, then $(x,g(x,y))\in R$,
which means $(\langle x,\langle f,g\rangle\rangle,
\langle g(x,y),0\rangle)\in R\wedge (R\to_k S)$
by the definition of $\wedge$. This gives
$(z,G(z,y))\in R\wedge (R\to_k S)$, completing the proof
in the case where $G(z,y)$ returns $\langle g(x,y),0\rangle$.

It remains to handle the case where $z$ is well-formatted and
$G(z,y)$ returns $\langle\langle x,y\rangle,1\rangle$.
If $f(x)$ does not halt within $p_k(|x|+|\langle f\rangle|)$
steps or
$g(x,y)$ does not halt within $p_k(|x|+|y|+|\langle g\rangle|)$ 
steps, then by definition of $R\to_k S$ we have
$(\langle f,g\rangle,\langle x,y\rangle)\in (R\to_k S)$.
Then by definition of $\wedge$ we have
$(\langle x,\langle f,g\rangle\rangle,
\langle\langle x,y\rangle,1\rangle)\in R\wedge (R\to_k S)$,
which means $(z,G(z,y))\in R\wedge (R\to_k S)$ as desired.

Finally, we handle the case in which
$z=\langle x,\langle f,g\rangle\rangle$, $f(x)$ and $g(x,y)$
terminate within their budget, and $G(z,y)$ returns
$\langle \langle x,y\rangle,1\rangle$. This can only
happen when $V_R(x,g(x,y))$ rejects, so we know that
$(x,g(x,y))\notin R$.
Now, the condition $(F(z),y)\in S$ becomes $(f(x),y)\in S$.
We therefore conclude that $(f(x),y)\in S$ but
$(x,g(x,y))\notin R$, and by definition that gives us
$(\langle f,g\rangle,\langle x,y\rangle)\in (R\to_k S)$
for any $k$. By definition of $\wedge$, we then get
$(\langle x,\langle f,g\rangle\rangle,
\langle \langle x,y\rangle,1\rangle)\in R\wedge (R\to_k S)$,
completing the proof.
\end{proof}

\begin{lemma}[Graded deduction]\label{lem:GradedDeduction}
Let $R,S,T$ be search problems and suppose that
$R\wedge T\le S$. Then there exists $k\in\mathbb{N}$ such that
$T\le (R\to_k S)$.
\end{lemma}

\begin{proof}
The idea is as follows: if we have a reduction from
$R\wedge T$ to $S$ which runs in polynomial time,
and we are given a $T$ input, we can use currying
to get a reduction from $R$ to $S$ (giving
an input to $R\to_k S$). Refuting that $R$-to-$S$
reduction forces larger reduction (from $R\wedge T$ to $S$)
to correctly solve the $T$ input, so we can convert valid
outputs for the $R\to_k S$ problem back into valid outputs
for the $T$ input. We now go through the details.

Let $(F',G')$ reduce $R\wedge T$ to $S$. We define a reduction
$(F,G)$ from $T$ to $R\to_k S$, where we choose $k\in\mathbb{N}$
later. Define
$F(x)=\langle f_x,g_x\rangle$ with $f_x(u)=F'(\langle u,x\rangle)$
and $g_x(u,v)=G'(\langle u,x\rangle,v)_-$, where the subscript
``$-$'' denotes the operation ``if the output string is
$\langle z,b\rangle$ with $b\in\{0,1\}$, remove the bit $b$ and return $z$''.
Let $G(x,\langle u,v\rangle)=G'(\langle u,x\rangle,v)_-$,
with $G(x,w)$ returning arbitrarily if $w$ is not of the form
$\langle u,v\rangle$ for strings $u$ and $v$.

Note that $F$ is efficient as it is just doing surgery on Turing
machines (it need not run them).
$G$ is efficient because $G'$ is assumed to be efficient.
The machine $f_x$ that $F(x)$ outputs simulates
$F'(\langle u,x\rangle)$ on input $u$. Note that $F'$
is efficient, so there is some $i$ such that it runs in $p_i(n)$
steps on all inputs of size $n$ and all $n\in\mathbb{N}$.
It is called by $f_x(u)$ on an input of size
$|\langle u,x\rangle|$,
which we can ensure is less than $|u|+|\langle f_x\rangle|$
by setting the encoding of $f_x$ to contain $x$ and some overhead.
The run $f_x(u)$ may be a bit slower than $F'(\langle u,x\rangle)$
due to overhead, but there is still some integer $i'\in\bN$
such that $f_x(u)$ always terminates within 
$p_{i'}(|u|+|\langle f_x\rangle|)$ steps, for all $u$ and $x$.
Similarly, $g_x(u,v)$ simulates $G'$ on input of size roughly
$|u|+|x|+|v|$, which can be made at most
$|u|+|v|+|\langle g_x\rangle|$ by encoding $g_x$ to contain the string $x$.
Since $G'$ is efficient, there is some integer $j$ such that
$g_x(u,v)$ terminates within $p_j(|u|+|v|+|\langle g_x\rangle|)$
steps for all $u$, $v$, and $x$. Taking $k=\max\{i',j\}$,
both $f_x$ and $g_x$ will always stay within the budget $p_k$
on all possible inputs.

We now show $(F,G)$ reduces $T$ to $R\to_k S$ for that choice
of $k$. Suppose $(F(x),y)\in (R\to_k S)$. Since
$F(x)=\langle f_x,g_x\rangle$ is correctly formatted,
the definition of $R\to_k S$ makes it so that $y$ must have
the formatting $y=\langle u,v\rangle$ for some strings $u$ and $v$.
Then $(\langle f_x,g_x\rangle,\langle u,v\rangle)\in (R\to_k S)$.
By our choice of $k$, $f_x(u)$ and $g_x(u,v)$ terminate within
the required number of steps, which means the only way for
$(\langle f_x,g_x\rangle,\langle u,v\rangle)\in (R\to_k S)$ to
be satisfied is to have $(f_x(u),v)\in S$ and
$(u,g_x(u,v))\notin R$.

The condition $(f_x(u),v)\in S$ gives
$(F'(\langle u,x\rangle),v)\in S$, and the reduction
property of $(F',G')$ from $R\wedge T$ to $S$ gives
$(\langle u,x\rangle,G'(\langle u,x\rangle,v))\in R\wedge T$.
By definition of $\wedge$, we know $G'(\langle u,x\rangle,v)$
has the form $\langle z,b\rangle$ for a string $z$ and bit $b$.

The condition $(u,g_x(u,v))\notin R$ gives
$(u,G'(\langle u,x\rangle, v)_-)\notin R$,
which means $(u,z)\notin R$; since we had
$(\langle u,x\rangle,\langle z,b\rangle)\in R\wedge T$,
this forces $b=1$ and $(x,z)\in T$.
Finally, since
$G(x,\langle u,v\rangle)=G'(\langle u,x\rangle,v)_-=z$,
we get $(x,G(x,y))\in T$, showing that $(F,G)$
is a reduction from $T$ to $R\to_k S$ as claimed.
\end{proof}

In any graded Heyting algebra, we can define
$\neg_k x\coloneqq (x\to_k \bot)$. For search problems, we define
things a little differently, but in a way that turns out to be
equivalent for $k\ge 2$. We will define $\neg S$ to be the
refutation problem.

\begin{definition}
Let $S$ be any search problem and let $k\in\mathbb{N}$.
Define $\neg_k S$ to be the following search problem.
If the input $z$ is not formatted like $\langle A\rangle$
for a Turing machine $A$, all outputs are accepted
(i.e. $(z,y)\in \neg_k S$ for all $y$). If $z=\langle A\rangle$,
we say $(\langle A\rangle,x)\in \neg_k S$ if either
$A(x)$ fails to halt within $p_k(|x|+|\langle A\rangle|)$ steps
(where $p_k(n)=n^k+k$), or else $(x,A(x))\notin S$.
If both conditions fail, i.e. if $A(x)$ halts in time and solves
$x$ for $S$, we define $(\langle A\rangle,x)\notin \neg_k S$.
\end{definition}

\begin{lemma}\label{lem:GradedNegation}
Let $S$ be any search problem, and let $k\ge 2$.
then $\neg_k S$ is equivalent to $S\to_k \mathtt{triv}$,
where recall $\mathtt{triv}$ is the set of all pairs of strings.
\end{lemma}

This lemma should be intuitively obvious: a reduction from $S$
to the trivial problem is just an algorithm for $S$, so the task
of refuting such a reduction is equivalent to the task
of refuting an algorithm for $S$. The proof is a bit long
due to unimportant technicalities. One such technicality is
preserving the exact value of $k$ in these reductions
for $k\ge 2$. This is actually not necessary, since all our applications
always quantify over $k$ (so changing $k$ to $k+1$ does not affect
anything); however, it will be more notationally convenient to
keep the exact value of $k$ the same in $\neg_k S$
and $S\to_k \mathtt{triv}$.

\begin{proof}
We show reductions in both directions. First, we reduce
$S\to_k \mathtt{triv}$ to $\neg_{k} S$ using a reduction $(F,G)$
defined as follows. If $z$ is not of the form $\langle f,g\rangle$,
$F(z)$ behaves arbitrarily. Otherwise,
$F(\langle f,g\rangle)=\langle g'\rangle$,
where $g'$ is the Turing machine with $g'(x)=g(x,\epsilon)$.
We implement $g'$ by having it replace the input $x$
on its input tape with $\langle x,\epsilon\rangle$ before running
$g$; this requires only additive linear overhead in $|x|$.
Note that
$p_k(|x|+|\langle g'\rangle|)-p_k(|x|+|\langle g\rangle|)
\ge p_2(|x|+|\langle g'\rangle|)-p_2(|x|+|\langle g\rangle|)
\ge 2|x|(|\langle g'\rangle|-|\langle g\rangle|)$.
We ensure the code length $|\langle g'\rangle|$ is long
enough compared to $|\langle g\rangle|$
so that $2|x|(|\langle g'\rangle|-|\langle g\rangle|)$
is greater than the additive linear time overhead of the behavior
of $g'$, for all $x$. This ensures $g'(x)$ runs in time
$p_k(|x|+|\langle g'\rangle|)$
whenever $g(x,\epsilon)$ runs in time
$p_k(|x|+|\epsilon|+|\langle g\rangle|)$.
We then define
$G(z,w)=\langle w,\epsilon\rangle$.

Note that $F$ and $G$ run in polynomial time. ($F$ only
does surgery on Turing machines, and does not need to run $g$.)

Suppose
$(F(z),x)\in \neg_k S$. If $z$ is poorly formatted,
then $(z,w)\in (S\to_k\mathtt{triv})$ for all $w$, and hence
$(z,G(z,x))\in (S\to_k\mathtt{triv})$. If $z=\langle f,g\rangle$,
then $F(z)=\langle g'\rangle$, so our assumption gives
$(\langle g'\rangle,x)\in \neg_k S$. Since $\langle g'\rangle$
is formatted like the code of a Turing machine, this means
either $g'(x)$ exceeds its time budget of
$p_k(|x|+|\langle g'\rangle|)$, or else $(x,g'(x))\notin S$.
In the former case, the run $g(x,\epsilon)$ must exceed
its time budget of
$p_k(|x|+|\epsilon|+|\langle g\rangle|)$ by our construction
of $g'$.
This means $g$ exceeds its time budget on $(x,\epsilon)$
with respect to the task $S\to_k\mathtt{triv}$, so
$(\langle f,g\rangle,\langle x,\epsilon\rangle)\in (S\to_k \mathtt{triv})$.
Hence $(z,G(z,x))\in (S\to_k\mathtt{triv})$, as desired.

In the latter case, we have $(x,g'(x))\notin S$,
so $(x,g(x,\epsilon))\notin S$. Now, if $f(x)$ fails to halt
in $p_k(|x|+|\langle f\rangle|)$ steps, then $\langle x,\epsilon\rangle$
is a valid output pair for $\langle f,g\rangle$ with respect to
$S\to_k\mathtt{triv}$. Otherwise, the string $f(x)$ is well-defined
(that is, the run $f(x)$ halts), and
then $(f(x),\epsilon)\in\mathtt{triv}$
(since all pairs of strings are in $\mathtt{triv}$).
In this case the pair $\langle x,\epsilon\rangle$ is 
counterexample to the reduction $\langle f,g\rangle$ from
$S$ to $\mathtt{triv}$, so it is a valid output pair for
$S\to_k \mathtt{triv}$. In all cases,
$(\langle f,g\rangle,\langle x,y\rangle)\in (S\to_k \mathtt{triv})$,
so $(z,G(z,x))\in (S\to_k\mathtt{triv})$, as desired.

We now give a reduction $(F,G)$ in the reverse direction,
from $\neg_k S$ to $S\to_k \mathtt{triv}$.
Define $F(\langle g\rangle)=\langle f,g'\rangle$ where $f$ is any
function that runs very fast (in time $p_2(|x|+|\langle f\rangle|)$
on all inputs $x$) and $g'$ is defined by $g'(x,y)=g(x)$.
We implement $g'$ such that $g'(x,y)$ it runs in time at most
$p_k(|x|+|y|+|\langle g'\rangle|)$ whenever $g(x)$ runs in time
at most $p_k(|x|+|\langle g\rangle|)$;
this is possible because $g'(x,y)$ just needs to delete $y$
from the input (which takes linear time) and then run $g$,
and the code length of $g'$ can be an additive constant larger than
that of $g$, and also we are assuming $k\ge 2$.

Define $G(z,\langle x,y\rangle)=x$, with $G$ acting arbitrarily
if its second argument is not of the form $\langle x,y\rangle$.
It is clear that $F$ and $G$ run in polynomial time ($F$ only does
surgery on Turing machines and does not need to run $g$).

Suppose $(F(z),w)\in (S\to_k \mathtt{triv})$. If $z$ is not formatted
like $\langle g\rangle$, then $(z,x)\in \neg_k S$ for any $x$,
so in particular $(z,G(z,w))\in \neg_k S$.
Otherwise, $F(z)=\langle f,g\rangle$ for that constant value of $f$,
and our assumption becomes
$(\langle f,g'\rangle,w)\in (S\to_k \mathtt{triv})$.
This means $w$ must be formatted like $\langle x,y\rangle$
for strings $x$ and $y$. Moreover, by the choice of $f$,
$f(x)$ can never exceed its runtime budget of
$p_k(|x|+|\langle f\rangle|)$ steps; the condition
$(\langle f,g\rangle,\langle x,y\rangle)\in (S\to_k \mathtt{triv})$
then implies that either $g'(x,y)$ exceeds the time budget
of $p_k(|x|+|y|+|\langle g\rangle|)$, or else
$(f(x),y)\in \mathtt{triv}$ and $(x,g'(x,y))\notin S$.

In the former case, it must be the case that $g(x)$ exceeds
its time budget of $p_k(|x|+|\langle g\rangle|)$ by our construction
of $g'$. In this case, $(\langle g\rangle,x)\in\neg_k S$,
which means $(z,G(z,w))\in \neg_k S$.

In the latter case, we get $(x,g'(x,y))\notin S$, which means
$(x,g(x))\notin S$. Then we once again have
$(\langle g\rangle,x)\in\neg_k S$, so in all cases
$(z,G(z,w))\in \neg_k S$, as desired.
\end{proof}

The following lemma will be useful.

\begin{lemma}\label{lem:GradedArrowMonotonicity}
In a graded Heyting algebra, if
$a\le b$, then for all $k\in\mathbb{N}$
there exists $\ell\in\mathbb{N}$ such that
$(b\to_k x) \le (a\to_\ell x)$
and $(x\to_k a)\le (x\to_\ell b)$.
\end{lemma}

\begin{proof}
$a\wedge (b\to_k x)$ is below $(b\to_k x)$
and below $a\le b$, so it is below the meet of
$b\to_k x$ and $b$, i.e.
$a\wedge (b\to_k x)\le b\wedge(b\to_k x)\le x$.
But then there exists $\ell$ such that
$(b\to_k x)\le (a\to_\ell x)$, giving the first result.

The second follows from $x\wedge (x\to_k a)\le a\le b$,
which implies the existence of $\ell$
with $(x\to_k a)\le (x\to_\ell b)$.
\end{proof}

\subsection{Heyting arrows in the ideal completion are
refutation classes}

We now characterize the Heyting operations of arrow
and negation for the ideal completion of
$\mathsf{Verifiable}$. These operations turn out to be
complexity-class level version of the metacomplexity
refutation tasks $\to_k$ and $\neg_k$ from the
previous section.

Recall that for two ideals $\mathsf{A},\mathsf{B}$
of $\mathsf{Verifiable}$, the Heyting arrow
$\mathsf{A}\to\mathsf{B}$ was defined as the largest
ideal $\mathsf{C}$ with the property that
$\mathsf{C}\cap\mathsf{A}\subseteq\mathsf{B}$.
Recall also that $\neg\mathsf{A}$ was defined as
$\mathsf{A}\to\bot$, where the bottom element
$\bot$ was $\mathsf{FP}$; thus
$\neg\mathsf{A}$ is the largest ideal $\mathsf{C}$
with the property that $\mathsf{C}\cap\mathsf{A}=\mathsf{FP}$.
We now give characterizations of these in terms
of metacomplexity refutation tasks.

For a set of verifiable search problems $\mathsf{A}$, we use
$\downcl\mathsf{A}$ to denote its downwards closure under
mapping reductions. For a single search problem $R$,
we abuse notation and write $\downcl R$ for $\downcl\{R\}$;
this is the complexity class for which $R$ is a
complete problem.

We characterize the Heyting arrow of principal ideals first
(that is, of complexity classes with complete problems).
Then we characterize the Heyting arrow of general
ideals. Both of the following lemmas amount to saying
that the Heyting arrow of two ideals
consists of all verifiable search problems that reduce
to a refutation task for refuting a reduction between
a problem in the first ideal to a problem in the
second.

\begin{lemma}\label{lem:PrincipalHeytingArrowChar}
Let $R,S\in\mathsf{Verifiable}$. Then
$\downcl R\to\downcl S=\downcl\{R\to_k S:k\in\mathbb{N}\}$.
\end{lemma}

\begin{proof}
First note that
$\downcl\{R\to_k S:k\in\mathbb{N}\}
\cap\downcl R\subseteq\downcl S$.
This is because if $T\in \downcl\{R\to_k S:k\in\mathbb{N}\}
\cap\downcl R$, then $T\le R\to_k S$ for some $k\in\mathbb{N}$
and $T\le R$; but then by the meet property
$T\le R\wedge (R\to_k S)$, and by \cref{lem:GradedModusPonens}
and transitivity we get $T\le S$, which gives
$T\in\downcl S$.

Next let $\mathsf{C}$ be an ideal of $\mathsf{Verifiable}$
with $\downcl R\cap \mathsf{C}\subseteq\downcl S$.
If $T\in\mathsf{C}$, then $\downcl T\subseteq\mathsf{C}$,
so $\downcl R\cap\downcl T\subseteq\downcl S$.
But $\downcl R\cap\downcl T=\downcl(R\wedge T)$ by the
meet property, so we get
$\downcl(R\wedge T)\subseteq \downcl S$ and hence
$R\wedge T\le S$. Then \cref{lem:GradedDeduction}
gives some $k\in\mathbb{N}$ such that
$T\le R\to_k S$, which means
$T\in \downcl\{R\to_k S:k\in\mathbb{N}\}$.
Since $T\in\mathsf{C}$ was arbitrary, we get
$\mathsf{C}\subseteq \downcl\{R\to_k S:k\in\mathbb{N}\}$,
which means $\downcl\{R\to_k S:k\in\mathbb{N}\}$ is the
largest ideal whose intersection with $\downcl R$
is contained in $\downcl S$.
\end{proof}

\begin{lemma}\label{lem:HeytingArrowChar}
Let $\mathsf{A}$ and $\mathsf{B}$ be ideals of
$\mathsf{Verifiable}$. Then the Heyting arrow
$\mathsf{A}\to\mathsf{B}$
(which was defined as the maximum ideal $\mathsf{C}$
of $\mathsf{Verifiable}$ with the property that
$\mathsf{C}\cap\mathsf{A}\subseteq\mathsf{B}$)
is equal to \[\bigcap_{R\in\mathsf{A}}\bigcup_{S\in\mathsf{B}}
(\downcl R\to\downcl S).\]
In other words, $\mathsf{A}\to\mathsf{B}$
consists of all verifiable search problems $T$ with the
property that for each $R\in\mathsf{A}$ there exists
$S\in\mathsf{B}$ and $k\in\mathbb{N}$ such that $T$
reduces to $R\to_k S$ (the task of refuting an
$O(n^k)$-time reduction from $R$ to $S$).
\end{lemma}

\begin{proof}
Let $\mathsf{C}=\bigcap_{R\in\mathsf{A}}\bigcup_{S\in\mathsf{B}}
(\downcl R\to\downcl S)$.
We first show $\mathsf{C}\cap\mathsf{A}\subseteq\mathsf{B}$,
which implies that
$\mathsf{C}\subseteq (\mathsf{A}\to\mathsf{B})$.
Let $T\in \mathsf{C}\cap\mathsf{A}$. Then
$T\in \mathsf{A}$ and for each $R\in\mathsf{A}$
there exists $S\in\mathsf{B}$ such that
$T\in(\downcl R\to\downcl S)$. Picking $R=T$,
there exists $S$ such that $T\in(\downcl T\to\downcl S)$.
Since $\downcl T\to\downcl S$ is downwards closed
(being an ideal), we get
$\downcl T\subseteq (\downcl T\to\downcl S)$.
Then
$\downcl T=\downcl T\cap (\downcl T\to\downcl S)\subseteq\downcl S\subseteq \mathsf{B}$,
which gives $T\in\mathsf{B}$, as desired.

We now show the reverse inclusion, that
$(\mathsf{A}\to\mathsf{B})\subseteq \mathsf{C}$.
Let $T\in(\mathsf{A}\to\mathsf{B})$.
Then $\downcl T$ is below $(\mathsf{A}\to\mathsf{B})$
in the ideal completion order, so we must have
$\downcl T\cap\mathsf{A}\subseteq \mathsf{B}$.
In particular, for any $R\in\mathsf{A}$, we have
$\downcl T\cap\downcl R\subseteq\mathsf{B}$.
Moreover, there is a single search problem $S\in\mathsf{B}$
such that $\downcl T\cap\downcl R\subseteq\downcl S$
(indeed, we can take $S=T\wedge R$, and then by the meet
property
$\downcl S=\downcl T\cap\downcl R\subseteq\mathsf{B}$,
which implies $S\in\mathsf{B}$).
By the definition of the Heyting arrow, we get
$\downcl T\subseteq (\downcl R\to\downcl S)$.
This shows that for each $R\in\mathsf{A}$ there exists
$S\in\mathsf{B}$ such that
$\downcl T\subseteq (\downcl R\to\downcl S)$,
from which we conclude that
$\downcl T\subseteq\bigcap_{R\in\mathsf{A}}
\bigcup_{S\in\mathsf{B}}(\downcl R\to\downcl S)$.
Thus $T\in\mathsf{C}$, as desired.
\end{proof}

We next characterize the Heyting negations in terms
of refutation tasks.

\begin{lemma}\label{lem:HeytingNegationChar}
Let $R\in\mathsf{Verifiable}$. Then
$\neg \downcl R=\downcl\{\neg_k R:k\in\mathbb{N}\}$,
where $\neg \downcl R$ is the Heyting negation and
$\neg_k R$ is the $O(n^k)$ refutation task.

Moreover, let $\mathsf{C}$ be an ideal of 
$\mathsf{Verifiable}$. Then
$\neg\mathsf{C}=\bigcap_{R\in\mathsf{C}}\neg\downcl R$.
\end{lemma}

\begin{proof}
The Heyting negation is defined by
$\neg\mathsf{C}=(\mathsf{C}\to\bot)$, and in our ideal
completion lattice we have
$\bot=\mathsf{FP}=\downcl\mathtt{triv}$.
Using \cref{lem:PrincipalHeytingArrowChar}
and \cref{lem:GradedNegation}, we get
\[\neg\downcl R=(\downcl R\to\downcl\mathtt{triv})
=\downcl\{R\to_k\mathtt{triv}:k\in\mathbb{N}\}
=\downcl\{\neg_k R:k\in\mathbb{N}\}.\]

Using \cref{lem:HeytingArrowChar}, we get
\[\neg\mathsf{C}=(\mathsf{C}\to\downcl\mathtt{triv})
=\bigcap_{R\in\mathsf{C}}\bigcup_{S\le \mathtt{triv}}
\bigcup_{k\in\mathbb{N}}\downcl(R\to_k S).\]
By \cref{lem:GradedArrowMonotonicity},
for all $S\le \mathtt{triv}$ and all $k$ we have
some $\ell$ such that
$(R\to_k S)\le (R\to_\ell\mathtt{triv})$.
It follows that
\[\neg\mathsf{C}=\bigcap_{R\in\mathsf{C}} 
\bigcup_{\ell\in\mathbb{N}} \downcl (R\to_\ell\mathtt{triv})
=\bigcap_{R\in\mathsf{C}}\neg\downcl R,\]
as desired.
\end{proof}

\cref{thm:main} now follows from \cref{lem:PrincipalHeytingArrowChar},
\cref{lem:HeytingArrowChar}, and \cref{lem:HeytingNegationChar}, together with the meet
and join properties of an ideal completion.

\subsection{Discussion: was this just symbol pushing?}

Some of the proofs in this section may have had the feel
of being abstract symbol-pushing. We want to clarify the
main insights.

Essentially the entire proof of \cref{thm:main}
follows from two lemmas: the graded modus ponens and deduction
properties of the lattice of verifiable search problems
(\cref{lem:GradedModusPonens} and \cref{lem:GradedDeduction}).
Nearly everything else so far can be viewed as formalism
surrounding those two technical lemmas.

However, we want to emphasize that a key contribution
required for \cref{thm:main}
is the correct setup of the definitions. We needed
the following insights:
\begin{enumerate}
\item Considering refutation tasks for \emph{reductions}
in the first place (not just the commonly-studied
refutation tasks for algorithms)
\item Setting everything up for search problems rather than
decision problems, and for Turing machines instead of circuits
\item Requiring verifiability, but not totality; handling
non-total tasks by allowing impossible problems instead
of via promises
\item Using time-bounded refutation tasks so that there's
a \emph{sequence} of refutation tasks (one for each $k$)
instead of a single one
\item Considering the ideal completion of the lattice
of search problems instead of the more common
Dedekind-MacNeille completion
\end{enumerate}
Though some of these might sound simple in hindsight,
all of these insights were difficult to come by, and it
took us several years to formalize the clean exposition
of \cref{thm:main} even
after having the intuition that something
like it should hold.

We also note that AI played no role
in the results of this section or the formulation of
\cref{thm:main}, except for older models like
ChatGPT o3 mini or later Gemini 3 pro
which we used primarily to help us learn
about order theory and intuitionistic logic;
we also tried brainstorming a good definitional setup
using these models, but this was not successful.
Any ``overly abstract'' notation
that the reader may perceive
in this section was purely the fault of the human authors
trying to manually
find good abstractions and should not be viewed as AI slop.

\paragraph{Benefits of \cref{thm:main}.}
Once we have \cref{thm:main}, we immediately get to use
intuitionistic logic to get results about refutation tasks.
Examples were given in \cref{sec:IntroLogicResults},
and include non-intermediacy results like
``the refutation tasks of end-of-line are either in
$\mathsf{FP}$ or outside of $\mathsf{PPAD}$''
(see \cref{cor:nonitermediacy} for a full statement
as well as the one-line proof).
In our view, the Heyting structure of metacomplexity
refutation tasks is both elegant and clarifying.

\section{Prime ideals}

We now give some prime ideals of $\mathsf{Verifiable}$.
These are ideals $\mathsf{C}$ which are proper
and have the property that
if $R\wedge S\in\mathsf{C}$ then either $R\in\mathsf{C}$
or $S\in\mathsf{C}$; in particular, since
$R\wedge \neg_k R\in\mathsf{FP}$ for all $k\in\mathbb{N}$,
and since $\mathsf{FP}\subseteq\mathsf{C}$ for each
ideal $\mathsf{C}$, a prime ideal satisfies the property
that for each verifiable search problem $R$ we either have
$R\in\mathsf{C}$ or else for all $k$,
$\neg_k R\in\mathsf{C}$. If $\mathsf{C}$ is a class of
``tractable'' problems, this says every search problem
is either tractable or has tractable refutation tasks.

\subsection{Preliminaries: nonprincipal ultrafilters}
\label{sec:nonprincipal}

In order theory, a filter is an ideal in the reverse order.
That is, it is nonempty, an upper set (upwards closed),
and closed under meets. A filter is called an ultrafilter
if it is a maximal proper filter (that is, it is not the
entire lattice and is not contained in any other proper
filter). A filter $F$ is called principal if it is generated
by a single element (that is, if there is some $x$
such that $F=\{y:x\le y\}$). A filter which is not
principal is called nonprincipal.

In a bounded lattice, Zorn's lemma can be used to show that
each proper filter is contained in an ultrafilter.

We will be interested in ultrafilters on the natural numbers.
The lattice in question will be the set of all subsets of
$\mathbb{N}$, ordered by $\subseteq$. A filter is then
a nonempty collection of subsets of $\mathbb{N}$ which
is closed under supersets and under finite intersection;
the sets in the filter can be viewed as ``large'' subsets
of $\mathbb{N}$, and we require that a superset of a large
set is a large set, an intersection of two large sets is large,
and the collection of large sets is nonempty (so $\mathbb{N}$
itself is large). Note that a principal filter on the
natural numbers would be a filter that says ``a set is large
iff it is a superset of $A$''
for some fixed $A\subseteq\mathbb{N}$.
An example of a nonprincipal filter would be
the collection of all cofinite subsets of $\mathbb{N}$
(sets whose complement is finite).

It is not hard to show that any \emph{ultrafilter} $\mathcal{U}$
on $\mathbb{N}$ must have the property that for each
$S\subseteq\mathbb{N}$, either $S\in \mathcal{U}$ or else
$\mathbb{N}\setminus S\in \mathcal{U}$. Moreover,
we can construct a nonprincipal ultrafilter $\mathcal{U}$
by starting with the filter of cofinite subsets of $\mathbb{N}$
and extending it to an ultrafilter via Zorn's lemma. This is
``construction'' is inherently nonconstructive, as it uses
the axiom of choice. The resulting ultrafilter $\mathcal{U}$,
if viewed as a collection of ``large'' subsets of $\mathbb{N}$,
has the following properties:
\begin{enumerate}
\item No finite set is large
\item For each $A\in\mathbb{N}$, either $A$ is large or else
$\mathbb{N}\setminus A$ is large
\item a superset of a large set is large
\item the intersection of two large sets is large.
\end{enumerate}

We will need a nonprincipal ultrafilter $\mathcal{U}$ on
$\mathbb{N}$ is the constructions that follow.

\subsection{A prime ideal inside infinitely-often-(FP/poly)}

Fix a nonprincipal ultrafilter $\mathcal{U}$ on $\mathbb{N}$.
Recall the notation $p_k(n)=n^k+k$.

Let $\mathsf{(FP/poly)_{\mathcal{U}}}$ denote all verifiable
search problems
$S$ with the following property: for each $k\in\mathbb{N}$,
there exists $U\in\mathcal{U}$ and $\ell\in\mathbb{N}$ such
that for each $n\in U$ and $m\le p_k(n)$, there is a circuit
$C_{n,m}$ taking $m$ input bits and having size at most $p_{\ell}(n)$
such that $C_{n,m}$ solves $S$ on all inputs of size $m$.
In other words, we are requiring not just that a family
of circuits solves $S$ on all input sizes in a large set
$U\in\mathcal{U}$ of natural numbers,
but also that for any polynomial growth
rate $p_k$, we solve $S$ on input sizes which are within
$p_k$ of a number in that large set $U$. The order of quantifiers
lets us choose the large set $U$ and the polynomial growth rate
$p_\ell$ of the circuit size after we know the slack rate $p_k$
on the input size that we are required to handle.

We will show $\mathsf{(FP/poly)_{\mathcal{U}}}$ is a prime
ideal contained in $\mathsf{(FP/poly)_{io}}$. For this, we need
to show it is nonempty, proper, downwards closed under 
reductions, closed under binary joins, ``meet prime''
(i.e. if it contains a meet it contains
one of the two arguments), and contained in
$\mathsf{(FP/poly)_{io}}$.

Clearly, $\mathtt{triv}\in\mathsf{(FP/poly)_{\mathcal{U}}}$,
so the set is nonempty. Additionally,
$\mathtt{imp}\notin \mathsf{(FP/poly)_{\mathcal{U}}}$, so
the set is proper. If $S\in\mathsf{(FP/poly)_{\mathcal{U}}}$,
then fixing any $k$, like $k=1$, we get some $U\in\mathcal{U}$
and $\ell\in\mathbb{N}$ such that for all $n\in U$,
$C_{n,n}$ solves $S$ on input size $n$ and has size at most
$p_{\ell}(n)$; since no set in the ultrafilter $\mathcal{U}$
can be finite, this means we get a family of circuits
of size at most $p_\ell$ which solves $S$ on the infinite
set of input sizes $U$, so $S\in\mathsf{(FP/poly)_{io}}$.
Thus
$\mathsf{(FP/poly)_{\mathcal{U}}}\subseteq \mathsf{(FP/poly)_{io}}$.

We show the remaining three properties: that this set
is downwards closed under reductions, closed under joins,
and meet prime.

\begin{lemma}
$\mathsf{(FP/poly)_{\mathcal{U}}}$ is downwards closed.
\end{lemma}

\begin{proof}
Let $S\in \mathsf{(FP/poly)_{\mathcal{U}}}$ and $R\le S$.
Let $(F,G)$ be the reduction from $R$ to $S$, and let
$p_t(n)$ bound its polynomial runtime. Fix $k\in\mathbb{N}$,
and let $k'$ be such that $p_{k'}(n)\ge p_t(p_k(n))$.
Let $U\in\mathcal{U}$ be the ultrafilter set that exists
for $S$ for the parameter $k'$, so that we have a family
of circuits $C_{n,m}$ for all $m\le n\in U$ that compute
$S$, with $C_{n,m}$ taking inputs of size $m$ and having size
$p_{\ell}(n)$ (for a fixed $\ell\in\mathbb{N}$ dependent on $k'$).

Fix $n\in U$ and $m\le n$; we design a circuit
$C'_{n,m}$ for computing $R$. On input $x$ of length $m$,
$C'_{n,m}$ applies a circuit version of $F$ to $x$
to get $F(x)$, where $|F(x)|\le p_t(m)\le p_{k'}(n)$. Then
it applies $C_{n,|F(x)|}$ to $F(x)$ to get an output string $y$
such that $(F(x),y)\in S$, where $|y|\le p_\ell(n)$.
Finally, it applies
a circuit version of $G$ to the pair $(x,y)$
(an input of size at most a constant more than $m+p_\ell(n)$,
so $O(p_{\ell}(n))$) and returns the result. Since the runtime
of $G$ is bounded by $p_t$, the final circuit $C'_{n,m}$
is of size at most $O(p_{t\ell}(n))$. In particular, there exists
$\ell'$ (dependent only on $t$ and $\ell$, not $m$ or $n$)
such that $C'_{n,m}$ is of size at most $p_{\ell'}(n)$.

Now, by the reduction property of $(F,G)$ and the correctness
of $C_{n,m}$, we get that $C'_{n,m}$ must solve $R$
on all inputs of size $m$, whenever $n\in U$ and $m\le p_{k}(n)$.
Moreover, $C'_{n,m}$ is of size at most $p_{\ell'}(n)$ for all
$m$ and $n$. This means $R\in \mathsf{(FP/poly)_{\mathcal{U}}}$,
as desired.
\end{proof}

\begin{lemma}
$\mathsf{(FP/poly)_{\mathcal{U}}}$ is closed under joins.
\end{lemma}

\begin{proof}
Let $S_0,S_1\in \mathsf{(FP/poly)_{\mathcal{U}}}$. To show
$S_0\vee S_1\in \mathsf{(FP/poly)_{\mathcal{U}}}$,
let $k\in\mathbb{N}$.
Let $U_0,U_1\in\mathcal{U}$ and $\ell_0,\ell_1\in\mathbb{N}$
be the parameters that exist for $S_0$ and $S_1$ respectively,
so that $S_0$ has circuits $C^0_{n,m}$ of size $p_{\ell_0}(n)$
for $m\le p_{k}(n)$ and $S_1$ has circuits $C^1_{n,m}$ of size
$p_{\ell_1}(n)$ for $m\le p_{k}(n)$.

Let $U=U_0\cap U_1$. Then $U\in\mathcal{U}$. Let $n\in U$
and let $m\le p_k(n)$. An input to $S_0\vee S_1$ looks like
$\langle x,b\rangle$ with $b\in \{0,1\}$. Construct a circuit
family $C_{n,m}$ on inputs of size $m$ as follows:
$C_{n,m}$ checks formatting, outputting arbitrarily if the input
is poorly formatted; if the input is $\langle x,b\rangle$,
it then applies $C^b_{n,|x|}$ to $x$ and outputs the result.
The formatting check requires size dependent on $m\le p_k(n)$
but no other parameters. The fork on $b$ is constant overhead,
and other than that we need circuit size 
$|C^0_{n,|x|}|+|C^1_{n,|x|}|\le p_{\ell_0}(n)+p_{\ell_1}(n)$.
Pick $\ell$ large enough so that this circuit $C_{n,m}$
is of size at most $p_\ell(n)$ for all $n$;
this depends on $k,\ell_0,\ell_1$ but not $n$ or $m$.

Since $C^0_{n,|x|}$ and $C^1_{n,|x|}$ succeed for $S_0$ and $S_1$
respectively (since $n\in U_0\cap U_1$ and $|x|\le m\le p_k(n)$),
the circuit $C_{n,m}$ computes $S_0\vee S_1$ for all $m\le p_k(n)$
and has size at most $p_\ell(n)$ whenever $n\in U$. We conclude
$S_0\vee S_1\in\mathsf{(FP/poly)_{\mathcal{U}}}$ as desired.
\end{proof}

\begin{lemma}
$\mathsf{(FP/poly)_{\mathcal{U}}}$ is meet-prime.
\end{lemma}

\begin{proof}
Suppose by contradiction that
$S_0\wedge S_1\in\mathsf{(FP/poly)_{\mathcal{U}}}$
yet neither $S_0$ nor $S_1$ is in $\mathsf{(FP/poly)_{\mathcal{U}}}$.
Let $k_0$ and $k_1$ be such that the respective membership conditions
for $S_0$ and $S_1$ fail, and let $k=\max\{k_0,k_1\}$;
then the membership conditions for $S_0$ and $S_1$ both fail
for $k$. On the other hand, for $k'>k$ to be chosen later,
the membership for $S_0\wedge S_1$ it succeeds,
so let $U\in\mathcal{U}$ and $\ell\in\mathbb{N}$ be such that
whenever $n\in U$ and $m\le p_{k'}(n)$ we have $C_{n,m}$
computing $S_0\wedge S_1$ on input size $m$ and circuit size
$p_\ell(n)$.

Pick $n\in U$ and consider the circuit subfamily $C_{n,m}$
for the fixed $n$ and all values of $m\le p_{k'}(n)$; this computes
$S_0\wedge S_1$ on all input sizes at most $p_{k'}(n)$.
Ensure $k'$ was picked such that when $x_0$ and $x_1$
are of size at most $p_k(n)$, the string $\langle x_0,x_1\rangle$
is of size at most $p_{k'}(n)$ (this $k'$ depends only on $k$,
not on $n$).
Now for any $x_0$ and $x_1$ of size at most $p_{k}(n)$,
the $\{C_{n,m}\}_m$ subfamily correctly computes
$\langle x_0,x_1\rangle$ for $S_0\wedge S_1$. By definition
of this meet, this means the circuit family outputs
$\langle y,b\rangle$ on such an input (this is a deterministic
function of $\langle x_0,x_1\rangle$).

Consider the matrix $M$
where the rows are indexed by strings $x_0$ and the columns
by strings $x_1$, both of size at most $p_k(n)$,
and the entry $M[x_0,x_1]$ is defined to be the bit $b$
from the output $\langle y,b\rangle$ that the circuit subfamily
gives on input $\langle x_0,x_1\rangle$. View this as a payoff
matrix in a $2$-player game; the row player wants the outcome $1$
(i.e. that $x_1$ gets solved) and the column player wants the outcome
$0$. In other words, each player is picking an input in the hopes
that the circuit subfamily chooses to solve the other player's input.

When both players are allowed mixed strategies,
the minimax theorem says that the expected payoff does not
depend on who chooses their strategy first; in other words,
the game has a value. without loss of generality, let's suppose
that the column player wins with probability at least $1/2$.
This means she has a mixed strategy --
a probability distribution $\mu$
over choice of $x_1$ -- against which for \emph{every} $x_0$,
the circuit subfamily chooses to solve $x_0$ when $x_1\sim \mu$
with probability at least $1/2$.
We have now a (not-necessarily-samplable)
randomized circuit family computing $x_0$
for $S_0$, as follows: for each possible size $m'$ of $x_0$,
the circuit family at size $m'$, denoted $C'_{n,m'}$,
samples $x_1\sim\mu$, constructs $\langle x_0,x_1\rangle$,
runs the correct member of $\{C_{n,m}\}_m$ on
$\langle x_0,x_1\rangle$ to get an output
$\langle b,y\rangle$, and outputs $y$, succeeding in the event
where $b=0$ (which happens with probability at least $1/2$
on all $x_0$ of size $m'$). In other words, $C'_{n,m'}$
is a probability distribution over circuits
of size $p_\ell(n)$ plus overhead,
which take inputs of size $m'$; and this distribution
over circuits has worst-case success
probability at least $1/2$ (and it can tell when it succeeded via
the bit $b$).

Amplify $C'_{n,m'}$ by repeating it $m'+1$ times and outputting
the first successful run (if any) to reduce the error probability
below $2^{-m'}$. This can be done in the
probability-distribution-over-circuits model of randomized
circuit computation (i.e. we do not require samplability).
Now, when the error is that low, there must exist
some random seed for which the amplified circuit succeeds
on all $2^{m'}$ possible values of $x_0$. Hard code this random
seed, and denote the resulting deterministic circuit by $C''_{n,m'}$.
Its input size is still $m'$. Its circuit size is now computed
by starting with at most $p_\ell(n)$, adding overhead for the
randomized circuit pre- and post-processing, and multiplying
by $m'+1$, possibly with some additional overhead.
Now, $m'\le p_k(n)$ and the overheads are fixed-degree polynomials
in $m'$ at the most, so there exists some $\ell'$ dependent
on $k,\ell$ (but not $m,m',n$) such that $p_{\ell'}(n)$
upper bounds the size of $C''_{n,m'}$.

We conclude that for all $m'\le p_k(n)$, $C''_{n,m'}$
computes $S_0$ and has size at most $p_{\ell'}(n)$. This is for
one specific $n\in U$. For other values of $n\in U$, it is possible
that we instead get a circuit with those same parameters that
computes $S_1$ instead of $S_0$. Let $U_0$ be the subset
of $U$ consisting of values of $n$ for which we get circuits
for $S_0$, and let $U_1$ be the subset of $U$ consisting of
values of $n$ for which we get circuits for $S_1$.
Then $U_0\cup U_1=U$. If $U_0\in\mathcal{U}$, this contradicts
the failure of the membership condition of
$\mathsf{(FP/poly)_{\mathcal{U}}}$ for $S_0$ (with parameter $k$).
If $U_0\notin \mathcal{U}$, then $\mathbb{N}\setminus U_0$ must
be in $\mathcal{U}$ due to the ultrafilter property.
Then $(\mathbb{N}\setminus U_0)\cap U\in\mathcal{U}$,
yet $(\mathbb{N}\setminus U_0)\cap U\subseteq U_1$,
so we must have $U_1\in\mathcal{U}$. This contradicts the failure
of the membership condition of
$\mathsf{(FP/poly)_{\mathcal{U}}}$ for $S_1$ (with parameter $k$).
In both cases, we get a contradiction, as desired.
\end{proof}

These lemmas give the conclusion that
$\mathsf{(FP/poly)_{\mathcal{U}}}$ is a prime ideal contained
in $\mathsf{(FP/poly)_{io}}$, proving \cref{thm:FPpolyPrime}.
We note that this ideal was constructively defined in terms
of the ultrafilter $\mathcal{U}$, which itself is not
constructive as it requires the axiom of choice.

The existence of the prime ideal in $\mathsf{(FP/poly)_{io}}$
means that for each verifiable search problem $S$,
either $S$ has infinitely-often polynomial sized circuits,
or else each refutation task $\neg_k S$ for $k\in\mathbb{N}$
has infinitely-often polynomial sized circuits
(see \cref{cor:FPio}).

\subsection{A prime ideal inside infinitely-often-HeurBPP}

We give another example of a tractable-looking prime ideal
by showing there is a prime ideal in
$\mathsf{HeurBPP_{io}}$. We first define the precise version
of $\mathsf{HeurBPP}$ that we will use.

We define a \emph{samplable distribution family} to be a probabilitistic
Turing machine $D$ which runs in polynomial time and, on input
$0^n$, samples a string in $\{0,1\}^*$. Note that since $D$ runs
in polynomial time, all samples from $D(0^n)$ will always be at most
polynomially sized in $n$, but we do not require the samples to be
of size exactly $n$ (and smaller string sizes are allowed).

For a search problem $S$, we say that a probabilistic
Turing machine $M$ succeeds in solving $S$ against $D$
with error rate $k$ on the $n$-th distribution if
\[\Pr_{x\sim D(0^n)}[(x,M(\langle x,0^n\rangle))\in S]
\ge 1-\frac{1}{p_k(n)}.\]
In other words, the machine $M$ takes in an input string $x$
to the problem $S$, with $x$ sampled from $D(0^n)$; $M$ also
has access to the parameter $n$ that was used to sample $x$,
and the algorithm may also be chosen based on the desired error
decline rate $p_k(n)$ (recall that $p_k(n)=n^k+k$).
$M$ then produces
an output string $y$ which should satisfy $(x,y)\in S$ except
with a small probability of error $1/p_k(n)$. This probability
is taken over both the internal randomness of $M$ and the sample
$x\sim D(0^n)$. Note that this success condition applies to one
specific choice of $n\in\mathbb{N}$.

Let $\mathcal{U}\subseteq\mathcal{P}(\mathbb{N})$ be any collection
of sets of natural numbers. We say that the algorithm $M$
succeeds in solving $S$ against distribution family $D$
with parameter $k$ with respect to success condition $\mathcal{U}$
if the set of $n\in\mathbb{N}$
for which $M$ succeeds on the $n$-th distribution (as defined above)
is one of the sets in the collection $\mathcal{U}$. For example,
we could pick $\mathcal{U}=\{\mathbb{N}\}$ if we want to require
success for all $n$, or we can pick
$\mathcal{U}=\{Q\subseteq\mathbb{N}:|Q|=\infty\}$ if we want
to require the algorithm to succeed infinitely often.

We say that the search problem $S$ is \emph{heuristically solvable
by $\mathsf{BPP}$ algorithms} (with respect to success condition
$\mathcal{U}$) if for all samplable distribution familes
$D$ and all $k\in\mathbb{N}$
there exists a probabilistic Turing machine $M$ such that
$M$ succeeds in solving $S$ against distribution $D$ (with respect
to the success condition $\mathcal{U}$).

In other words, the order of quantifiers is: we fix
$S$ and $\mathcal{U}$ (no quantifiers), and we say $S$ is heuristically
solvable with respect to success condition $\mathcal{U}$
if the following holds. For all samplable distribution families
$D$ and all $k$, there exists a $\mathsf{BPP}$ algorithm $M$
and some $U\in\mathcal{U}$, such that for all $n\in U$,
$M$ succeeds in solving $S$ against distribution $D(0^n)$
with error at most $1/p_k(n)$.\footnote{Note that we allow the
algorithm $M$ to depend nonuniformly on the error growth rate $k$;
other definitions of $\mathsf{HeurBPP}$ often require the algorithm
to be uniform in $k$ (by taking the target error $\epsilon$
as an input). The reason for allowing nonuniformity is to get the proof
of \cref{lem:HeurJoins} to work; with a uniformity requirement,
we would still get a prime lower set, but $\mathsf{HeurBPP_{ultra}}$
may no longer be an ideal.}

We let $\mathsf{HeurBPP_{io}}$ be the class
$\mathsf{HeurBPP_{\mathcal{U}}}$ where $\mathcal{U}$ is the set
of all infinite subsets of $\mathbb{N}$; this means a verifiable
search problem $S$ is in $\mathsf{HeurBPP_{io}}$ if
for each $k\in\mathbb{N}$ and samplable distribution family
$D$, there is a uniform algorithm $M$
which solves $S$ to error $1/p_k(n)$ against
$D(0^n)$ for infinitely many different values of $n$.

Our prime ideal will be
$\mathsf{HeurBPP_{\mathcal{U}}}$ where $\mathcal{U}$ is a
nonprincipal ultrafilter on $\mathbb{N}$. We must show
this set of search problems is nonempty, proper,
downwards closed, closed
under joins, meet prime, and contained in
$\mathsf{HeurBPP_{io}}$.

It is easy to see that 
$\mathtt{triv}\in\mathsf{HeurBPP_{\mathcal{U}}}$
and that $\mathtt{imp}\notin \mathsf{HeurBPP_{\mathcal{U}}}$
(since $D(0^n)$ can sample a single impossible input of
fixed size for all $n$, and a nonprincipal ultrafilter
does not contain the empty set).
The property $\mathsf{HeurBPP_{\mathcal{U}}}\subseteq
\mathsf{HeurBPP_{io}}$ follows from the fact that a
nonprincipal ultrafilter $\mathcal{U}$ contains only infinite
sets of natural numbers.

It remains to show that this class is downwards closed,
closed under joins, and meet prime.

\begin{lemma}\label{lem:HeurClosed}
For nonprincipal ultrafilter $\mathcal{U}$,
$\mathsf{HeurBPP_{\mathcal{U}}}$ is
downwards closed under reductions.
\end{lemma}

\begin{proof}
Let $S\in\mathsf{HeurBPP_{\mathcal{U}}}$ and suppose $R\le S$
with $R\in\mathsf{Verifiable}$.
Let $(F,G)$ be the reduction from $R$
to $S$. We wish to show $R\in \mathsf{HeurBPP_{\mathcal{U}}}$, 
so let $D$ be any samplable distribution family and fix $k\in\mathbb{N}$.
Consider the samplable distribution family
$D'$ defined by setting $D'(0^n)$ to sample $x$ from $D(0^n)$,
apply $F$ to get $F(x)$, and return $F(x)$. Since $F$ is polynomial
time and $D$ is samplable, $D'$ is also clearly samplable. Since
$S\in\mathsf{HeurBPP_{\mathcal{U}}}$, there is a $\mathsf{BPP}$ machine
$M'$ which solves $S$ against $D'$ (with the error parameter $k$)
with respect to success condition
$\mathcal{U}$. This means
$M'$ succeeds at solving $S$ against $D'(0^n)$ to error
$1/p_k(n)$ for all $n\in U$,
for some set of natural numbers $U\in\mathcal{U}$.

We construct a $\mathsf{BPP}$ machine $M$ for solving $R$ against $D$
to error $1/p_k(n)$. We will set $M(\langle x,0^n\rangle)$
to output $G(x,M'(\langle F(x),0^n\rangle))$.
Let $n\in U$. When $x\sim D(0^n)$,
we have $F(x)\sim D'(0^n)$ by definition; since $M'$ succeeds
in solving against $D'$ to error $1/p_k(n)$
when $n\in U$, we conclude that
$M'(\langle F(x),0^n\rangle)$ is a valid output for $F(x)$ with respect
to $S$ except with probability $1/p_k(n)$. By the reduction
property of $(F,G)$, since $(F(x),M'(\langle F(x),0^n\rangle))\in S$
we get $(x,G(x,M'(\langle F(x),0^n\rangle)))\in R$, so
$(x,M(\langle x,0^n\rangle))\in R$. Thus $M$ correctly solves $x$
with respect to $R$ except with probability $1/p_k(n)$.
Since this succeeds for all $n\in U$,
and since $U\in\mathcal{U}$,
we conclude that $M$ solves $R$ against $D$ with error parameter $k$
with respect to success condition $\mathcal{U}$, completing the proof.
\end{proof}

\begin{lemma}\label{lem:HeurJoins}
When $\mathcal{U}$ is a nonprincipal ultrafilter,
$\mathsf{HeurBPP_{\mathcal{U}}}$ is closed under joins.
\end{lemma}

\begin{proof}
Let $S_0,S_1\in\mathsf{HeurBPP_{\mathcal{U}}}$. We wish to show
$S_0\vee S_1\in\mathsf{HeurBPP_{\mathcal{U}}}$, so consider
a samplable distribution family $D$ for $S_0\vee S_1$,
and fix the parameter $k$.

We will use the following terminology. An input
$z$ to $S_0\vee S_1$ will be called a $0$-input if it has
the form $z=\langle x,0\rangle$. It will be called a $1$-input
if it has the form $z=\langle x,1\rangle$. If it has neither form
(so it is poorly formatted), we call it a $-1$-input.

Define the samplable distribution family
$D_0$ where $D_0(0^n)$ samples from $D(0^n)$ repeatedly
until it sees a sample that is either a $0$-input or
a $-1$-input; when it finds such a sample $z$,
if it is a $-1$-input, it outputs $z$, and if it is
a $0$-input $z=\langle x,0\rangle$,
it outputs $x$. $D_0(0^n)$ does so
for $2p_k(n)$ samples, and if it failed to halt at that
point, it gives up and outputs an arbitrary string.
Define $D_1$ analogously, except $D_1(0^n)$ looks for samples
from $D(0^n)$ which are either $-1$-inputs or $1$-inputs
(instead of $-1$-inputs or $0$-inputs).
It is clear that $D_0$ and $D_1$ are samplable, since $D$ is
samplable and we repeat $D$ polynomially many times.

View $D_0$ as a samplable distribution family over inputs to $S_0$,
and consider the parameter $4k$. Since
$S_0\in\mathsf{HeurBPP_{\mathcal{U}}}$,
there exists a $\mathsf{BPP}$
machine $M_0$ and a set of naturals $U_0\in\mathcal{U}$
such that $M_0$ succeeds in solving $S_0$ against $D_0$
for $n\in U_0$, except with probability $1/p_{4k}(n)$.
Similarly, there exist $M_1$ and $U_1\in\mathcal{U}$
such that $M_1$ succeeds in solving $S_1$ against $D_1$
for $n\in U_1$ except with probability $1/p_{4k}(n)$.

We now construct an algorithm $M$ for $S_0\vee S_1$
which should succeed with respect to $D$ and $k$
for all $n\in U_0\cap U_1$. On poorly formatted inputs, $M$
outputs arbitrarily. Otherwise,
$M(\langle \langle x,b\rangle,0^n\rangle)$
runs $M_b(\langle x,0^n\rangle)$ and returns its output.

For fixed $n\in U_0\cap U_1$, we analyze the probability
that $M$ succeeds on an input sampled from $D(0^n)$.
Let $P_0$ be the probability that a sample from $D(0^n)$
is a $0$-input, let $P_1$ be the probability it is a $1$-input,
and let $P_{-1}=1-P_0-P_1$ be the probability it is a $-1$-input.
For simplicity of analysis, let's assume that on $-1$-inputs,
$M$ actually makes calls to both $M_0$ and $M_1$ and only
succeeds if both succeed; this will only make $M$ perform
worse, so upper bounding this new error rate will upper bound
the true error rate.

Then the (modified) probability of error of $M$ against $D(0^n)$ is
$(P_0+P_{-1})Q_0+(P_1+P_{-1})Q_1=(1-P_1)Q_0+(1-P_0)Q_1$,
where $Q_0$ is the probability
that $M_0$ errs on the distribution $D_0'$ defined
as $D(0^n)$ conditioned on non-$1$-inputs, and $Q_1$
is the probability that $M_1$ errs on the distribution
$D_1'$ defined as $D(0^n)$ conditioned on non-$0$-inputs.

There are two sources of error for $M_0$ on $D_0'$.
One of them is the total variation distance between $D_0'$
and $D_0(0^n)$. The other is the probability that $M_0$
errs on $D_0(0^n)$. The former is just the probability
that the algorithm which tries to sample a non-$1$-input
from $D(0^n)$ in the definition of $D_0(0^n)$ fails to find
one. This probability is $P_1^{2p_k(n)}\le e^{-2(1-P_1)p_k(n)}$.
The latter is the error rate of $M_0$ against $D_0(0^n)$,
which we know is at most $1/p_{4k}(n)$. By the union bound,
the error $Q_0$ is at most $1/p_{4k}(n)+e^{-2(1-P_1)p_k(n)}$.
We similarly bound $Q_1$, and then the error of $M$ against
$D(0^n)$ is at most
\[(1-P_1)(1/p_{4k}(n)+e^{-2(1-P_1)p_k(n)})
 +(1-P_0)(1/p_{4k}(n)+e^{-2(1-P_0)p_k(n)})\]
\[\le \frac{2}{p_{4k}(n)}
    +(1-P_1)e^{-2(1-P_1)p_k(n)}+(1-P_0)e^{-2(1-P_0)p_k(n)}.\]
We use the (very lossy but universally holding) bound
$p_{4k}(n)\ge 4p_k(n)$ to upper bound $2/p_{4k}(n)\le 1/2p_k(n)$.
Also, using basic calculus, one can check that
the maximum value of $x e^{-tx}$ occurs at $x=1/t$, where the
value is $1/et$. Therefore, $xe^{-tx}\le 1/et$ for all $x$ and $t>0$.
Taking $t=2p_k(n)$ and $x=(1-P_1)$ or $x=(1-P_0)$, the error
of $M$ is upper bounded by $1/2p_k(n)+1/ep_k(n)<1/p_k(n)$.

Thus $M$ succeeds on $n$ for all $n\in U_0\cap U_1$. Since
$\mathcal{U}$ is closed under intersections, $M$ succeeds on all
$n$ from a set in $\mathcal{U}$, and since this held for any
samplable distribution family and any $k$ we conclude
$S_0\vee S_1\in\mathsf{HeurBPP_{\mathcal{U}}}$.
\end{proof}

\begin{lemma}\label{lem:HeurPrime}
When $\mathcal{U}$ is a nonprincipal ultrafilter,
$\mathsf{HeurBPP_{\mathcal{U}}}$ is meet prime.
\end{lemma}

\begin{proof}
Let $S_0,S_1\in\mathsf{Verifiable}$, and suppose that
$S_0\notin\mathsf{HeurBPP_{\mathcal{U}}}$ and
$S_1\notin\mathsf{HeurBPP_{\mathcal{U}}}$. We will show
$S_0\wedge S_1\notin\mathsf{HeurBPP_{\mathcal{U}}}$.

Since $S_0\notin\mathsf{HeurBPP_{\mathcal{U}}}$ and
$S_1\notin\mathsf{HeurBPP_{\mathcal{U}}}$, let $D_0$ and $D_1$
be samplable distributions that are hard for $S_0$ and $S_1$
respectively. Let $k$ be large enough so that no algorithm
succeeds with error $1/p_k(n)$ on a set in $\mathcal{U}$
for either $S_0$ (against $D_0$) or $S_1$ (against $D_1$).

Consider the product distribution $D(0^n)=D_0(0^n)\times D_1(0^n)$
over pairs $\langle x_0,x_1\rangle$ that we get by sampling
$x_0\sim D_0(0^n)$ and $x_1\sim D_1(0^n)$ independently.
Then $D$ is clearly a samplable distribution. Use the parameter
$k'$ (to be determined later)
and consider a randomized algorithm $M$ which attempts to solve
$S_0\wedge S_1$ against distribution $D(0^n)$ to error $1/p_{k'}(n)$;
let $U\subseteq\mathbb{N}$ be the set of natural numbers $n$
for which $M$ succeeds in this goal. We wish to show
$U\notin\mathcal{U}$.

Note that we may assume $M$ always produces
outputs of the form $\langle y,b\rangle$ for $b\in\{0,1\}$:
if the input to $M$ is poorly formatted, all outputs are valid, so
$M$ may as well produce such an output, and if the input is
well-formatted, outputs not of this form are never accepted, so
there is never any advantage to deviating from the form
$\langle y,b\rangle$.

Define the algorithm $M_0$ for $S_0$ as follows. On input
$x_0$, the algorithm $M_0(\langle x_0,0^n\rangle)$ repeatedly samples
$x_1\sim D_1(0^n)$ and runs
$M(\langle \langle x_0,x_1\rangle,0^n\rangle)$ to
produce an output $\langle y,b\rangle$. If $b=0$, the algorithm
$M_0$ outputs $y$. If $b=1$, the algorithm $M_0$ instead
resamples a new $x_1\sim D_1(0^n)$ and tries again.
After trying $\lceil 8p_{k}(n)\ln(4p_k(n))\rceil$ attempts,
if a $b=0$ case has not been found,
$M_0$ gives up and declares failure.
We define $M_1$ analogously: it receives $x_1$ as input
and repeatedly samples $x_0\sim D_0(0^n)$.
It is clear that $M_0$ and $M_1$ run in probabilistic polynomial time.

By the choice of $D_0$ and $k$, we know that the $\mathsf{BPP}$
algorithm $M_0$ only succeeds in solving $S_0$ against $D_0(0^n)$
to error $1/p_k(n)$ for a set $V_0$ of $n$ that is not in $\mathcal{U}$.
Since $\mathcal{U}$ is an ultrafilter,
the complement $U_0=\overline{V_0}$
of $V_0$ must be in $\mathcal{U}$; in other words, the set $U_0$
of $n$ on which $M_0$ fails is a set in the ultrafilter.
Similarly, the set $U_1$ of $n$ on which $M_1$ fails to solve
$S_1$ against $D_1(0^n)$ (to error $1/p_k(n)$) is also in the
ultrafilter $\mathcal{U}$. Since $\mathcal{U}$ is closed under
intersections, $U_0\cap U_1\in\mathcal{U}$; this is the set of $n$
on which both algorithms fail.

Fix $n\in U_0\cap U_1$. We will show that $M$ fails to solve
$S_0\wedge S_1$ against $D(0^n)$ to error $1/p_{k'}(n)$ for
this value of $n$. This will mean that $M$ only succeeds on a subset
of the complement of $U_0\cap U_1$, and since $U_0\cap U_1$ is in
$\mathcal{U}$, this subset of the complement cannot be in
$\mathcal{U}$. That will complete the proof.

Since $M_0$ fails to against $D_0(0^n)$ to error $1/p_k(n)$,
there is at least a $1/2p_k(n)$ fraction of the samples
(weighted by their probabilities under $D_0(0^n)$) such that
the error (due to internal randomness of $M_0$)
on those samples is at least $1/2p_k(n)$.
Let $A_0$ be the set of these high-error samples from $D_0(0^n)$.
Similarly, for $M_1$ there is a set $A_1$ of samples from
$D_1(0^n)$ with total probability mass at least $1/2p_k(n)$
on which its error is at least $1/2p_k(n)$.

Consider $D(0^n)$ (the product distribution of $D_0(0^n)$
and $D_1(0^n)$) conditioned on the set $A_0\times A_1$.
On this conditional distribution, $M$ returns pairs
$\langle y,b\rangle$ and it either holds that $b=0$
with probability at least $1/2$ or that $b=1$ with probability
at least $1/2$. Without loss of generality, suppose it holds that
$b=0$ with probability at least $1/2$. Then conditioned on only
$x_0\in A_0$, the probability of seeing $x_1\in A_1$ and also
$b=0$ is at least $1/4p_k(n)$. By Markov's inequality
for the distribution $D_0(0^n)$ conditioned on $A_0$,
there is a subset $B_0\subseteq A_0$ with probability at least
$1/8p_k(n)$ (in this conditional distribution) such that
for all $x_0\in B_0$, the probability of seeing $b=0$
is at least $1/8p_k(n)$.

Consider the behavior of $M_0$ when it receives an input
$x_0$ from $B_0$. Recall that it samples $x_1\sim D_1(0^n)$
and runs $M(\langle\langle x_0,x_1\rangle,0^n\rangle)$
to check if $b=0$
for a total of at least $8p_k(n)\ln(4p_k(n))$ tries.
Each time it samples $x_1$ like this, its probability of not seeing
$b=0$ is at most $1-1/8p_k(n)$, so the probability of not
seeing $b=0$ all of those times is at most
\[(1-1/8p_k(n))^{8p_k(n)\ln(4p_k(n))}\le \frac{1}{4p_k(n)}.\]
Since $B_0\subseteq A_0$, the probability of error
of $M_0$ on the input $x_0$ is at least $1/2p_k(n)$. Even
if it errs whenever it fails to find $b=0$, that only happens
with probability at most $1/4p_k(n)$, so it must also err with
probability at least $1/4p_k(n)$ when $b=0$.

This means
that $M$ errs with probability at least $1/4p_k(n)$ when
conditioned on $x_0\in B_0$ and on $b=0$. Recall that
the probability of $b=0$ conditioned on $B_0$ was at least
$1/8p_k(n)$, the probability of $B_0$ conditioned on $A_0$
was at least $1/8p_k(n)$, and the unconditioned probability of $A_0$
was at least $1/2p_k(n)$; multiplying these, we get that $M$
errs with probability at least $\frac{1}{512p_k(n)^4}$ overall.
Picking $k'$ such that $p_{k'}(n)$ is always larger than
$512p_k(n)^4$, we get that $M$ fails to achieve the required
success $1-1/p_{k'}(n)$ for any $n\in U_0\cap U_1$. Since
$U_0\cap U_1\in\mathcal{U}$, any set disjoint from it is not
in $\mathcal{U}$, so the set of $n$ on which $M$ succeeds
in solving $S_0\wedge S_1$ against $D(0^n)$ to error
$1/p_{k'}(n)$ is not an ultrafilter set. This completes the proof.
\end{proof}

\section{Closure properties of regular classes}
\label{sec:TopologyApplications}

\subsection{Preliminaries: dense prenuclei}
\label{sec:PointlessPrelim}

We need the following definition.

\begin{definition}
A \emph{dense weak prenucleus} for a Heyting algebra
is an operator $\nu$ acting on its elements which satisfies
\begin{enumerate}
\item Inflation: $a\le \nu(a)$
\item Meet property: $\nu(a)\wedge b\le \nu(a\wedge b)$
\item Density: $\nu(\bot)=\bot$
\end{enumerate}
for all elements $a$ and $b$ of the Heyting algebra.
\end{definition}

The ``prenuceus'' terminology comes from pointless topology; see \cite{PP12} for a reference.
Our definition is a little weaker than the standard notion
of a dense prenucleus, but suffices for our purposes;
we therefore call the object a dense weak prenucleus.

We need the following lemma.

\begin{lemma}\label{lem:DenseWeakPrenucleus}
A dense weak prenucleus $\nu$ satisfies $\nu(\neg a)=\neg a$
for all $a$.
\end{lemma}

\begin{proof}
We have $\neg a\le \nu(\neg a)$ by inflation. Now,
\[\nu(\neg a)\wedge a\le \nu(\neg a\wedge a)=\nu(\bot)=\bot,\]
where we used the meet property and density property,
as well as the property $\neg a\wedge a=\bot$
(recall that $\neg a$ is by definition the maximum element
whose meet with $a$ is $\bot$).
We conclude that the meet of $\nu(\neg a)$ with $a$
is $\bot$, which means $\nu(\neg a)$ is below
$\neg a$ (since $\neg a$ is by definition the largest
element whose meet with $a$ is $\bot$). Thus
$\nu(\neg a)\le \neg a$, so $\nu(\neg a)=\neg a$ as desired.
\end{proof}

We note that the negated elements in a Heyting algebra
are called the regular elements. That is, $x$ is regular
if and only if $x=\neg y$ for some $y$, which happens
if and only if $\neg\neg x=x$.
The lemma says dense weak prenuclei preserve
the regular elements. For our purposes, the elements
are the ideals (i.e. complexity classes of verifiable search
problems), and the regular elements are those classes
$\mathsf{C}$ with $\neg\neg\mathsf{C}=\mathsf{C}$:
these classes are closed under double refutation.

Dense weak prenuclei can be used to
provide closure properties of these regular ideals.

\subsection{Closure under Turing reductions}
\label{sec:TuringReductionClosure}

\begin{definition}[$\mathsf{FP}^S$]
Let $S$ be a search problem. We define the class
$\mathsf{FP}^S\subseteq\mathsf{Verifiable}$ to include all
polytime verifiable search problems $R$ which \emph{Turing}
reduce to $S$. By this we mean that there is an oracle
Turing machine $A$ such that whenever it is given access to
an oracle function $h$ which solves $S$
(meaning $(z,h(z))\in S$ for all $z$),
we have $(x,A^h(x))\in R$ for all $x$.
\end{definition}

Note that in the above definition, we have allowed $h$
to output super-polynomially large solutions for $S$
on an input $x$. To fully define Turing reductions,
we must rectify this by allowing the algorithm $A$
to use an amount of computation between queries that is polynomial
in both the input size and the size of any oracle outputs
received so far. More formally, $A$ is polynomial time if there
exists a polynomial growth rate $p_k$ such that $A(x)$
makes at most $p_k(|x|)$ queries, each query has size
at most $p_k(|x|)$, and the amount of computation
between any two queries is at most $p_k(n)$, where $n$
is the total length of the input and all oracle outputs received
so far. Note that since we require $A$ to succeed for
\emph{all} oracle functions $h$ which solve $S$,
it in particular needs to succeed for all polynomially-growing
$h$; our definition adds the condition that the algorithm
still needs to succeed (albeit with extra resources) even
when $h$ is not polynomially bounded. We view this as a technical
detail that only matters for problems outside of $\mathsf{TFNP}$.

\begin{lemma}
For any $S\in\mathsf{Verifiable}$, the class $\mathsf{FP}^S$
is nonempty, downwards closed under mapping reductions,
and closed under binary joins, making it an ideal of
$\mathsf{Verifiable}$.
\end{lemma}

\begin{proof}
It is not hard to see that if $R\le T$ and $T$ Turing reduces
to $S$, then $R$ also Turing reduces to $S$; just use
the mapping reduction $(F,G)$ from $R$ to $T$ and the
Turing reduction $A$ from $T$ to $S$ to construct
a new Turing reduction which simulates $F(x)$, then feeds that
into $A$, then applies $G(x,A(F(x)))$ to the output.
This shows that $\mathsf{FP}^S$ is downwards closed.
It also clearly contains $\mathsf{FP}$, so is nonempty.
If $R$ and $T$ both Turing reduce to $S$, then the problem
$R\vee T$ (whose nontrivial inputs are $\langle x,0\rangle$ or
$\langle x,1\rangle$, with the bit indicating whether to
solve $x$ for $R$ or for $T$) is clearly also Turing reducible
to $S$, just by simulating either the reduction from $R$ to $S$
or from $T$ to $S$, depending on the input. Hence $\mathsf{FP}^S$
is an ideal of $\mathsf{Verifiable}$.
\end{proof}

\begin{definition}
For a class of problems $\mathsf{C}\subseteq\mathsf{Verifiable}$,
define \[\mathsf{FP}^\mathsf{C}
\coloneqq\bigvee_{S\in\mathsf{C}}\mathsf{FP}^S.\]
Note that this join is well-defined in the ideal completion
of $\mathsf{Verifiable}$, since it is a complete lattice.
This operator therefore maps ideals of $\mathsf{Verifiable}$
to ideals of $\mathsf{Verifiable}$.
\end{definition}

\begin{lemma}\label{lem:prenucleus}
The function $\nu(\mathsf{C})=\mathsf{FP}^\mathsf{C}$
is a dense weak prenucleus.
\end{lemma}

\begin{proof}
It is clear that $S\in\mathsf{FP}^S$, and hence also
$\mathsf{C}\subseteq\mathsf{FP}^\mathsf{C}$ for any ideal 
$\mathsf{C}$. This shows the inflation condition 
$\mathsf{A}\subseteq\nu(\mathsf{A})$.
Density is also easy: if $S\in\mathsf{FP}$, then $\mathsf{FP}^S$
is clearly just $\mathsf{FP}$, so
\[\mathsf{FP}^\mathsf{FP}=\bigvee_{S\in\mathsf{FP}}\mathsf{FP}^S
=\bigvee_{S\in\mathsf{FP}}\mathsf{FP}=\mathsf{FP}.\]

It remains to show that
$\nu(\mathsf{A})\cap \mathsf{B}\subseteq\nu(\mathsf{A}\wedge\mathsf{B})$.
Recall that $\nu(\mathsf{A})=\bigvee_{S\in\mathsf{A}}\mathsf{FP}^S$,
and that in the complete Heyting algebra,
finite meets distribute over infinite joins.
Additionally, meet for our ideal lattice is just the
intersection of the classes. We therefore get
\[\nu(\mathsf{A})\cap\mathsf{B}
=\bigvee_{S\in\mathsf{A}}(\mathsf{FP}^S\cap\mathsf{B}),
\qquad\nu(\mathsf{A}\cap\mathsf{B})
=\bigvee_{R\in\mathsf{A}\cap\mathsf{B}}\mathsf{FP}^R.\]
To show the inequality, it suffices to show that every
$\mathsf{FP}^S\cap B$ is below
$\bigvee_{R\in \mathsf{A}\cap\mathsf{B}}\mathsf{FP}^R$.
Since the infinite join is the closure of the union under finite 
joins and downwards under reductions, it suffices
to show $\mathsf{FP}^S\cap \mathsf{B}$ is a subset of
$\bigcup_{R\in\mathsf{A}\cap\mathsf{B}}\mathsf{FP}^R$
for any $S\in\mathsf{A}$. Fix such $S$, and
let $T\in\mathsf{FP}^S\cap B$. Then $T\in\mathsf{B}$,
so $S\wedge T\in \mathsf{A}\cap\mathsf{B}$.
Taking $R=S\wedge T$, it suffices to show that
$T\in\mathsf{FP}^R$.

We give a Turing
reduction from $T$ to $R=S\wedge T$;
this will complete the proof.
Recall that $T\in\mathsf{FP}^S$, so there is some Turing reduction
$M$ from $T$ to $S$. We define a Turing reduction $M'$
from $T$ to $S\wedge T$ which simulates $M$, and whenever
$M(x)$ makes a query of the form $z$ to the oracle for $S$,
the algorithm $M'(x)$ instead makes a query of the form
$\langle z,x\rangle$ to the oracle for $S\wedge T$.
The oracle answer must solve either $z$ or $x$ (and this is labeled).
If it solves $x$ (with respect to $T$),
the algorithm $M'(x)$ terminates and returns this answer.
If it solves $z$, the oracle continues the simulation of $M(x)$.
Eventually, if $M'(x)$ did not terminate early, the oracle
algorithm $M(x)$ produces a solution for $x$ with respect to $S$,
which $M'(x)$ then returns. It is not hard to check
that this reduction is valid, showing $T\in\mathsf{FP}^{S\wedge T}$
and completing the proof.
\end{proof}

\begin{corollary}
If $\mathsf{C}\subseteq\mathsf{Verifiable}$ is an ideal,
then $\neg\mathsf{C}$ is closed under Turing reductions,
i.e. $\mathsf{FP}^{\neg\mathsf{C}}=\neg\mathsf{C}$.
In particular, since $\mathsf{C}\subseteq\neg\neg \mathsf{C}$,
we also have $\mathsf{FP}^\mathsf{C}\subseteq\neg\neg \mathsf{C}$
for any ideal $\mathsf{C}$.
\end{corollary}

\begin{proof}
This follows from \cref{lem:prenucleus} and \cref{lem:DenseWeakPrenucleus}.
\end{proof}

\paragraph{Complexity-theoretic interpretation.}
For comparison, we give a complexity-only proof sketch.
Let's assume for simplicity that $\mathsf{C}$ has a complete
problem $S$; then $\neg\mathsf{C}$ is the downwards closure
of $\{\neg_k S\}$, but
we'll pretend that it has a single complete problem $\neg S$, which is
``given as input a polynomial time algorithm for $S$, find
an input where it fails''.

We want to show that anything that Turing reduces to $\neg S$
actually mapping reduces to $\neg S$. Consider a Turing reduction
$A$ from verifiable search problem $R$ to $\neg S$. The queries
$A$ makes to $\neg S$ are of the form ``find a flaw in this
polynomial time algorithm for $S$''. We construct a mapping reduction
$(F,G)$ from $R$ to $\neg S$, as follows. On input $x$,
$F(x)$ constructs the algorithm $\langle B_x\rangle$ which
behaves as follows. $B_x(z)$ simulates $A(x)$.
Recall that the queries
$A(x)$ makes are of the form ``find a counterexample to this
algorithm $M$ for $S$''. For each such query $M$, the algorithm
$B_x(z)$ checks whether $(z,M(z))\in S$ using the
verifier for $S$; if not, it returns $z$ as a solution to that
query. If ever a query $M$ is such that $M(z)$ correctly solves
$z$, $B_x(z)$ halts and returns this solution $M(z)$. Otherwise,
it returns the output of $A(x)$ when all the oracle answers
were $z$.

As mentioned, $F(x)$ returns $\langle B_x\rangle$, the code
of $B_x$. The algorithm $G(x,y)$ simulates $B_x(y)$ and returns
the result. Now, if $(F(x),y)\in \neg S$,
it means that $y$ is a successful refutation for the algorithm
$B_x$, which means that all the oracle queries $M$ made by
$A(x)$ are successfully refuted by the string $y$. In that case,
$B_x(y)$ outputs the output of $A(x)$ when $A(x)$ is given
access to a true oracle for $\neg S$ (which happens to always
returns $y$ on the questions asked by $A(x)$).
This means that $A(x)$ solves $x$ for $R$, since $A$ was
a Turing reduction from $R$ to $\neg S$. Hence
$G(x,y)=A(x)$ is a valid solution for $x$, and $(x,G(x,y))\in R$,
as desired.

\paragraph{Discussion.}
The complexity proof is nontrivial even for the simplified setting
in which everything has a complete problem.
The formalism of Heyting algebra, which was required
for \cref{lem:DenseWeakPrenucleus},
was again useful for learning about
refutation problems. Indeed, it was locale-theoretic
considerations (thinking about sublocales and dense sublocales
of the ideal completion of the search problem lattice)
which lead us to consider such closure properties
in the first place. We view this as an instance in which
thinking about pointless
topology has lead to insights into metacomplexity
refutation tasks.

\subsection{Characterization by hard problems}

We will now need to consider
two types of search problems: polynomial time
verifiable search problems,
and decidably verifiable search problems.
The former was denoted by the set $\mathsf{Verifiable}$;
let's denote the latter by $\mathsf{DecidablyVerifiable}$.
Both may contain impossible search problems.
Both $\mathsf{Verifiable}$ and
$\mathsf{DecidablyVerifiable}$ form lattices
under the reducibility partial order, with
$\mathsf{Verifiable}$ a subset of $\mathsf{DecidablyVerifiable}$.

Let $\nu$ be an operator on the ideals of the reducibility
lattice for $\mathsf{Verifiable}$, acting as follows:
\[\nu(\mathsf{C})=\{S\in\mathsf{Verifiable}:
    \forall T\in\mathsf{DecidablyVerifable}\;
    (\forall R\in \mathsf{C}\; R\le T)\Rightarrow S\le T\}.\]
In other words, $\nu(\mathsf{C})$ contains all
verifiable search problems $S$ which reduce to each hard
problem for $\mathsf{C}$, where we allow the hard problems
to be decidably verifiable rather than just verifiable.
More generally, in this section we require problems in our ideals
to be (polytime) verifiable,
but use decidably verifiable problems as hard problems
(upper bounds) for these ideals.

\begin{lemma}
$\nu(\mathsf{C})$ is an ideal of $\mathsf{Verifiable}$ for each
ideal $\mathsf{C}$ of $\mathsf{Verifiable}$.
\end{lemma}

\begin{proof}
Since $\mathtt{triv}$ reduces to all search problems,
it reduces to all decidably verifiable upper bounds on $\mathsf{C}$,
so $\mathtt{triv}\in\nu(\mathsf{C})$ and $\nu(\mathsf{C})$ is
nonempty. If $S\in\nu(\mathsf{C})$ and $R\le S$,
then for any upper bound $T$ on $\mathsf{C}$, we have
$S\le T$ and thus $R\le T$ by transitivity. Thus
$R\in\nu(\mathsf{C})$, making it downwards closed under reductions.

Finally, if $R,S\in\nu(\mathsf{C})$, we wish to show
$R\vee S\in\nu(\mathsf{C})$. Let $T$ be any decidably verifiable
upper bound on $\mathsf{C}$. Then $R\le T$ and $S\le T$.
By the definition of join, we have $R\vee S\le T$. Since
$T$ was arbitrary, $R\vee S\in\nu(\mathsf{C})$, as desired.
\end{proof}

\begin{lemma}
$\nu$ is inflationary and dense. In addition, it is monotone,
satisfying $\nu(\mathsf{A})\subseteq\nu(\mathsf{B})$
whenever $\mathsf{A}\subseteq\mathsf{B}$.
\end{lemma}

\begin{proof}
If $S\in\mathsf{C}$ and $T$ is any decidably verifiable upper
bound on $\mathsf{C}$, then by definition we have $S\le T$;
hence $S\in\nu(\mathsf{C})$,
and so $\mathsf{C}\subseteq\nu(\mathsf{C})$, giving
the inflation property.

Observe that the search
problems in $\mathsf{FP}$ all reduce to $\mathtt{triv}$, so
$\mathtt{triv}$ is an upper bound on $\mathsf{FP}$. Thus
each problem in $\nu(\mathsf{FP})$ reduces to $\mathtt{triv}$,
so $\nu(\mathsf{FP})=\mathsf{FP}$ as desired.

Finally, suppose $\mathsf{A}\subseteq\mathsf{B}$, and let
$S\in\nu(\mathsf{A})$. Let $T$ be a decidably verifiable upper bound
on $\mathsf{B}$. Since $\mathsf{A}\subseteq\mathsf{B}$,
$T$ is an upper bound on $\mathsf{A}$, so $S\le T$. Hence
$S\in\nu(\mathsf{B})$, so $\nu(\mathsf{A})\subseteq\nu(\mathsf{B})$.
\end{proof}

\begin{lemma}
$\nu(\mathsf{A})\cap\mathsf{B}
\subseteq\nu(\mathsf{A}\cap\mathsf{B})$.
\end{lemma}

\begin{proof}
Let $S\in\nu(\mathsf{A})\cap\mathsf{B}$. Then $S\in\nu(\mathsf{A})$
and $S\in\mathsf{B}$. Let $T$ by any decidably verifiable
upper bound on $\mathsf{A}\cap\mathsf{B}$. We wish to show
$S\le T$.

For any $R\in \mathsf{A}$, since we have $S\in\mathsf{B}$,
we get $R\wedge S\in \mathsf{A}\cap\mathsf{B}$ and
hence $R\wedge S\le T$. \cref{lem:GradedDeduction}
then gives $R\le (S\to_k T)$ for some $k\in\mathbb{N}$
(despite the non-polytime-verifiability of $T$). This is true
for any $R\in\mathsf{A}$, so we get
$\mathsf{A}\subseteq\downcl\{S\to_k T\}_k$,
and thus $\mathsf{A}\subseteq\downcl\{S\to_k T\}_k\cap\mathsf{Verifiable}$.
Applying the monotonicity of $\nu$ gives
\[\nu(\mathsf{A})\subseteq\nu(\downcl\{S\to_k T\}_k\cap\mathsf{Verifiable})
=\downcl\{S\to_k T\}_k)\cap\mathsf{Verifiable},\]
where we used the theorem below. Since $S\in\nu(\mathsf{A})$,
we get $S\in\downcl\{S\to_k T\}_k$, which means there exists
$k\in\mathbb{N}$ with $S\le (S\to_k T)$. Then
$S\le (S\wedge (S\to_k T))$, and \cref{lem:GradedModusPonens}
gives $(S\wedge (S\to_k T))\le T$
(despite the non-polytime-verifiability of $T$).
Thus $S\le T$, as desired.
\end{proof}

The only missing ingredient is the following theorem:

\begin{theorem}\label{thm:STGalois}
$\nu(\downcl\{S\to_k T\}_k\cap\mathsf{Verifiable})
=\downcl\{S\to_k T\}_k\cap\mathsf{Verifiable}$.
\end{theorem}

For ease of notation, we will define
$G_{ST}=\downcl\{S\to_k T\}_k\cap\mathsf{Verifiable}$ when
$S$ is polytime verifiable and $T$ is decidably verifiable;
the theorem then says $\nu(G_{ST})=G_{ST}$.

Unfortunately, the proof is quite long.
We prove it in the next section, after developing some theory
which may be of independent interest.

Once this theorem is established, the dense weak prenucleus
properties of $\nu$ give us the conclusion that the
regular ideals of $\mathsf{Verifiable}$ are characterized
by their decidably-verifiable hard problems
(that is, they are closed under $\nu$). Moreover,
for any ideal $\mathsf{C}$ of $\mathsf{Verifiable}$,
the double refutation class $\neg\neg\mathsf{C}$
contains all verifiable search problems that reduce to each
decidably-verifiable hard problem for $\mathsf{C}$.

\section{The diagonalization argument}

\subsection{A detour on complexity class types}

We now turn to categorizing the syntactic specification level
of a complexity class of search problems. We will deal only
with polytime verifiable ($\mathsf{Verifiable}$)
problems here, though
much of the discussion can be generalized.

\paragraph{Clocked Turing machines.}
We require the following definition: a \emph{clocked Turing machine}
is formally a pair $(M,p)$ of a Turing machine $M$ and a polynomial
$p$, but we will often treat it as a single machine $V$. When
we run $V$ on an input $x$, we actually simulate $M(x)$ for up to
$p(|x|)$ steps, and then halt it and reject if $M(x)$ did not accept
within that number of steps. In other words,
a clocked Turing machine is a Turing machine that can be syntactically
seen to always halt within $p(n)$ steps for an immediately visible
polynomial $p$.

For an ideal $\mathsf{C}$ of $\mathsf{Verifiable}$
as well as a Turing machine
$M$ (possibly with an oracle), we say that $M$ enumerates $\mathsf{C}$
if it enumerates clocked Turing machines $V_1,V_2,\dots$
such that $\{L(V_i):i\in\mathbb{N}\}=\mathsf{C}$. Here
$L(V_i)$ denotes the language of $V_i$, which is the search problem
the clocked machine $V_i$ verifies. (We assume for simplicity
that $M$ also visibly modifies these clocked machines
to always reject inputs that are not pairs of strings;
such poorly-formatted inputs to the verifier
won't matter for our purposes.)

\paragraph{A syntactic hierarchy.}
We classify ideals of $\mathsf{Verifiable}$
into the following syntactic hierarchy.
\begin{itemize}
\item Level 0: principal ideals. These are classes that have a complete
problem: a single search problem $S\in\mathsf{C}$ such that everything
in $\mathsf{C}$ reduces to $S$.
\item Level 1: syntactic classes. These are
ideals $\mathsf{C}$ of $\mathsf{Verifiable}$
such that there is some enumerable set of search
problems in $\mathsf{C}$ with the property that every search problem
in $\mathsf{C}$ reduces to something in the set.
\item Level 2: ``tamely semantic'' classes. These are
ideals of $\mathsf{Verifiable}$
which can be enumerated using a Halting oracle.
In other words, there is an oracle Turing machine $M$ with a Halting
oracle which enumerates polynomial time verifiers $V_i$
such that $L(V_i)\in\mathsf{C}$ for all $i\in\mathbb{N}$ and such that
each $S\in\mathsf{C}$ occurs as $L(V_i)$ for at least one $i\in\mathbb{N}$.
\item Higher levels: we can define the third level as classes
enumerable with additional Turing jumps; for example, we give a
$\emptyset''$ oracle
(oracle to the Halting-problem-for-TMs-with-Halting-oracles)
to the enumerator for level 3 classes, and a $\emptyset'''$ for
level 4 classes, and so on.
\end{itemize}

\paragraph{Some technicalities.}
In the definition of syntactic classes, we don't require
the enumerator to enumerate the whole class,
just a set which is jointly
complete for the class. However, for the tamely semantic classes
and higher levels,
we require the enumerator to enumerate the whole class. This turns
out not to matter, because a Halting oracle is powerful enough
to verify reductions: if there is an enumerator (with Halting oracle)
for a set of problems that is jointly complete for the class $\mathsf{C}$,
then there is also an enumerator for the entire class $\mathsf{C}$
of verifiable search problems. (We omit the proof of this fact,
but see \cref{lem:TamelySemantic} for a similar argument.)

We give some examples of the different types of classes.

\begin{itemize}
\item Level 0 (complete problems): the classic $\mathsf{TFNP}$
subclasses such as $\mathsf{PPAD}$, $\mathsf{PPP}$, and $\mathsf{PLS}$
live here, since they have complete problems. Any
complexity class with a complete
problem is automatically an ideal and belongs to this level.
\item Level 1 (syntactic): if $S,T\in\mathsf{Verifiable}$,
the class $G_{ST}=\downcl\{S\to_k T\}$ is at this level.
This is because we can easily enumerate the verifiers of $S\to_k T$
using the verifiers of $S$ and $T$.
\item Level 2 (tamely semantic): many common semantic classes are at this
level. $\mathsf{TFNP}$ is at this level;
we can enumerate it by first enumerating clocked Turing
machines $V$, then
using the Halting oracle to check if there is
some $x\in\{0,1\}^*$ such that no string $y$ with $|y|\le p(|x|)$
has $V(x,y)$ accepting. By similar arguments,
$\mathsf{TFNP}\cap \mathsf{FBPP}$ (which is sometimes denoted
$\mathsf{TFZPP}$)
and $\mathsf{TFNP}\cap \mathsf{FBQP}$
are at this level.
\item Higher levels: the non-impossible problems in 
$\mathsf{Verifiable}$
(a wider class than just the $\mathsf{TFNP}$ problems)
are in level $3$,
as they require $3$ alternating quantifiers: there exists a verifier $V$
such that for all inputs $x$ there exists an output $y$ such that
$V(x,y)$ accepts. The heuristically solvable problems in
$\mathsf{FNP}$ are in level $4$: there exists a verifier $V$ such
that for all samplable distributions $D$ there exists an algorithm
$A$ such that for all input sizes $n$, $V$ accepts the solutions of $A$
on inputs of size $n$ sampled from $D$ with high probability.
\end{itemize}

We also note that each level is closed under finite joins,
which can be verified by recalling the join of two ideals is given by
$\mathsf{C}_1\vee \mathsf{C}_2
=\downcl\{R\vee S:R\in \mathsf{C}_1, S\in\mathsf{C}_2\}$,
and closed under finite intersections.\footnote{The
claim that level $0$ is closed under intersections
may be surprising to readers used to classes of decision problems;
it follows from the existence of meets in the lattice of search
problems.}

\begin{lemma}\label{lem:TamelySemantic}
If $S$ and $T$ are decidably verifiable, then
$\{R\in\mathsf{Verifiable}: \exists k\in\mathbb{N}\; R\le(S\to_k T)\}$
is tamely semantic. In particular, $G_{ST}$ is tamely semantic
for $S\in\mathsf{Verifiable}$ and $T\in\mathsf{DecidablyVerifiable}$.
\end{lemma}

\begin{proof}
Since $S$ and $T$ are decidably verifiable, fix verifiers
$V_S$ and $V_T$ for them, which always halt on every input.

Enumerate $(V,F,G,k,p)$ where $V$ is a Turing machine
representing a verifier, $F$ and $G$ are Turing machines
attempting to reduce the search problem defined by the verifier $V$
to $S\to_k T$, and $p$ is a polynomial
meant to upper bound the runtimes of these machines.
Then modify $V$ so that it always rejects inputs that are not
pairs of strings, and has a runtime upper bound of $p(n)$
(rejecting if that runtime is exceeded). Call the new machine $V'$.

Next, construct the verifier $V_{S\to_k T}$
for $S\to_k T$ by modifying the code of $V_S$ and $V_T$;
this takes finite time for each $k$
(and hence for each tuple we're enumerating). The new verifier
$V_{S\to_k T}$ always halts on every input, but not necessarily
in polynomial time.

We will now use the Halting oracle. To do this, we construct
a machine $M$ for which we want to know the
halting behavior. the machine $M$ cycles through
all pairs of strings $(x,y)$, and for each one,
it first checks if $F(x)$ fails to halt in time $p(|x|)$
or $G(x,y)$ fails to halt in time $p(|x|+|y|)$;
in that case, $M$ halts. Otherwise, $M$ runs
$V_{S\to_k T}(F(x),y)$ (which always halts in finite time)
as well as $V'(x,G(x,y))$ (which likewise always halts in finite time).
If the former accepted and the latter rejected, $M$
halts. Otherwise, $M$ proceeds to the next tuple $(x,y)$
of strings.

The enumerator over the tuples $(V,F,G,k,p)$ constructs
this $M$ and calls the Halting oracle on it. It then prints out
the clocked polynomial time verifier $V'$
if $M$ does not halt, and discards it if $M$ halts.

Since every language in the class is polytime verifiable
and reduces to $S\to_k T$ for some $k$, there is some valid
verifier $V$, reduction $(F,G)$, value of $k$, and polynomial
upper bound $p$ on the runtimes, such that the tuple $(V,F,G,k,p)$
does not have any counterexample pairs $(x,y)$; then $M$ fails
to halt on this tuple, and the enumerator outputs $V'$.
The language of $V'$ is the same as the language of $V$,
so all languages in the class are enumerated.

Conversely, if the enumerator prints out $V'$,
then by construction this is a polynomial time verifier, and since
$M$ failed to halt for the corresponding tuple $(V,F,G,k,p)$,
it necessarily means that $(F,G)$ are a polynomial time mapping reduction
from $L(V')$ to $S\to_k T$. This means $L(V')$ is in the class,
completing the proof.
\end{proof}

To show that $G_{ST}$ is preserved by $\nu$
(that is, \cref{thm:STGalois}), we will actually prove
the more general result that all tamely semantic classes
are preserved by $\nu$.

\subsection{The Galois closure theorem}

We say an ideal $\mathsf{C}$ is Galois closed if
$\nu(\mathsf{C})=\mathsf{C}$, where recall $\nu(\mathsf{C})$
is the ideal of all polytime verifiable search problems $S$
which reduce to each $\mathsf{C}$-hard decidably verifiable search
problem.

\begin{theorem}\label{thm:GaloisClosed}
Let $\mathsf{C}\subseteq\mathsf{Verifiable}$
be a tamely semantic ideal.
Then $\mathsf{C}$ is Galois closed.

In other words, let $S\in\mathsf{Verifiable}$ be a polytime
verifiable search problem. Suppose that for all decidably verifiable
search problems $T$ that are hard for $\mathsf{C}$ we have
$S\le T$. Then $S\in\mathsf{C}$.
\end{theorem}

To prove this theorem, we will require Shoenfield's limit lemma,
a standard tool in recursion theory.

\begin{lemma}[Shoenfield's limit lemma \cite{Sho59}]\label{lem:Shoenfield}
A function $r$ is computable with a Halting oracle if and only
if there is a computable function $\hat{r}$ such that
\[r(x)=\lim_{s\to\infty} \hat{r}(x,s),\]
where the limit notation means that for any fixed $x$ and
all sufficiently large
$s\in\mathbb{N}$, the value of $\hat{r}(x,s)$ is constant
and equal to $r(x)$.
\end{lemma}

\paragraph{The idea behind the diagonalization.}
The gist of our argument for proving
\cref{thm:GaloisClosed} will be as follows:
consider an enumeration of the search problems in $\mathsf{C}$,
say $S_1,S_2,\dots$. Since $\mathsf{C}$ is an ideal,
the join of the first $k$ search problems in $\mathsf{C}$ is still
in $\mathsf{C}$. However, if we take a function 
$\alpha\colon\mathbb{N}\to\mathbb{N}$ which tends to infinity,
and consider the search problem $\mathsf{C}_\alpha$
which, on input size $n$,
behaves like the join of the first $\alpha(n)$ search problems
$S_1\vee S_2\vee \dots\vee S_{\alpha(n)}$,
then this search problem will be an upper bound
on $\mathsf{C}$: every $S_k$ will reduce to it, because on large
enough input sizes $n$, we will have $\alpha(n)\ge k$ and so
$\mathsf{C}_\alpha$ will behave like a join of a set of search problems
which includes $S_k$.

We'll choose $\alpha$ which tends to infinity extremely
slowly, too slowly for any polynomial time algorithm to notice;
Then if $S$ is any problem that reduces to all upper bounds on
$\mathsf{C}$, then $S$ reduces to $\mathsf{C}_\alpha$ for this
slow-growing function, but the polynomial time reduction cannot
notice that $\alpha$ is not constant and is forced to reduce
$S$ to some finite join $S_1\vee\dots\vee S_k$ of search problems
in $\mathsf{C}$; since $\mathsf{C}$ is an ideal, we'll get 
$S\in\mathsf{C}$.

To implement this strategy, we start with the following formal
definition of $\mathsf{C}_\alpha$.

\begin{definition}
Let $\mathsf{C}\subseteq\mathsf{Verifiable}$ be a tamely semantic
ideal, and using Shoenfield's limit lemma (\cref{lem:Shoenfield})
let $\hat{r}$ be a computable function taking inputs in
$\mathbb{N}\times\mathbb{N}$ such that the limit function
$r(i)=\lim_{s\to\infty} \hat{r}(i,s)$ exists for all $i$
and the value of $r(i)=V_i$ is the $i$-th polynomial time verifier
for a language in $\mathsf{C}$ by some fixed enumeration.

For any function $\alpha\colon\mathbb{N}\to\mathbb{N}$,
define the search problem $\mathsf{C}_\alpha$ as follows.
Given a pair of strings $(z,w)$, we call the input part $z$
``good'' if the following all hold:
\begin{enumerate}
\item $z$ has the form $z=\langle x,i,0^n\rangle$ for a string
$x$ and positive integers $i$ and $n$,
\item $i\le \alpha(n)$, and
\item $V_{i,|z|}=\hat{r}(i,|z|)$ is formatted as a valid clocked Turing
machine.
\end{enumerate}
If $z$ is a bad input, we define $(z,w)\in \mathsf{C}_\alpha$
for every string $w$. If $z$ is a good input, we define 
$(z,w)\in\mathsf{C}_\alpha$ if $w$ has one of the following two forms:
\begin{enumerate}
\item $w=\langle y,0\rangle$ for a string $y$ such that
$V_{i,|z|}(x,y)$ accepts, or
\item $w=\langle m,1\rangle$ for a positive integer $m$ such
that $m\ge |z|$ and $\hat{r}(i,m)\ne \hat{r}(i,|z|)$.
\end{enumerate}
If $z$ is a good input and $w$ has neither of the above
forms, then $(z,w)\notin \mathsf{C}_\alpha$.
\end{definition}

\begin{lemma}\label{lem:CAlphaDecidablyVerifiable}
If $\alpha\colon\mathbb{N}\to\mathbb{N}$ is computable,
then the search problem $\mathsf{C}_\alpha$ is decidably
verifiable.
\end{lemma}

\begin{proof}
We describe a verifier for $\mathsf{C}_\alpha$ which
takes a pair of strings $(z,w)$ and decides if it is in
$\mathsf{C}_\alpha$.

The verifier first checks if $z$ is good. This requires it
to check the string formatting $z=\langle x,i,0^n\rangle$,
which is easy; it is then required to check $i\le \alpha(n)$,
which is possible because $\alpha$ is computable;
then it is required to check whether $V_{i,|z|}$ is formatted
like a Turing machine, which is possible since $\hat{r}(i,|z|)$
is computable. If $z$ is bad,
the verifier will accept $\langle z,w\rangle$ regardless of $w$.

Next, if $z$ is good, the verifier can check if $w$
is formatted like $\langle y,0\rangle$ or $\langle m,1\rangle$;
if not, the verifier will reject $(z,w)$. If $w=\langle y,0\rangle$,
the verifier will run
$V_{i,|z|}(x,y)$ as a clocked Turing machine; this terminates
after finitely many steps, due to the polynomial clock
that is part of the definition of $V_{i,|z|}$. The verifier
will accept $\langle z,y\rangle$ if $V_{i,|z|}(x,y)$ accepts
and reject if it rejects. If $w=\langle m,1\rangle$,
the verifier will check that $m\ge |z|$ and
will then run $\hat{r}(i,m)$, which terminates in finitely
many steps, to get $V_{i,m}$; it will then check if
$V_{i,|z|}=V_{i,m}$ and reject if equality holds while accepting
if it doesn't hold.
\end{proof}

\begin{lemma}\label{lem:CAlphaUnbounded}
If $\alpha\colon\mathbb{N}\to\mathbb{N}$ is such that
$\limsup_{n\to\infty} \alpha(n)=\infty$, then
$S\le \mathsf{C}_\alpha$ for any $S\in\mathsf{C}$.
\end{lemma}

\begin{proof}
Consider such $\alpha$ and fix $S\in\mathsf{C}$.
Let $i\in\mathbb{N}$ be such that
$V_i=\lim_{n\to\infty} \hat{r}(i,n)$ is a (clocked)
verifier for $S$.
Let $n_0$ be such that $\hat{r}(i,n)=V_i$ for all $n\ge n_0$,
and let $n\ge n_0$ be such that $\alpha(n)\ge i$
(this exists by the $\limsup$ condition on $\alpha$).
We define a reduction $(F,G)$ from $S$ to $\mathsf{C}_\alpha$
by defining $F(x)=\langle x,i,0^n\rangle$ and $G(x,\langle y,0\rangle)=y$,
with $G(x,w)$ behaving arbitrarily if $w$ is not of the form
$\langle y,0\rangle$.

Since $i$ and $n$ are constants independent of $x$,
it is clear that $F$ and $G$ run in polynomial time.
Suppose now that $(F(x),w)\in \mathsf{C}_\alpha$.
This means $(\langle x,i,0^n\rangle,w)\in \mathsf{C}_\alpha$.
The string $\langle x,i,0^n\rangle$ is good
for $\mathsf{C}_\alpha$ since by construction $i\le \alpha(n)$
and $\hat{r}(i,|z|)=V_i$ is a clocked Turing machine (noting
that $|z|\ge n\ge n_0$).
Since $(\langle x,i,0^n\rangle,w)\in\mathsf{C}_\alpha$
for a good input $\langle x,i,0^n\rangle$, it follows
that either $w=\langle y,0\rangle$ with $V_{i,|z|}(x,y)$ accepting
or $w=\langle m,1\rangle$ with $m\ge |z|$ and
$\hat{r}(i,m)\ne \hat{r}(i,|z|)$.
However, since $|z|\ge n\ge n_0$ and $m\ge |z|$, the latter cannot
occur, since $\hat{r}(i,\cdot)$ is constant on values greater
than $n_0$. Hence $w=\langle y,0\rangle$ with
$V_{i,|z|}=V_i$ accepting the pair $(x,y)$. Since $V_i$
is a clocked verifier for $S$, we must have $(x,y)\in S$.
By definition of $G$, we get
$(x,G(x,w))\in S$, which proves the reduction condition.
\end{proof}

\begin{lemma}\label{lem:CAlphaImpossible}
If $\mathsf{C}$ does not contain impossible problems, then
for every function $\alpha\colon\mathbb{N}\to\mathbb{N}$,
the search problem $\mathsf{C}_\alpha$ is not impossible.
\end{lemma}

\begin{proof}
We must show that for every string $z$ there exists a string
$w$ such that $(z,w)\in \mathsf{C}_\alpha$. If $z$ is bad,
then $(z,w)\in\mathsf{C}_\alpha$ for every value of $w$,
so suppose $z=\langle x,i,0^n\rangle$ is good.

If $\hat{r}(i,|z|)=r(i)$ is equal to its limiting value,
then its value $\hat{r}(i,|z|)$ is a clocked verifier
$V_i$ for some search problem in $\mathsf{C}$; in that case,
since the class does not contain impossible search problems,
we know that $L(V_i)$ is possible, so there must exist
some string $y$ such that $(x,y)\in L(V_i)$. Then
$V_i(x,y)$ accepts, which means $w=\langle y,0\rangle$
is a valid output for $z$ and $(z,w)\in\mathsf{C}_\alpha$.

If $\hat{r}(i,|z|)=V_{i,|z|}$ does not equal $r(i)=V_i$,
then $\hat{r}(i,\cdot)$ did not yet reach its limiting
value at $|z|$, so there is some $m\ge |z|$ with
$\hat{r}(i,m)=V_i\ne \hat{r}(i,|z|)$. In that case,
$w=\langle m,1\rangle$ is a valid output for $z$,
and $(z,w)\in\mathsf{C}_\alpha$.
\end{proof}

\begin{lemma}\label{lem:join4}
The following is a valid finite join of search problems
$S_1,S_2,\dots,S_n$: on input $\langle x,i\rangle$
with $i\le n$, output $y$ such that $(x,y)\in S_i$
(on inputs not of the form $\langle x,i\rangle$ with
$i\le n$, output anything). This join preserves verifiability.
\end{lemma}

We already saw that binary joins exist, so finite joins
exist too; this lemma just gives a convenient equivalent
form of the finite join.

\begin{proof}
The proof is identical to the binary case; we omit it.
\end{proof}

\begin{lemma}\label{lem:CAlphaBounded}
If $\alpha\colon\mathbb{N}\to\mathbb{N}$ is bounded above,
then for any polytime verifiable
search problem $R$ with $R\le \mathsf{C}_\alpha$
we have $R\in \mathsf{C}$.
\end{lemma}

\begin{proof}
First, recall every search problem is reducible to an
impossible search problem; since $\mathsf{C}$ is downwards
closed under reductions, if it contains an impossible search problem
then it contains
all polytime verifiable search problems, and so $R\in\mathsf{C}$.
Suppose then that $\mathsf{C}$ does not contain impossible
problems; by \cref{lem:CAlphaImpossible}, $\mathsf{C}_\alpha$
is not impossible for any $\alpha$.

Let $\alpha$ be bounded above, and let $k$ be its maximum
value. Let $n_0$ be large enough that for all $n\ge n_0$
and all $i=1,2,\dots,k$, we have $\hat{r}(i,n)=r(i)$
(i.e. $\hat{r}(i,\cdot)$ reached its limiting value for all
$n\ge n_0$ and all $i\le k$).

Consider the first $k$ search problems
$S_1,S_2,\dots,S_k\in\mathsf{C}$ whose (clocked)
verifiers are printed out by $r(1),r(2),\dots,r(k)$
(this list of search problems is allowed to have repetitions).
Since $\mathsf{C}$ is an ideal, we have
$S_1\vee S_2\vee\dots\vee S_k\in \mathsf{C}$.
Fix polytime verifiable $R$ such that $R\le\mathsf{C}_\alpha$.
Let $(F',G')$ be the reduction from $R$ to $\mathsf{C}_\alpha$.
We give a reduction $(F,G)$ from $R$
to $S_1\vee\dots\vee S_k$.

Define $F(z)$ as follows. If $F'(z)=\langle x,i,0^n\rangle$
for some $i,n$ with $i\le k$, then $F(z)$
outputs $\langle x,i\rangle$. Otherwise, $F(z)$ outputs
arbitrarily. Since $F'$ runs in polynomial time, so does $F$.

Define $G(z,y)$ as follows. It first runs $F'(z)$
and checks whether $|F'(z)|\ge n_0$. If so,
$G(z,y)$ outputs $G'(z,\langle y,0\rangle)$.
Otherwise, if $|F'(z)|<n_0$, then $F'(z)$ is one of finitely
many strings; for each of those strings, the code of $G$
stores a valid output for $\mathsf{C}_\alpha$
(which is guaranteed to exist as $\mathsf{C}_\alpha$
is not impossible). $G(z,y)$ then does a lookup to find
a valid output $s$ such that $(F'(z),s)\in\mathsf{C}_\alpha$.
It outputs $G'(z,s)$. Since $F'$ and $G'$ run in polynomial
time, and since $n_0$ is an absolute constant, $G'$ runs
in polynomial time.

We now show $(F,G)$ is a valid reduction from $R$
to $S_1\vee\dots\vee S_k$.
Suppose $(F(z),y)\in S_1\vee\dots\vee S_k$.
If $F'(z)$ is a bad input for $\mathsf{C}_\alpha$,
then $(F'(z),w)\in\mathsf{C}_\alpha$
for $w=\langle y,0\rangle$
(since that holds for any output string $w$ on poorly formatted inputs).
In that case, the reduction property of $F',G'$
gives $(z,G'(z,w))\in R$, and so $(z,G(z,y))\in R$ as desired.

Suppose then that $F'(z)=\langle x,i,0^n\rangle$ is good
for $\mathsf{C}_\alpha$. In particular, $i\le \alpha(n)$
(by the definition of good inputs) and $\alpha(n)\le k$
(since $k$ is the maximum of $\alpha$), and the definition of $F$
gives $F(z)=\langle x,i\rangle$.
Our assumption was $(F(z),y)\in S_1\vee\dots\vee S_k$,
which gives $(\langle x,i\rangle,y)\in S_1\vee \dots\vee S_k$.
By definition of the finite join in \cref{lem:join4},
this gives $(x,y)\in S_i$, and so $V_i(x,y)$ accepts.

We split into two cases: either $|F'(z)|\ge n_0$ or
$|F'(z)|<n_0$. Consider the former case first. Then
since $i\le k$ and $|F'(z)|\ge n_0$, we have
$\hat{r}(i,|F'(z)|)=V_i$ for the limiting value
$V_i=r(i)$ (this is by the choice of $n_0$).
Since this machine accepts $(x,y)$, by definition of
$\mathsf{C}_\alpha$ we get that $w=\langle y,0\rangle$
is a valid output for the input $F'(z)=\langle x,i,0^n\rangle$,
meaning $(F'(z),w)\in\mathsf{C}_\alpha$. By the reduction
property of $F',G'$, we then get $(z,G'(z,w))\in R$,
so $(z,G(z,y))\in R$, as desired.

Finally, suppose that $|F'(z)|<n_0$. Then $G(z,y)$
looks up a string $s$ such that $(F'(z),s)\in \mathsf{C}_\alpha$.
The reduction property of $F',G'$ then gives
$(z,G'(z,s))\in R$, which means $(z,G(z,y))\in R$, as desired.
\end{proof}

Recall the properties of $\mathsf{C}_\alpha$ collected so far:
if $\mathsf{C}$ has no impossible problems, then $\mathsf{C}_\alpha$
is not impossible; if $\alpha$ is computable, $\mathsf{C}_\alpha$
is decidably verifiable; if $\alpha$ is bounded, then
verifiable problems which reduce to $\mathsf{C}_\alpha$ are in
$\mathsf{C}$; and if $\alpha$ is not bounded, then every problem
in $\mathsf{C}$ reduces to it. To complete the proof of
\cref{thm:GaloisClosed}, we just need to pick $\hat{\alpha}$
which is computable and straddles the line between bounded
and unbounded such that no polynomial time reduction can tell which;
this forces $\mathsf{C}_\alpha$ to have both the property that
it upper bounds $\mathsf{C}$ and the property that everything
below it is in $\mathsf{C}$ at the same time. (Actually, we can't
quite do this; instead, our definition of this straddling $\hat{\alpha}$
will depend on the search problem $S$ that we want to show is in
$\mathsf{C}$.)

\begin{proof}[Proof of \cref{thm:GaloisClosed}]
Fix a polytime verifiable search problem $S$ which is reducible
to every decidably verifiable $\mathsf{C}$-hard search problem.
Then in particular, $S$ reduces to $\mathsf{C}_\alpha$
whenever $\alpha$ is a computable, unbounded function.
We now construct a particular
such $\hat{\alpha}$ which essentially diagonalizes the reductions
that go from $S$ to problems of the form $\mathsf{C}_\alpha$.

Define $\hat{\alpha}(n)$ via the following computation.
For each $i=1,2,\dots$ do the following.
Enumerate the first $i$ tuples $(F,G,p)$, and for each,
attempt to disprove the claim that $F$ and $G$ are polynomial
time reductions (running in $p(n)$ steps on inputs of size $n$)
from $S$ to $\mathsf{C}_i$, where here $\mathsf{C}_i$ is the
search problem $\mathsf{C}_\alpha$ for the constant
function $\alpha(n)=i$. The attempted disproof works by
enumerating all strings pairs $(z,w)$ with $z$ and $w$
of length at most $\log_2 n$, and for each pair, first checking
the runtime of $F(z)$ and $G(z,w)$ (if it exceeds the bound
set by $p$, the tuple is disproved). If this check passes,
the tuple $(F,G,p)$ may still be disproved by the pair $(z,w)$
if the reduction property fails; we check this by
running $V_S(z,G(z,w))$ to test if $(z,G(z,w))\in S$,
and running the decidable verifier for $\mathsf{C}_i$
on $(F(z),w)$ to test if $(F(z),w)\in \mathsf{C}_i$
(note that we can computably do this since this search problem
is decidably verifiable by \cref{lem:CAlphaDecidablyVerifiable}).
If we successfully eliminated all of the first $i$ tuples
$(F,G,p)$ using pairs of strings of length at most $\log_2 n$,
we increment $i$ and repeat the process.
Eventually, either $i$ exceeds $n$ or else we find the first
$i$ for which this process fails to eliminate the first $i$
reduction tuples $(F,G,p)$. If $i$ exceeds $n$, we output $n$;
if we get stuck at $i$, we output $i-1$ (the last value of $i$ for
which we've succeeded).

This function $\hat{\alpha}(n)$ is computable by construction.
There are two cases: either
$\limsup_{n\to\infty}\hat{\alpha}(n)=\infty$, or else
$\hat{\alpha}$ is bounded.

Suppose $\hat{\alpha}$ is bounded above by $k$.
Then for all values of $n$, we failed to eliminate the first
$k+1$ reductions in their claim of reducing $S$ to
$\mathsf{C}_{k+1}$ using strings of length at most $n$.
In particular, there must be at least one reduction tuple
$(F,G,p)$ within the first $k+1$ tuples which successfully
reduces $S$ to $\mathsf{C}_{k+1}$. Since the constant function
$\alpha(n)=k+1$ is bounded and $S\le \mathsf{C}_{k+1}$,
\cref{lem:CAlphaBounded} says that $S$ must be in $\mathsf{C}$,
as desired.

Suppose then that $\limsup_{n\to\infty}\hat{\alpha}(n)=\infty$.
In that case, by \cref{lem:CAlphaUnbounded}, we get
that $\mathsf{C}_{\hat{\alpha}}$ is hard for the
class $\mathsf{C}$.
By our assumption on $S$, we then get
$S\le \mathsf{C}_{\hat{\alpha}}$.
This means there is some tuple $(F,G,p)$ which is a valid
reduction from $S$ to $\mathsf{C}_{\hat{\alpha}}$; let $i$ be the
index of this tuple in our enumeration. Recall
$\limsup_{n\to\infty}\hat{\alpha}(n)=\infty$.
Let $n_1$ be the first time a value greater than $i$
occurs in the sequence $\alpha(n)$, so $\alpha(n_1)=i_1>i$
and $\alpha(n)\le i$ for all $n<n_1$. Then let $n_2$
be the first time a value greater than $i_1$ occurs in the sequence,
so $\alpha(n_2)=i_2>i_1$ and $\alpha(n)\le i_1$ for all
$n<n_2$. Continue this way for $n_3,n_4,\dots$
and $i_3,i_4,\dots$. Set $i_0=i$.

For each value of $k\ge 1$,
let $(z_k,w_k)$ be a pair of strings of length at most
$\log_2 n_k$ which disproves the claim that the reduction
$(F,G,p)$ reduces $S$ to $\mathsf{C}_{i_k}$.

Note that by our assumption, $F$ and $G$ run in time $p$
on all inputs. The condition that $(z_k, w_k)$ disproves
the reduction therefore amounts to the claim that
$(F(z_k),w_k)\in \mathsf{C}_{i_k}$ but
$(z_k,G(z_k,w_k))\notin S$; this must hold for all $k\ge 1$.
However, since $(F,G)$ is a reduction from $S$ to 
$\mathsf{C}_{\hat{\alpha}}$, the contrapositive of the reduction
condition gives $(F(z_k),w_k)\notin \mathsf{C}_{\hat{\alpha}}$
for all $k$ (since $(z_k,G(z_k,w_k))\notin S$).
Note that the way valid outputs
are defined in the definition of $\mathsf{C}_\alpha$ does not
depend on $\alpha$ if the input is good, and also not if the
input is bad; $\alpha$ only affects whether the input is good or bad.
Since $(F(z_k),w_k)\notin \mathsf{C}_{\hat{\alpha}}$
and $(F(z_k),w_k)\in \mathsf{C}_{i_k}$,
the only possibility is that $F(z_k)$ is bad for
$\mathsf{C}_{i_k}$ and good for $\mathsf{C}_{\hat{\alpha}}$.
Examining the definition of a good input reveals that the only
possibility is $F(z_k)=\langle x_k, j_k,0^{\ell_k}\rangle$
with $j_k\le \hat{\alpha}(\ell_k)$ and $j_k> i_k$.
In particular, $\hat{\alpha}(\ell_k)>i_k=\alpha(n_k)$.
Since $\alpha(\ell_k)>i_k$ and the first time a value
greater than $i_k$ occurs in the $\alpha(n)$ sequence
is $n_{k+1}$, we have $\ell_k\ge n_{k+1}>n_k$.
In particular, $|F(z_k)|>n_k$. However, we had $|z_k|\le \log_2 n_k$.
Moreover, since $\ell_k>n_k$ and $n_k$ increases with $k$,
there must be infinitely many distinct strings $z_k$.
Hence on infinitely many strings $z$ we have
$|F(z)|\ge 2^{|z|}$. This means $F$ must take exponential time,
which is a contradiction.
\end{proof}

\cref{thm:STGalois} now follows from \cref{thm:GaloisClosed}
and \cref{lem:TamelySemantic}. Note that we proved a much more general
result: all tamely semantic ideals (including many common semantic
classes, such as $\mathsf{TFNP}$, $\mathsf{FBQP}\cap\mathsf{TFNP}$,
etc.) are characterized by their decidably-verifiable hard problems.
We think this structural complexity result is of independent interest.

\ifanon
\else
\phantomsection\addcontentsline{toc}{section}{Acknowledgements}
\section*{Acknowledgements}
This research is supported in part by the Natural Sciences and
Engineering Research Council of Canada (NSERC), DGECR-2019-00027 and
RGPIN-2019-04804.\footnote{Cette recherche a été financée par le
Conseil de recherches en sciences naturelles et en génie du Canada
(CRSNG), DGECR-2019-00027 et RGPIN-2019-04804.}
\fi


\printArxivBibliography

\end{document}